\documentclass[reprint,aps,pra,notitlepage,nofootinbib,longbibliography,floatfix,superscriptaddress]{revtex4-2}

\usepackage[utf8]{inputenc}
\usepackage{float}
\usepackage{graphicx}
\usepackage{adjustbox}
\usepackage{comment}
\usepackage{physics}
\usepackage{esint}
\usepackage{amssymb}  
\usepackage{mathtools}
\usepackage{amsmath}
\usepackage{array}
\usepackage{amsthm}
\usepackage{mathrsfs}
\usepackage{booktabs}
\usepackage{multirow}
\usepackage[dvipsnames]{color,xcolor,colortbl}
\usepackage[colorlinks=true,linkcolor=NavyBlue,citecolor=NavyBlue,urlcolor=NavyBlue,plainpages=false,pdfpagelabels]{hyperref}
\usepackage{bm}
\usepackage[normalem]{ulem}
\usepackage{tikz}
\usetikzlibrary{tikzmark,calc}
\usepackage{cancel}

\makeatletter 

\renewcommand\onecolumngrid{%
	\do@columngrid{one}{\@ne}%
	\def\set@footnotewidth{\onecolumngrid}%
	\def\footnoterule{\kern-6pt\hrule width 1.5in\kern6pt}%
}

\renewcommand\twocolumngrid{%
	\def\footnoterule{
		\dimen@\skip\footins\divide\dimen@\thr@@
		\kern-\dimen@\hrule width.5in\kern\dimen@}
	\do@columngrid{mlt}{\tw@}
}%

\makeatother    

\DeclareMathOperator{\sgn}{sgn}

\newcommand{\intsum}{\,\,\,\,\,\mathclap{\displaystyle\int_{\boldsymbol{\nu}}}\mathclap{\textstyle\sum\,}\,\,\,}

\newcommand{\intsumTEXT}
{\,\,\,\,\,\mathclap{\scalebox{.8}{$\displaystyle\int$}}\hspace{0.5pt}\mathclap{\scalebox{.85}{$\textstyle\sum$}\,}\,\,}

\newcommand{\intsumtwo}{\,\,\,\,\,\mathclap{\displaystyle\int_{\boldsymbol{\nu'}}}\mathclap{\textstyle\sum\,}\,\,\,}

\renewcommand{\dd}{\mathrm{d}}
\newcommand{\ee}{\mathrm{e}}
\newcommand{\ci}{\mathrm{i}}
\newcommand{\bn}{\mathbf{n}}

\newcommand{\bnu}{\boldsymbol{\nu}}

\newcommand{\lr}[1]{\left({#1}\right)}
\newcommand{\lrsq}[1]{\left[{#1}\right]}

\renewcommand{\ev}[1]{\langle{#1}\rangle}

\begin{document}
 
\title{Collective excitations in quantum gravity condensates}

\author{Andrea Calcinari}
\email{andrcalc@ucm.es}
\affiliation{Departamento de F\'isica Te\'orica and IPARCOS, Facultad de Ciencias F\'isicas, Universidad Complutense de Madrid, Plaza de Ciencias 1, 28040 Madrid, Spain}

\author{Adri\`a Delhom}
\email[Corresponding author: ]
{adria.delhom@lkb.upmc.fr}
\affiliation{Departamento de F\'isica Te\'orica and IPARCOS, Facultad de Ciencias F\'isicas, Universidad Complutense de Madrid, Plaza de Ciencias 1, 28040 Madrid, Spain}
\affiliation{Laboratoire Kastler Brossel, Sorbonne Université, CNRS, ENS-Université PSL, Collège de France, 4 Place Jussieu, 75005 Paris, France}

\author{Daniele Oriti}
\email{doriti@ucm.es}
\affiliation{Departamento de F\'isica Te\'orica and IPARCOS, Facultad de Ciencias F\'isicas, Universidad Complutense de Madrid, Plaza de Ciencias 1, 28040 Madrid, Spain}

\begin{abstract}

A central open problem in quantum gravity is to understand how continuum spacetime emerges from quantum-geometric degrees of freedom in a background-independent setting. A many-body perspective suggests that spacetime emerges as a hydrodynamic phase of many atoms of quantum geometry. This idea underlies several approaches to quantum gravity, and it has been explicitly realised in the group field theory formalism. However, quantum fluctuations beyond the mean-field regime remain largely unexplored. We fill this gap by importing Bogolyubov theory to quantum gravity condensates, showing that leading beyond-mean-field effects manifest as collective excitations, in direct analogy with phonons in laboratory Bose--Einstein condensates. We implement the construction in a tractable group field theory model, where condensates of quantum-geometric atoms reproduce nonsingular expanding cosmologies, and derive the leading beyond-mean-field corrections to the emergent Friedmann dynamics. These results identify a new class of quantum-gravity excitations and establish a controlled bridge between microscopic quantum-gravitational dynamics, many-body collective phenomena, and signatures of spacetime emergence.
\end{abstract}

\maketitle

\section{Introduction}

The problem of quantum gravity is often posed as the task of merging general relativity with quantum mechanics \cite{Carlip:2001wq}. This task involves a large number of interrelated technical and conceptual challenges \cite{deBoer:2022zka}, as well as the goal of predicting new phenomenology \cite{Addazi:2021xuf}. As a result, a variety of approaches are being developed by a growing community \cite{Buoninfante:2024yth}. These diverge both in mathematical formalisms and foundational perspectives: from attempts to quantise general relativity within standard field-theoretic approaches \cite{Donoghue:1994dn,Donoghue2024,Eichhorn:2018yfc}, to radical strategies based on novel kinds of fundamental structures, often of discrete type \cite{Ashtekar:2021kfp,Loll:2019rdj,ThiemannBook,Surya:2019ndm}. In the latter case, the smooth spacetime and fields we are accustomed to are viewed as emergent properties of more fundamental frameworks \cite{Oriti:2006ar,Oriti:2018dsg,Oriti:2013jga}. 

A central challenge in quantum gravity is not only to define the microscopic constituents of the theory, but also to identify and control the collective regime in which an effective spacetime description becomes valid \cite{Hu:2005ub,Sindoni:2011ej,BenAchour:2024gir}.

In many ways, the problem of spacetime emergence is structurally analogous to a many-body problem. One seeks to describe the macroscopic, hydrodynamic regime of a system whose microscopic constituents need not resemble the continuum they collectively produce \cite{Anderson:1972pca,CollectivePhenomenaBook,Morrison_2012}. In ordinary many-body physics, the relevant large-scale observables often describe collective regimes rather than properties of individual constituents. Examples range from fluids and superfluids to ordered phases of matter, where effective continuum descriptions emerge only after coarse graining many microscopic degrees of freedom \cite{Altland:2006si,Coleman_2015}. The same lesson suggests that continuum spacetime, if emergent, should be understood as a collective phase of an underlying quantum system of non-spatiotemporal nature. However, unlike in ordinary many-body systems, no background geometry is available \emph{a priori} to understand the fundamental degrees of freedom or to guide the coarse-graining procedure in quantum gravity \cite{BenAchour:2024gir}.

Existing constructions of emergent spacetime in quantum gravity focus on the leading mean-field hydrodynamic regime \cite{Ambjorn:2004qm,Taveras:2008ke,Steinacker:2010rh,Alesci2,Alesci3,GFTquantumST_Oriti,GFTcosmoLONGpaper,Gielen_2016}. This is the natural first step: one identifies macroscopic observables of collective nature, and derives their effective dynamics as a mean field equation. In cosmological setups, for instance, this collective regime is naturally described by hydrodynamics on minisuperspace, which describes cosmological dynamics as a fluid evolving in the space of possible homogeneous geometries and matter field configurations \cite{Oriti:2024qav}.

However, a genuine many-body system is not exhausted by its mean field. The leading corrections are often not microscopic single-particle excitations, but collective excitations of the whole system \cite{Altland:2006si,Coleman_2015}. Understanding these collective degrees of freedom is essential both for controlling the validity of the continuum approximation, and for identifying possible traces of the microscopic theory in the emergent---continuum---spacetime regime. {These degrees of freedom have never been identified within background-independent quantum gravity: their spectrum, their dynamics, and their imprint on macroscopic observables are all unknown. Filling this gap is the purpose of the present work.}

Paradigmatic systems to draw lessons from are Bose--Einstein condensates (BECs), where leading corrections to the mean-field dynamics are described in terms of collective Bogolyubov excitations rather than excitations of individual atoms \cite{Bogolyubov:1947zz,pitaevskii2016bose}. If spacetime is similarly a condensate-like quantum many-body phase, it is natural to ask whether its quantum fluctuations admit an analogous collective description. 

In this work, we leverage tools from many-body physics to address this issue within background-independent quantum gravity. In particular, we import{---for the first time---}the Bogolyubov framework into quantum gravity approaches where spacetime emerges as a mean-field condensed phase of a many-body system of quantum-geometric atoms. 
 
Our main result is that interactions among quantum gravity atoms reorganise leading corrections to the mean-field dynamics into collective excitations of the emergent spacetime. We also show that interactions induce quantum depletion: even the state with no collective excitations contains out-of-condensate quantum-geometric atoms. Both results are analogous to phonons and depletion in atomic Bose gases \cite{pitaevskii2016bose}.

To illustrate the framework, we apply it to a concrete background-independent quantum gravity model in which a many-body formulation is fully developed, known as group field theory (GFT) \cite{Oriti:2006se,FreidelGFT}. GFTs are formulated as quantum field theories (QFTs) defined on abstract group manifolds, rather than on spacetime. They typically involve non-local interactions, and their quanta admit an interpretation in terms of building blocks of quantum geometry \cite{Oriti_GFTandLQG}, with the corresponding Feynman diagrams being dual to discrete spacetime histories \cite{DePietri:1999bx,Reisenberger_2000}. GFT can be viewed as a second quantisation of loop quantum gravity \cite{Oriti_GFT2ndLQG} and a completion of spin foam models \cite{Perez:2012wv,Perez:2003vx,Oriti:2011jm} and of lattice gravity path integrals \cite{Bonzom:2009hw,Baratin:2011tx,Baratin:2011hp,Hamber:2009mt,Loll:1998aj}, {providing a prescription for summing over all the discrete structures on which transition amplitudes are defined}.

In this framework, continuum spacetime emerges as a mean-field hydrodynamic phase of a system of many interacting GFT quanta \cite{GFTquantumST_Oriti,GFTcosmoLONGpaper,Gielen_2016}. This many-body perspective has motivated the study of renormalisation, phase structure, and collective behaviour in GFT, leading to the understanding that the physically relevant macroscopic regime of the theory is a condensate phase \cite{ORITI2017235,Oriti:2024qav}. Explicit results exist in cosmological setups, where GFT condensate cosmology describes homogeneous universes as many-body states in which a single quantum-geometric mode is macroscopically occupied \cite{GFTcosmoLONGpaper,Gielen_2016,Oriti_2016}. The relational evolution of the corresponding volume observable reproduces effective Friedmann dynamics at large volume and replaces the classical Big Bang singularity with a quantum bounce \cite{Oriti_2016,BOriti_2017,relham_Wilson_Ewing_2019}. {The framework has since been developed in several directions---anisotropies, accelerated expansion, cosmological perturbations and inhomogeneities \cite{Aniso,OritiPang,Marchetti:2021gcv,Jercher:2023kfr,Gielen:2025jcb}---but always as refinements of the mean field itself, leaving the collective excitations of the underlying many-body state, and their dynamics, unexplored.}

GFT thus supplies precisely the ingredients needed for a controlled many-body analysis of emergent spacetime: microscopic quanta, a Fock-space structure, condensate states, and macroscopic geometric observables. 

We work within a simplified but fully controlled GFT setting: a deparametrised Abelian model with local interactions, coupled to a free massless scalar field used as relational clock. We focus on a regime where one GFT mode is macroscopically occupied while interactions remain perturbative. We expand around the condensate background and truncate the dynamics at quadratic order in the non-condensed modes. We diagonalise the resulting Hamiltonian and identify the corresponding normal modes as collective excitations of the emergent spacetime. We then derive their relational dynamics, and identify the conditions under which the beyond-mean-field corrections remain perturbatively controlled. To conclude, we analyse how these collective excitations modify macroscopic observables within a GFT condensate cosmology setup. 

{Our results provide the first beyond-mean-field description of an emergent spacetime: we identify the collective sector, derive its relational dynamics, and compute its imprint on a macroscopic continuum spacetime observable.} Thus, we establish a concrete bridge between microscopic quantum-gravity dynamics, many-body collective excitations, and spacetime emergence. 

The paper is organised as follows. In section \ref{sec:QGcosmo} we introduce the framework in general terms, focusing on the emergence of homogeneous cosmology as the simplest hydrodynamic regime of a many-body quantum gravity system. We then discuss how GFT naturally accommodates our framework, allowing to perform concrete analytic calculations within a specific model of emergent condensate cosmology. In section \ref{sec:BogoGFT} we include weak interactions, perform the quadratic expansion around the condensate background, and diagonalise the fluctuation Hamiltonian of the model in terms of collective modes. In section \ref{sec:EvolutionofBogolons} we study the relational dynamics of these collective excitations, including adiabatic and non-adiabatic effects, and analyse the regime of validity of the Bogolyubov approximation. In section \ref{sec:GFTbogocosmo} we derive the corresponding beyond-mean-field corrections to the effective Friedmann dynamics. Appendix~\ref{app:GFTAppendix} provides an overview on GFT, and Appendices~\ref{app:AppDiag} and~\ref{app:ApproximationVaidity} contain the technical details of the diagonalisation procedure and the regime of validity of the Bogolyubov approximation.

\section{Quantum gravity condensates and emergent cosmology}\label{sec:QGcosmo}

Several quantum gravity approaches see the spacetime continuum as emergent in a coarse-grained regime of a more fundamental background-independent theory. Examples include loop quantum gravity \cite{ThiemannBook,Ashtekar:2021kfp}, spin foam models \cite{Perez:2012wv,Perez:2003vx}, group field theory \cite{Oriti:2006se,FreidelGFT}, tensor models \cite{DiFrancesco:1993cyw,Gurau:2011xp}, causal dynamical triangulations \cite{Ambjorn:2004qm,Loll:2019rdj}, causal sets \cite{Surya:2019ndm} or graphity \cite{Graphity} (see also \cite{Trugenberger:2015xma,Trugenberger:2016vhs,Trugenberger:2016viw} for a combinatorial approach to quantum gravity). In most cases, the building blocks are quantum-geometric degrees of freedom, and the coarse graining is realised through a mean field approximation. 

As in any coarse-graining procedure, the amount of macroscopic variables needed to describe such continuum limit depends on the complexity of the physical regime one intends to capture. For a homogeneous cosmological sector, the result of this coarse graining is naturally described by hydrodynamics in minisuperspace, where the relevant continuum geometric degrees of freedom are encoded in a small set of collective variables \cite{Oriti:2024qav}. From the emergent spacetime point of view, the most coarse-grained description corresponds to a homogeneous and isotropic continuum universe, characterised only by its volume.  

A simple way of making this idea more concrete is to assume that the microscopic theory admits an observable $\hat V$ which, in the coarse-grained regime, corresponds to a volume observable. In the emergent spacetime regime, the macroscopic volume of a homogeneous universe is then identified with the expectation value
\begin{equation}
    V \coloneqq \langle \hat V\rangle \,,
\end{equation}
evaluated on a suitable condensate-like state, in which relative fluctuations of $\hat V$ are sufficiently small to allow a meaningful interpretation of $V$ as a continuum geometric observable. It is in this sense that cosmological spacetime emerges as a condensed phase in a coarse-grained regime of the fundamental theory \cite{Hu:2005ub,Sindoni:2011ej}. The resulting continuum dynamics has recently been formulated as hydrodynamics on minisuperspace \cite{Oriti:2024qav}. 

Within the microscopic theory, the quantum-geometric atoms can be thought of as \textit{carrying} spatial volume. The volume observable is then a generic one-body operator
\begin{equation}
\hat V
=
\sum_{ij}
\mathfrak{v}_{ij}\,
\hat a_i^\dagger\hat a_j\,,
\label{eq:GenVolOp}
\end{equation}
where the indices $i, j$ label a complete basis of the single-particle Hilbert space and $\mathfrak{v}_{ij}=\mathfrak{v}_{ji}^*$ are the matrix elements of the volume operator in that basis. Physically, \eqref{eq:GenVolOp} accounts for the geometric contribution carried by quantum-gravity atoms, with possible coherences between different one-particle states. A simple recipe to take the coarse-graining limit is then given by
\begin{equation}
 V
=
\sum_{ij}
\mathfrak{v}_{ij}\,
\langle\hat a_i^\dagger\hat a_j\rangle\,.
\label{eq:general_volume_operator}
\end{equation}
This expresses the macroscopic volume of the quantum gravity condensate as an extensive collective observable, built from the contributions of many microscopic quantum-geometric degrees of freedom. 

In order to describe cosmological dynamics in terms of the mean-field volume \eqref{eq:general_volume_operator}, one needs to incorporate a notion of evolution into the model. Within background-independent approaches, this is often realised through a relational strategy \cite{Tambornino,Trinity,Goeller:2022rsx}. This approach consists of choosing internal degrees of freedom---typically massless scalar fields---to play the role of (quantum) reference frames, providing a {definition} of physical spacetime localisation. 

Isotropic and homogeneous cosmology is described via the Friedmann--Lema\^itre--Robertson--Walker (FLRW) metric. If we assume the existence of a massless scalar field $\chi$ which we choose as relational clock, the dynamics of the universe is then encoded in the volume observable\footnote{This is defined via a reference fiducial cell of volume $V_0$ and the scale factor characterising its expansion as $V=V_0 a^3(\chi)$.} $V(\chi)$. The relational cosmological dynamics of a FLRW universe are then given by
\begin{equation}
\label{eq:RelationalFLRW}
\lr{\frac{1}{V}\frac{\dd V}{\dd\chi}}^2=12\pi G + \text{extra content} \,,
\end{equation}
where the right-hand side can feature the contribution of extra fields \cite{Ladstatter:2025kgu}---or other gravitational degrees of freedom such as anisotropies \cite{Aniso}---with their energy density expressed with respect to the relational clock. Within classical general relativity, the expanding cosmologies described by this equation suffer from a singularity at early times, known as the Big Bang singularity. This is expected to be resolved by successful quantum gravity theories. There are examples of quantum gravity approaches which predict robust singularity resolution interpolating with the above FLRW dynamics; these include loop quantum cosmology \cite{Ashtekar:2006uz,Ashtekar:2007em,Bojowald:2001xe} and GFT \cite{Oriti_2016,BOriti_2017,toy,relham_Wilson_Ewing_2019} (see also \cite{Khoury:2001bz,Gasperini:2002bn,Cornalba:2002fi} for  string cosmology results).

We can now go back to the picture described above. Assuming that we formulate dynamics relationally in our quantum gravity model of choice, the simplest description of a viable emergent universe, in the coarse-grained sense we discussed, is attained if the evolution of $\langle \hat{V}\rangle$ is such that the Big Bang singularity is resolved, and it is well approximated by equation \eqref{eq:RelationalFLRW} for large volumes.

Crucially, the physical validity of a smooth geometric interpretation relies on the existence of a hydrodynamic regime, where quantum fluctuations over the mean field can be neglected. Taking lessons from condensed matter, a natural class of states achieving this are condensate states, namely states with a macroscopic occupation number in a single mode of the system. The transition from a discrete geometry to continuous spacetime is understood as a collective phenomenon, akin to macroscopic states in condensed matter physics. Large-scale spacetime here corresponds to a continuum, hydrodynamic phase of an underlying \textit{quantum fluid of geometry} \cite{Oriti:2024qav}.

At the same time, this perspective makes clear that the mean-field description cannot be exact. In interacting many-body systems, the leading quantum corrections to a condensate capture key aspects of collective physics: these are not arbitrary microscopic excitations, but collective modes of the system itself that survive coarse graining in the presence of interactions. The central questions we address are therefore whether fluctuations around a quantum-gravity condensate admit an analogous organisation, whether they remain under control during cosmological evolution, and what signatures they may imprint on the emergent continuum spacetime.

To answer such questions, we will focus on a tractable background-independent quantum gravity model in the following subsection. We will use the model to illustrate how quantum fluctuations in interacting quantum gravity condensates can be taken into account via collective Bogolyubov excitations, in analogy with phonons in BECs.

\subsection{A specific model: GFT condensate cosmology}\label{sec:GFTcosmo}

We now provide a concrete realisation of the paradigm discussed above in the group field theory (GFT) approach to quantum gravity. Here, we only discuss the main results of the formalism needed to understand our work. For an overview of the relevant aspects we refer the interested reader to Appendix \ref{app:GFTAppendix} and references therein. 

GFT provides a field-theoretic framework for quantum gravity in which the fundamental variables are fields over group manifolds rather than spacetime itself \cite{Oriti:2006se,FreidelGFT}. The formalism combines ideas from tensor models \cite{DiFrancesco:1993cyw,Gurau:2011xp} and spin foam models \cite{DePietri:1999bx,Reisenberger_2000}, yielding a background-independent description of discrete quantum geometries and their collective dynamics. For appropriate choices of group data and non-local interaction kernels, the perturbative expansion of the theory reproduces simplicial gravity path integrals \cite{Perez:2012wv,Perez:2003vx,Oriti:2011jm}. An advantage of GFT is that its QFT-like formulation provides both the systematic tools to handle collective dynamics of quantum-geometric degrees of freedom, and a natural prescription for summing over discrete spacetime histories, {so that the continuum limit need not be imposed through an external refinement prescription but becomes an internal dynamical question---to be addressed by renormalisation and phase structure studies of the model \cite{BenGeloun:2011jnm,BenGeloun:2013mgx,Carrozza:2016vsq,Marchetti:2022igl,Marchetti:2022nrf}.}

In the present work we adopt the canonical, second-quantised perspective on GFT \cite{Oriti_GFTandLQG}, particularly suited for the study of quantum gravity condensates and the use of many-body techniques. In this setting, the fundamental excitations admit an interpretation as elementary building blocks or \textit{quanta of geometry}, closely related to spin network states of canonical loop quantum gravity \cite{Oriti_GFT2ndLQG}. Continuum spacetime is then understood as a condensate of underlying quantum gravity degrees of freedom, making GFT a natural arena to illustrate the ideas discussed above.

Using the tools of relational dynamics \cite{Tambornino,Trinity,Goeller:2022rsx}, GFT condensates reproduce effective cosmological evolution directly from a full quantum gravity framework \cite{Oriti_2016,BOriti_2017,relham_Wilson_Ewing_2019,relhamadd}. We introduce a relational notion of evolution by coupling a free, massless scalar field $\chi$, which plays the role of a relational clock within a deparametrised\footnote{Relational dynamics can be introduced in two main ways in GFT \cite{Marchetti2021,Gielen:2024sxs,PWGFT,Marchetti:2024nnk}: using techniques similar to deparametrisation, or working in a timeless setting without choosing a clock \textit{a priori}. Both provide a Hilbert space formalism for GFTs, the former resting on canonical quantisation while the latter formulated via kinematical Fock space structures \cite{Gielen:2024sxs}. See Appendix \ref{app:GFTAppendix} for details.} framework (leaving extensions to additional and more realistic matter content for future work).

While realistic models are usually formulated on groups such as $SU(2)$ (the gauge group of the Ashtekar--Barbero formulation of general relativity \cite{barbero}) or $SL(2,\mathbb C)$ \cite{Perez:2003vx,Jercher:2021bie,GFTcosmoLONGpaper}, we employ a simplified but analytically tractable Abelian model based on $U(1)$ (see, e.g., \cite{Carrozza:2012uv,BenGeloun:2011rc,Lahoche:2018oeo,Lahoche:2018vun,Lahoche:2018ggd,Lahoche:2018hou} for related models), discussing only the ingredients relevant for our results. Our simplified model still consistently captures the emergence of non-singular bouncing Friedmann--Lema\^itre--Robertson--Walker (FLRW) dynamics once the GFT quanta are given a geometric interpretation, in analogy with more realistic constructions (see Appendix \ref{app:GFTAppendix} for details).

Our GFT is then defined by a real-valued field $\varphi (g_I,\chi)$,
\begin{equation}\label{eq:GFTfield}
		\varphi\,:\, U(1)^4 \times \mathbb{R} \rightarrow \mathbb{R} \,,
\end{equation}
with $g_I\in U(1)^4$ and $\chi\in\mathbb{R}$, governed by an action
\begin{equation}\label{eq:GFTaction}
	S [\varphi] = \frac{1}{2}\int \dd g_I \, \dd g_I' \, \dd \chi \; \varphi (g_I,\chi) K(g_I,g_I') \varphi (g_I',\chi) + V[\varphi] \,,
\end{equation}
where integration is with respect to the normalised Haar measure. Following the literature on GFT cosmology \cite{Oriti_2016,BOriti_2017,Gielen_2016,relham_Wilson_Ewing_2019}, phase transitions \cite{Marchetti:2022igl,Marchetti:2022nrf}, and renormalisation \cite{BenGeloun:2011jnm,BenGeloun:2013mgx,Carrozza:2016vsq}, we take a kinetic kernel invariant under group translations (specifically, the left group action) of the form
\begin{equation}\label{eq:genericK}
	K(g_I,g_I') = K^{(0)}(g_I^{-1}g_I') + K^{(2)} (g_I^{-1}g_I') \partial_\chi^2 \,,
\end{equation}
and leave the interaction potential $V[\varphi]$ general for now. 

Expanding the field in Peter-Weyl modes labelled by 4 integers $\bn=(n_1,...,n_4)$---the GFT analogue of Fourier modes---and performing a Legendre transform, we obtain the relational Hamiltonian \cite{relham_Wilson_Ewing_2019,relhamadd}
\begin{equation}\label{eq:freeham}
	H= -\frac{1}{2} \sum_\mathbf{n} \left( \frac{\pi_\mathbf{n}(\chi) \pi_\mathbf{-n}(\chi)}{K^{(2)}_\mathbf{n}} +K^{(0)}_\mathbf{n} \varphi_\mathbf{n}(\chi) \varphi_{-\mathbf{n}}(\chi)\right) - V[\varphi]\,,
\end{equation}
where $\pi_\mathbf{n} (\chi)= - K^{(2)}_\mathbf{n} \partial_\chi \varphi_\mathbf{-n} (\chi)$ is the conjugate momentum of the field (see appendix \ref{app:GFTAppendix} for details). Note that the field modes satisfy $\varphi_\bn^*=\varphi_{-\bn}$ due to the reality of the field \eqref{eq:GFTfield}; this means that the labels $\pm \bn$ describe the same physical degree of freedom. The Hamiltonian \eqref{eq:freeham} describes evolution with respect to relational time $\chi$, and allows to canonically quantise our GFT. 

The classical Poisson structure in phase space is mapped onto commutators in the algebra of quantum observables
\begin{equation}\label{eq:originalCCR}
	[\hat{\varphi}_\mathbf{n}(\chi) \, ,\, \hat{\pi}_\mathbf{m}(\chi)] = \ci\delta_{\mathbf{n,m}} \,,
\end{equation}
and the Heisenberg equation dictates the relational evolution for any observable as ${\rm i} \frac{\dd \hat{ O}}{\dd \chi} = [\hat{ O}, \hat{H}]+{\rm i} \partial_\chi \hat{O}$. As usual, we can define creation and annihilation operators satisfying $[\hat{a}_\mathbf{n}\,,\, \hat{a}_\mathbf{m}^\dagger]= \delta_\mathbf{n,m}$ for each mode as
\begin{equation}
\label{eq:aadag}
\begin{aligned}
    		\hat{a}_\mathbf{n} &= \frac{1}{\sqrt{2 \Omega_\mathbf{n}}} (\Omega_\mathbf{n} \, \hat{\varphi}_\mathbf{n} + {\mathrm i} \hat{\pi}_{-\mathbf{n}}) \,,\\
		\hat{a}_\mathbf{n}^\dagger &= \frac{1}{\sqrt{2 \Omega_\mathbf{n}}}( \Omega_\mathbf{n} \, \hat{\varphi}_{-\mathbf{n}} - {\mathrm i} \hat{\pi}_\mathbf{n}) \,,
        \end{aligned}
\end{equation}
where 
\begin{equation}\label{eq:Omegafree}
	\Omega_\mathbf{n} = \sqrt{\left|K_\mathbf{n}^{(0)} K_\mathbf{n}^{(2)}\right|} \,.
\end{equation}
These can be used to define a Fock vacuum as the state satisfying $\hat{a}_\mathbf{n}|0 \rangle=0$ for all $\mathbf{n}$, from which the Fock space is generated. As mentioned, inspired by formulations based on $SU(2)$ or $SL(2,\mathbb C)$, the 1-particle states of this Fock space will be \textit{assumed} to be quantum simplices with geometrical volume $\mathfrak{v}_\mathbf{n}$, given as a function of the four integers $\mathbf{n}$. We then define a volume operator in analogy with \eqref{eq:GenVolOp}, which accounts for the total volume contribution of the quanta:
\begin{equation}
\label{eq:volume}
	\hat{V}(\chi)= \sum_\mathbf{n}\hat{V}_{\bn}(\chi) =\sum_\mathbf{n} \mathfrak{v}_\mathbf{n} \hat{a}^\dagger_\mathbf{n}(\chi) \hat{a}_\mathbf{n} (\chi) \,.
\end{equation}
Interestingly, the Heisenberg dynamics of free GFTs naturally leads to a macroscopic occupation of a single quantum state, which behaves as a condensate. 

To understand this, we write the Hamiltonian \eqref{eq:freeham}---with $V[\varphi]=0$---in terms of GFT ladder operators \eqref{eq:aadag} as $\hat{H}=\sum_{\mathbf{n} } \hat{H}_\bn$ where, for each mode $\bn$, the specific form depends on the relative sign of $K^{(0)}_\mathbf{n}$ and $K^{(2)}_\mathbf{n}$:
\begin{equation}\label{eq:Hsumsinglemode} 
\begin{aligned}
    \hat{{H}}_\bn  =\begin{cases}
        \omega_\mathbf{n} \big(\hat{a}^\dagger_\mathbf{n} \hat{a}_\mathbf{n} + \frac{1}{2}\big) \,,  &\text{if }{\bn}\in \mathfrak{N}^{\text{HO}}\,,
        \\[1mm]
     \frac{1}{2}\omega_\mathbf{n} \big( \hat{a}^{\dagger}_\mathbf{n} \hat{a}^{\dagger}_\mathbf{-n}  + \hat{a}_\mathbf{n} \hat{a}_\mathbf{-n}  \big) \,, \qquad &\text{if }{\bn}\in \mathfrak{N}^{\text{SQ}}\,,
    \end{cases}
\end{aligned}
\end{equation}
where we have defined the mode-dependent parameter
\begin{equation}\label{eq:omegafree}
	\omega_\mathbf{n} = - \sgn\big(K^{(0)}_\mathbf{n}\big)\sqrt{\left|K^{(0)}_\mathbf{n}\big/K^{(2)}_\mathbf{n}\right|} \,,
\end{equation}
and
\begin{equation}\label{eq:setsfreetheory}
	\begin{aligned}
		&\mathbf{n}\in   \mathfrak{N}^{\text{HO}}\quad \Longleftrightarrow \quad\sgn \big(K^{(2)}_\mathbf{n}\big) = \sgn \big(K^{(0)}_\mathbf{n}\big)\,,\\
		&\mathbf{n}\in   \mathfrak{N}^{\text{SQ}} \quad \Longleftrightarrow \quad\sgn \big(K^{(2)}_\mathbf{n}\big) =- \sgn \big(K^{(0)}_\mathbf{n}\big)\,.
	\end{aligned}
\end{equation}
The dynamics of free GFT modes is then either that of a standard harmonic oscillator or an inverted one---described by a two-mode squeezing Hamiltonian \cite{TextSerafini}. Relational evolution is qualitatively different for each set of modes: while the number of quanta is conserved for the harmonic oscillator (HO) modes, it grows in time for the squeezing (SQ) modes, approaching exponential growth for $\chi\gtrsim (2\omega_\mathbf{n})^{-1}$.

Choices of kinetic term \eqref{eq:genericK} motivated by renormalisation studies \cite{BenGeloun:2011jnm,BenGeloun:2013mgx,Carrozza:2016vsq} lead to the existence of a mode $\mathbf{n}_0\in\mathfrak{N}^{\text{SQ}}$ for which the squeezing intensity rate is maximised \cite{Gielen_lowspin}. In that case, the $\bn_0$ mode is guaranteed to be macroscopically populated---as compared to the rest---at times $\chi\gtrsim(2\omega_{\mathbf{n}_0})^{-1}$, forming a GFT condensate. The Heisenberg evolution dictated by \eqref{eq:Hsumsinglemode} for the number operator $\hat{N}_{\rm c}\coloneqq(\hat{N}_{\bn_0}+\hat{N}_{-\bn_0})/2$---which counts physical condensed quanta---yields
\begin{align}\label{eq:condensateevolution}
    \hat{N}_{\rm c} (\chi)&=-\frac{1}{2}+ \lr{\hat{N}_{\rm c}(0)+\frac{1}{2}}\cosh(2\omega_{\bn_0}\chi)\\&+ \frac{{\rm i}}{2}  \left(\hat{a}_{\mathbf{n}_0}(0)\hat{a}_{-\mathbf{n}_0}(0)-\hat{a}^{\dagger}_{\mathbf{n}_0}(0)\hat{a}^{\dagger}_{-\mathbf{n}_0}(0)\right) \sinh(2\omega_{\mathbf{n}_0} \chi)\,. \notag
\end{align}
Then, the expectation value of $\hat{V}_{\rm c}(\chi)\coloneqq\mathfrak{v}_{\bn_0}\langle\hat{N}_{\rm c}(\chi)\rangle$ follows the mean-field equation
\cite{Oriti_2016,BOriti_2017,relham_Wilson_Ewing_2019}
\begin{equation}\label{eq:Friedmann}
	\left(\frac{1}{\langle\hat{V}_{\rm c}\rangle} \frac{{\mathrm d} \langle\hat{V}_{\rm c}\rangle}{{\mathrm d} \chi}\right)^2 = 4\omega_\mathbf{n_0}^2 \left( 1+\frac{\mathfrak{v}_\mathbf{n_0}}{ \langle\hat{V}_{\rm c}\rangle} -\frac{\mathfrak{v}_\mathbf{n_0}^2 \mathcal{I}_0}{ \langle\hat{V}_{\rm c}\rangle^2} \right)\,,
\end{equation}
where all the dependence on the choice of initial state is encoded in {the constant} $\mathcal{I}_0$.\footnote{\label{footnote:BounceCondition}We define $\mathcal{I}_0:=\langle \hat{N}_{\rm c}(0)\rangle\big(\langle \hat{N}_{\rm c}(0)\rangle+1\big)-C_0^2$, with $C_0:=\frac{\rm i}{2}\langle \hat{a}_{\mathbf{n}_0}(0)\hat{a}_{-\mathbf{n}_0}(0)-\hat{a}^{\dagger}_{\mathbf{n}_0}(0)\hat{a}^{\dagger}_{-\mathbf{n}_0}(0)\rangle$. Note that $\mathcal{I}_0\geq 0$ follows from the positivity of the bosonic covariance matrix.}

This equation recovers FLRW dynamics \eqref{eq:RelationalFLRW} at large volumes if we identify $\omega_\mathbf{n_0} ^2 =3\pi G$. Crucially, deviations from general relativity are incorporated in the inverse volume corrections, which induce a nonsingular bounce replacing the initial Big Bang singularity, reproducing results in loop quantum cosmology \cite{Ashtekar:2006uz,Ashtekar:2007em,Bojowald:2001xe}. This is a very robust finding that can be obtained within the GFT framework using different quantisation methods and for different GFT models  \cite{Oriti_2016,BOriti_2017,toy,relham_Wilson_Ewing_2019,relhamadd,Gielen_2020,Marchetti2021}, including extensions in which additional matter content is included \cite{Li:2017uao,Ladstatter:2025kgu}. The framework also accommodates cosmological anisotropies \cite{Mairi_2017,de_Cesare_2017,Aniso,Oriti:2023mgu}, natural mechanisms for early and late-time acceleration \cite{OritiPang,Pang:2025jtk,deCesare:2016rsf,Marchetti:2025jze}, and effective dynamics for cosmological perturbations \cite{Gielen:2017eco,Marchetti:2021gcv,Jercher:2023kfr,Jercher:2023nxa,Gielen:2025jcb}. 

The central message of this section is that continuum FLRW cosmology is recovered as a mean-field condensate regime of a fundamental background independent quantum theory of gravity. It is in this sense that spacetime emerges as a quantum-gravity fluid.

Let us finish this section by making the following remark. In free bosonic systems, idealised condensates at vanishing temperatures have all their atoms in the condensed state. However, if interactions are turned on, the mean field picture is corrected via quantum depletion and the presence of collective excitations known as phonons \cite{pitaevskii2016bose}. A condensed-matter perspective then motivates the following key questions: in the presence of interactions, how do quantum fluctuations correct the mean field dynamics of quantum gravity condensates? Or more generally, can the theory of collective excitations be developed in quantum gravity (and to what extent)?

\section{Collective excitations}
\label{sec:BogoGFT}

In this section we show how quantum fluctuations around interacting quantum gravity condensates contribute to macroscopic geometric observables as collective
quasiparticles. To apply it in quantum gravity, the argument requires only a bosonic many-body description of
quantum-geometric degrees of freedom, a condensate mean field, and a one-body
volume operator. 

While the mean field captures the hydrodynamic geometry of a
macroscopically occupied quantum state, the leading quantum corrections in an
interacting theory are not excitations of individual constituents. They are the
normal modes of the interacting condensate. We first formulate this mechanism
in model-independent terms, characterising the conditions for the existence of collective excitations. Then we discuss how they correct the expectation value of the corresponding volume observable. We will finally implement the construction in a
tractable GFT model, showing the results that will be needed to obtain the explicit corrections to the dynamics \eqref{eq:Friedmann} of cosmological condensates.

Let us assume a relational Hamiltonian many-body theory, with quantum-geometric excitations created and destroyed by operators $\{\hat{a}_i,\hat{a}_i^\dagger\}$, and let $c$ denote a macroscopically occupied quantum state such that the observables of interest admit a mean field expansion. The leading order of this expansion, or mean-field approximation, consists of replacing the corresponding creation and annihilation operators by their expectation value
\begin{equation}
\hat a_c
\longrightarrow
\langle\hat a_c\rangle\hat{\mathbb{I}}\,,
\label{eq:general_condensate_ladder_replacement}
\end{equation}
also known as order parameter. Expanding a generic Hamiltonian accordingly yields
\begin{equation}
\hat H
=
H_{0}\,\hat{\mathbb{I}}
+
\hat H_{2}
+
O(\hat a_i^3,\hat a_i^\dagger{}^3),
\qquad i\neq c,
\label{eq:general_H_expansion}
\end{equation}
where $\hat{H}_{n}$ is of order $n$ in out-of-condensate operators, and where $\hat{H}_{1}$ vanishes if the order parameter is chosen to satisfy the mean field equations, dictated by $H_{0}$. The leading
corrections to the mean field are then governed by the quadratic
Hamiltonian $\hat H_{2}$. This description applies in a perturbative regime, e.g., the weakly interacting or dilute regime of Bose gases, in which higher-order terms remain subleading. 

The most general quadratic Hamiltonian for the out-of-condensate modes has
Bogolyubov--de Gennes form \cite{pitaevskii2016bose}
\begin{equation}
\hat H_{2}
=
\sum_{i,j\neq c}
h_{ij}\,\hat a_i^\dagger\hat a_j
+
\frac{1}{2}
\sum_{i,j\neq c}
\left(
\Delta_{ij}\,\hat a_i^\dagger\hat a_j^\dagger
+ \text{h.c.}
\right)\,,
\label{eq:general_BdG_Hamiltonian}
\end{equation}
where $h$ and $\Delta$ depend on the specific model at hand---specifically, the mean-field configuration and the microscopic interaction coupling---and must satisfy $h_{ij}=h_{ji}^{*}$ and $\Delta_{ij}=\Delta_{ji}$ to guarantee hermiticity. While $h$ describes the free evolution of microscopic degrees of freedom, $\Delta$ describes an interaction between
non-condensed microscopic quanta and the condensate, inherited from the interaction among microscopic quanta. Heuristically, $\Delta$ can be thought of as the rate at which microscopic quanta are excited and decay in and out of the condensate. 

When $\Delta\neq0$, the Hamiltonian in general couples different degrees of freedom\footnote{Note that in the GFT free Hamiltonian \eqref{eq:Hsumsinglemode}, physical degrees of freedom are decoupled and no diagonalisation is needed.} and is not diagonal in the number basis of microscopic quanta. However, it can be diagonalised by a general Bogolyubov transformation
\begin{equation}
\hat b_\alpha
=
\sum_{i\neq c}
\left(
 U_{\alpha i}\hat a_i
+
 V_{\alpha i}\hat a_i^{\dagger}
\right)\,, 
\label{eq:general_Bogolyubov_transformation}
\end{equation}
where the coefficients $U$, $V$ must satisfy $U U^\dagger-V V^\dagger=
\mathbb I$ and $U V^T-V U^T=0$ to preserve canonical commutation relations, and $U^\dagger U-V^T V^*=\mathbb I$ and $U^\dagger V-V^T U^*=0$ to guarantee invertibility. The nontrivial step in any concrete application is to find the coefficients $U$ and $V$, which are determined by $h$ and $\Delta$.

We point out that for time-dependent condensates (such as for the quantum gravity model described in section \ref{sec:GFTcosmo}), $h$  and $\Delta$ in the corresponding quadratic Hamiltonian, and therefore the associated Bogolyubov coefficients, generally evolve with time, and can contribute additional terms to the dynamics of the collective excitations.

In terms of the new operators, $\hat{H}_2$ has a general structure which admits harmonic-oscillator like modes, but also modes that are squeezed as time evolves---these are called unstable Bogolyubov sectors \cite{Jain_2007,Peano:2016muy,Kustura:2019gic}. 

The new operators $\{\hat b_\alpha, \hat b_\alpha^\dagger\}$ create and annihilate
quanta in the normal modes of the system. The $V$-block mixes microscopic annihilation and creation operators,
implementing a multi-mode squeezing transformation. This indicates that the
degrees of freedom described by $\hat b_\alpha$ and $\hat b_\alpha^\dagger$ are
collective: they are not excitations of individual microscopic atoms of
geometry, but coherent combinations of them induced by the interacting
condensate.\footnote{In the non-interacting limit, the transformation \eqref{eq:general_Bogolyubov_transformation} becomes trivial and the collective modes reduce to the microscopic ones.}

This statement can be made sharp by the action of the annihilation operator for microscopic quanta on the vacuum for collective excitations
$|0\rangle_b$, known as Bogolyubov vacuum, defined by
\begin{equation}
\hat b_\alpha |0\rangle_b=0\,,
\qquad
\forall \alpha \,.
\label{eq:general_Bogolyubov_vacuum}
\end{equation}
Then, using the inverse of the transformation \eqref{eq:general_Bogolyubov_transformation} yields
\begin{equation}
\hat a_i |0\rangle_b
=
-
\sum_{\alpha}
 V_{\alpha i}\,
\hat b_\alpha^\dagger |0\rangle_b \,,
\label{eq:general_a_on_Bogolyubov_vacuum}
\end{equation}
which is non-vanishing unless $V=0$. This shows that the state with no collective excitations contains out-of-condensate quanta. This phenomenon, known as quantum depletion in the BEC \cite{pitaevskii2016bose}, predicts that the mean number of out-of-condensate quanta in the absence of collective excitations is given by
\begin{equation}
{}_b\langle 0|
\hat a_i^\dagger \hat a_i
|0\rangle_b
=
\sum_{\alpha}
 |V_{\alpha i}|^2\,,
\label{eq:general_depletion_number}
\end{equation}
and it also predicts the correlation structure 
\begin{equation}
{}_b\langle 0|
\hat a_i^\dagger \hat a_j
|0\rangle_b
=
\sum_{\alpha}
 V_{\alpha i}^* V_{\alpha j}\,,
\label{eq:general_depletion_density_matrix}
\end{equation}
between out-of-condensate microscopic quanta. The presence of this population is a purely quantum effect generated by the interactions that
define the collective modes themselves. We stress that in a time-dependent setting, depletion is accompanied by the dynamical production of collective excitations \cite{Fedichev:2003bv,Carusotto:2010mei,Jaskula_2012,Hung_2013,Eckel:2017uqx,Viermann:2022wgw,Sparn_2024,Gondret:2025eio}. This mechanism is the many-body analogue of particle creation of QFTs on expanding backgrounds \cite{Parker,Parker2}.

We now show how the contribution of collective excitations can leave imprints in macroscopic continuum-geometric observables. To that end, we focus on the volume observable defined in \eqref{eq:GenVolOp}. If
only the condensate mode has displacement, so that
$\langle \hat a_{i\neq c}\rangle=0$, 
then the expectation value of the volume splits into
\begin{equation}
V
=
\mathfrak{v}_{c}N_c
+
\Delta V\,,
\label{eq:general_volume_split}
\end{equation}
with $\mathfrak{v}_{c}\coloneqq \mathfrak{v}_{cc}$ and
\begin{equation}
\Delta V
\coloneqq
\sum_{i,j\neq c}
\mathfrak{v}_{ij}
\langle
\hat a_i^\dagger\hat a_j
\rangle \,.
\label{eq:general_volume_correction}
\end{equation}
The correction is therefore controlled by the out-of-condensate one-body
density matrix. 

Let us now discuss how both, depletion and collective excitations, induce a non-vanishing $\Delta V$. Substituting the inverse of the transformation
\eqref{eq:general_Bogolyubov_transformation} into
\eqref{eq:general_volume_correction}, we find
\begin{equation}
\begin{aligned}
\Delta V
&=
\sum_{\alpha\beta} \Big[
 \lr{A^{UU}_{\alpha\beta}+
 A^{VV}_{\alpha\beta}}
\langle
\hat b_\alpha^\dagger\hat b_\beta
\rangle
-
 A^{UV}_{\alpha\beta}
\langle
\hat b_\alpha^\dagger\hat b_\beta^\dagger
\rangle
\\  & \qquad \quad \quad -
 A^{UV}_{\beta\alpha}{}^*
\langle
\hat b_\alpha\hat b_\beta
\rangle \Big]
+
\sum_{\alpha} 
 A^{VV}_{\alpha\alpha}
\,,
\label{eq:general_volume_correction_Bogolyubov}
\end{aligned}
\end{equation}
where we have used the canonical commutation relations, and we have defined
\begin{equation}
\begin{aligned}
A^{UU}_{\alpha\beta}
&=
\sum_{i,j\neq c}\mathfrak{v}_{ij}
 U_{i\alpha}^{*}
 U_{j\beta}\,,
\\
A^{VV}_{\alpha\beta}
&=
\sum_{i,j\neq c}\mathfrak{v}_{ij}
 V_{i\alpha}^{*}
 V_{j\beta}\,,
\\
 A^{UV}_{\alpha\beta}
&=
\sum_{i,j\neq c}\mathfrak{v}_{ij}
 U_{i\alpha}^{*}
 V_{j\beta}\,.
\label{eq:general_VUU_VVV}
\end{aligned}
\end{equation}
Equation \eqref{eq:general_volume_correction_Bogolyubov} is the general
beyond-mean-field correction to the macroscopic volume. The first term is the
direct contribution of populated collective modes, the second and third terms capture anomalous collective correlations, and the last term is the correction from depletion. 

In section \ref{sec:GFTbogolons} we will explicitly compute how collective excitations and depletion modify a macroscopic geometric observable within a specific quantum gravity model. In particular, we will see how due to the time-dependence of the condensate cosmology described in section \ref{sec:GFTcosmo}, collective excitations can be dynamically produced. We note that in this picture, the excitations are tied to fundamental atoms of geometry, meaning spacetime itself---somewhat like matter---is subject to dynamical quantum production.

The conclusion is that the leading quantum corrections to an emergent volume
observable take the form of collective effects, particularly depletion and Bogolyubov excitations, for which we coin the term \textit{quantum-gravity bogolons}. 

\subsection{GFT bogolons}\label{sec:GFTbogolons}

In this section we will apply the framework outlined above to a specific background independent quantum gravity model, namely the GFT condensate cosmology discussed in \ref{sec:GFTcosmo}, with a non-vanishing interaction potential. We will show that, in a weakly interacting and dilute regime, leading order corrections to the mean field are described by depletion and collective Bogolyubov excitations of the GFT condensate. We will call them GFT \textit{bogolons}, in parallel to phonons in a BEC. 

In particular, we apply our framework to the deparametrised $U(1)$ model of section \ref{sec:GFTcosmo}, canonically quantised with respect to relational time $\chi$, and governed by the Hamiltonian \eqref{eq:freeham} with a quartic potential local in the group elements\footnote{An interaction local in the mode labels (e.g., $\sum_{\mathbf n}(\hat\varphi_{\mathbf n}\pm\hat\varphi_{-\mathbf n})^4$, extending the models of \cite{Gielen_2020,Gielen:2023han}) does not couple different degrees of freedom, with each mode experiencing self-interaction. {Such mode decoupling is not expected in generic GFTs for quantum gravity, where combinatorial matching patterns imposed by the interactions generically couple different labels.} Without coupling between modes, the mechanism underlying the Bogolyubov treatment is absent and no collective excitations arise.}
\begin{equation}\label{eq:inter}
	V [\varphi]=\frac{\lambda}{4!}
	\int \dd g_I \, \dd \chi \; \big[\hat{\varphi}(g_I, \chi)\big]^4 \,.
\end{equation}
The coupling $\lambda$ can be positive or negative. We will work in the weakly interacting regime, meaning $\lambda$ is assumed to be small enough so that the interactions produce perturbative corrections to the mean-field picture described in section \ref{sec:GFTcosmo} (see also appendix \ref{app:GFTAppendix}). 

Our results extend in a straightforward manner to any interaction potential local in the group elements. {While interacting kernels in GFTs with a natural geometric interpretation are non-local on the group manifold, we here restrict to local potentials as an illustration of our method. In particular, \eqref{eq:inter} carries no simplicial interpretation but it provides a natural benchmark, in analogy with the contact approximation commonly used to describe dilute Bose gases, which introduces inter-mode coupling in representation space without the full algebraic complexity of more realistic simplicial interactions.} Techniques to generalise Bogolyubov theory to non-local interaction potentials exist in weakly interacting Bose gases \cite{laghi2017excitations,bulakhov2018re,lewin2015bogoliubov,Ribeiro:2021zyz,Tang-Nonlocal}. We leave the adaptation of our framework to non-local interactions for future work.

Let us apply the framework developed above to GFT condensate cosmology. {Here the macroscopically occupied state $c$ is realised by the condensate degree of freedom $\pm\bn_0$, whose occupation is the symmetrised $\hat{N}_\text{c}$ introduced above \eqref{eq:condensateevolution}.} For our mean field regime, we will assume a Gaussian family of initial states characterised by the first moments 
\begin{equation}\label{eq:state}
\begin{aligned}
    &\expval{\hat{\varphi}_{\pm\bn_0}(0)}=\Phi_0\in\mathbb{R}\,,\\
    \expval{\hat{\pi}_{\pm\bn_0}(0)}=&\expval{\hat{\varphi}_{\neq\pm\bn_0}(0)}= \expval{\hat{\pi}_{\neq\pm\bn_0}(0)}=0 \,,
\end{aligned}    
\end{equation}
which implies 
\begin{equation}
    \expval{\hat{a}_{\pm\bn_0}(0)}=\langle\hat{a}^\dagger_{\pm\bn_0}(0)\rangle=\sqrt{\Omega_{\bn_0}/2}\,\Phi_0\,.
\end{equation}
The initial state is then coherently displaced by $\Phi_0$ only in the field quadrature of the dominant mode $\bn_0$. For the class of states considered here, this guarantees that the relative fluctuations over the mean
field remain small throughout relational cosmological evolution \cite{Gauss,AC_thesis}. 

Given that in the interacting GFT considered there are no known exact solutions of the mean-field equations, we will perform our mean-field expansion around the mean-field of the free theory. We will discuss the regime of GFT parameters in which the error in this approximation is subleading with respect to corrections by depletion and collective excitations. Thus, the expectation value for the condensate mode at any relational time {is
\begin{equation}\label{eq:PhiFree}
	\expval{\hat{\varphi}_{\pm\mathbf{n_0}}(\chi)}\coloneqq \Phi(\chi)=\Phi_0\cosh(\omega_{\bn_0}\chi)\,.
\end{equation}
This is exact since the free Hamiltonian is quadratic and first moments obey the classical equation of motion $\Phi''=\omega_{\bn_0}^2\Phi$, with $\Phi(0)=\Phi_0$ and $\Phi'(0)=0$---see \eqref{eq:state}. With $\Pi(\chi)\coloneqq\expval{\hat{\pi}_{\pm\bn_0}(\chi)}=-{\Omega_{\bn_0}}/{\omega_{\bn_0}}\Phi'(\chi)=-\Omega_{\bn_0}\Phi_0\sinh(\omega_{\bn_0}\chi)$, the expectation value of $\hat N_{\rm c}$ gives
\begin{equation}\label{eq:NcSplit}
	N_{\rm c}(\chi)=\frac{\Omega_{\bn_0}}{2}\Phi^2(\chi)+\frac{1}{2\Omega_{\bn_0}}\Pi^2(\chi)+\delta N_{\rm c}(\chi)\,,
\end{equation}
where $\delta N_{\rm c}$ collects the quantum fluctuations of the condensate mode (note $\delta N_{\rm c}\geq0$ by the Cauchy--Schwarz inequality). For the semiclassical states considered here, $\delta N_{\rm c}/N_{\rm c}=O(1/N_{\rm c}(0))$ at all times, so that $\delta N_{\rm c}$ can be consistently neglected whenever $\Phi^2$ is expressed in terms of $N_{\rm c}$. This is a crucial approximation underlying Bogolyubov's treatment \cite{Bogolyubov:1947zz}: under the prescription \eqref{eq:general_condensate_ladder_replacement}, $\hat N_{\rm c}\to\frac{\Omega_{\bn_0}}{2}\Phi^2+\frac{1}{2\Omega_{\bn_0}}\Pi^2$. Finally, since the free evolution conserves ${\Omega_{\bn_0}}\Phi^2-\Pi^2/\Omega_{\bn_0}={\Omega_{\bn_0}}\Phi_0^2 = 2 N_\text{c}(0)+O(\delta N_\text{c})$, one has
\begin{equation}\label{eq:PhiNc}
	\Phi^2(\chi)=\frac{N_{\rm c}(\chi)+N_{\rm c}(0)}{\Omega_{\bn_0}}+O(\delta N_\text{c})\,.
\end{equation}
Moreover, we can write the bound
\begin{equation}\label{eq:PhiNcRatio}
	\frac{\Phi^2(\chi)}{N_{\rm c}(\chi)}=\frac{1}{\Omega_{\bn_0}}\bigg(1+\frac{N_{\rm c}(0)}{N_{\rm c}(\chi)}+O(\delta N_\text{c})\bigg)\leq\frac{2}{\Omega_{\bn_0}}\,,
\end{equation}
which holds exactly, since $\Phi^2\leq\Phi_0^2\cosh(2\omega_{\bn_0}\chi)\leq 2N_{\rm c}/\Omega_{\bn_0}$ by \eqref{eq:NcSplit} and $\delta N_{\rm c}\geq0$. Thus $\Phi^2=2N_{\rm c}/\Omega_{\bn_0}+O(\delta N_\text{c})$ only at $\chi=0$, while $\Phi^2\to N_{\rm c}/\Omega_{\bn_0}$ at late times. We remark that the approximation \eqref{eq:PhiNc} is never needed to construct the collective excitations or to control their dynamics: the rest of this section and section~\ref{sec:EvolutionofBogolons} rely only on the exact relations \eqref{eq:PhiFree} and \eqref{eq:NcSplit} and on the bound \eqref{eq:PhiNcRatio}. Neglecting $\delta N_\text{c}$ in \eqref{eq:PhiNc} will be used in section~\ref{sec:GFTbogocosmo} to cast the corrected Friedmann equation as an expansion in inverse powers of the volume akin to \eqref{eq:Friedmann}.}

We now decompose our interaction potential \eqref{eq:inter} in GFT field modes as (see \eqref{eq:PWApp})
\begin{equation}\label{eq:Vstill4}
	V[\varphi] = \frac{\lambda}{4!} \sum_{\mathbf{n} \in \mathbb{Z}^4}\sum_{\mathbf{m} \in \mathbb{Z}^4}\sum_{\mathbf{p} \in \mathbb{Z}^4} \sum_{\mathbf{q} \in \mathbb{Z}^4} \delta_{\mathbf{n}+\mathbf{m}+\mathbf{p}+\mathbf{q},\, \mathbf{0}} \,\, \hat{\varphi}_{\mathbf{n}} \,\hat{\varphi}_{\mathbf{m}} \,\hat{\varphi}_{\mathbf{p}} \,\hat{\varphi}_{\mathbf{q}}\,,
\end{equation}
where $\delta_{\mathbf{n}+\mathbf{m}+\mathbf{p}+\mathbf{q}, \, \mathbf{0}} = \prod_{i=1}^4 \delta_{n_i+m_i+p_i+q_i, \, 0}$. With the full Hamiltonian written in terms of field modes, we are ready to perform a mean-field expansion as explained in section \ref{sec:BogoGFT}. Thus we write $\hat{H}=\hat{H}_0+\hat{H}_1+\hat{H}_2+  O({\lambda \Phi})$, obtaining
\begin{equation}
\label{eq:H1nondiag}
    \hat{H}_1=-\frac{\lambda}{6} \Phi^3 \lr{\hat\varphi_{3\bn_0}+\hat\varphi_{-3\bn_0}}\,,
\end{equation}
and
\begin{widetext}
\begin{align}
\label{eq:H2nondiag}
\hat{H}_2 = 	-\frac{1}{2}\sum_{\mathbf{n\neq \pm n_0}} \left[ 
	\left( \frac{\hat{\pi}_\mathbf{n} \hat{\pi}_\mathbf{-n}}{K^{(2)}_\mathbf{n}} +K^{(0)}_\mathbf{n} \hat{\varphi}_\mathbf{n} \hat{\varphi}_{-\mathbf{n}}\right) + \frac{\lambda}{2} \Phi^2 \left(2 \hat{\varphi}_\mathbf{n}\hat{\varphi}_{\mathbf{-n}} + \hat{\varphi}_\mathbf{n}\hat{\varphi}_{\mathbf{-n-2\mathbf{n_0}}}+ \hat{\varphi}_\mathbf{n}\hat{\varphi}_{\mathbf{-n+ 2 \mathbf{n_0}}}\right)
	\right] \,.
\end{align}
\end{widetext}
The reason why $\hat{H}_1$ does not vanish here is that the mean-field is a solution of the free theory only, not the full interacting theory. As a result, the out-of-condensate modes $\pm3\bn_0$ pick up a mean-field dependent source term in their evolution equations.

To illustrate our method with clarity within GFT, {we consider a regime where the linear perturbation term $H_1$ is negligible relative to the quadratic term $H_2$. Although $H_1$ is lower-order in fluctuations, its prefactor measures the deviation of the background from the classical equations of motion and vanishes for exact condensate solutions.} In Appendix \ref{app:ApproximationVaidity} we derive sufficient conditions for the existence of such a regime within the relational time interval $[0,\chi]$, which are given by (see \eqref{eq:DF} and below)
\begin{equation}
\label{eq:InteractionBound}
\left|\frac{1}{6}\frac{d_F^{3\bn_0}(\chi)}{\sqrt{\ev{\hat\varphi_{\pm3\bn_0}^2(\chi)}}}\lambda  \Phi^{3}(\chi)\right|\ll1 \,.
\end{equation}
When expanding around a free mean-field configuration, this condition is stronger than the usual diluteness criterion $|\lambda | \Phi^2 \ll1$. Since $\Phi(\chi)$, and with it $N_\text{c}(\chi)$, grows with time, these conditions set up a maximum relational time $\chi_*$ at which the interaction corrections that we are neglecting become comparable to the corrections due to collective excitations. Thus, our computations are relevant within the relational time interval $[0,\chi_*)$.

The quadratic Hamiltonian \eqref{eq:H2nondiag} describes the dynamics of the system when leading-order quantum corrections to the mean-field dynamics are taken into account. It is a concrete realisation of \eqref{eq:general_BdG_Hamiltonian}, where $h_{ij}$ and $\Delta_{ij}$ can be identified by writing it in terms of the GFT ladder operators \eqref{eq:aadag}---we work here with field and momentum quadratures for convenience.  

The next step is to find the Bogolyubov transformation \eqref{eq:general_Bogolyubov_transformation} which brings \eqref{eq:H2nondiag} into diagonal form.

\subsubsection*{Bogolyubov transformation for GFT collective excitations}\label{sec:bogotransGFT}

Diagonalising \eqref{eq:H2nondiag} requires a rather involved technical procedure that we describe in detail in Appendix \ref{app:AppDiag}. In what follows, we summarise the key aspects relevant to understand the results of the diagonalisation procedure.

The crucial feature of the Hamiltonian \eqref{eq:H2nondiag} is the specific structure of the coupling between mode labels. While the kinetic term shows trivial coupling between the $\mathbf{n}$ and $-\mathbf{n}$ labels (which describe the same physical degree of freedom), the interaction term contains explicitly non-diagonal pieces, coupling $\bn$ to $-\mathbf{n}-2\mathbf{n_0}$ and to $-\bn+2\mathbf{n_0}$. Because we sum over all $\bn\neq\mathbf{n_0}$, this leads to infinite chains of coupled modes and implies the existence of a Bloch--Floquet structure in momentum space, in which modes with different labels are coupled as long as they belong to the same chain \cite{Kuchment1993}. This structure implies that the space of modes partitions into disjoint, independent chains, one of which is depicted in figure \ref{fig:bi-chainMAIN} (see Appendix \ref{app:AppDiag} for all the details). 

\begin{figure}[h]
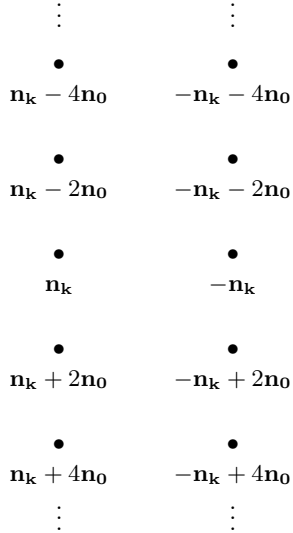

	\centering
\[
\begin{array}{c@{\qquad}c}
	\vdots & \vdots \\[0.5em]
	\tikzmarknode{A1}{\begin{array}{c}
			\bullet\\
			\mathbf{n_k} - 4\mathbf{n_0}
	\end{array}}
	&
	\tikzmarknode{B1}{\begin{array}{c}
			\bullet\\
			-\mathbf{n_k} - 4\mathbf{n_0}
	\end{array}}\\[2em]

	\tikzmarknode{A2}{\begin{array}{c}
			\bullet\\
			\mathbf{n_k} - 2\mathbf{n_0}
	\end{array}}
	&
	\tikzmarknode{B2}{\begin{array}{c}
			\bullet\\
			-\mathbf{n_k} - 2\mathbf{n_0}
	\end{array}}\\[2em]

	\tikzmarknode{A3}{\begin{array}{c}
			\bullet\\
			\mathbf{n_k}
	\end{array}}
	&
	\tikzmarknode{B3}{\begin{array}{c}
			\bullet\\
			-\mathbf{n_k}
	\end{array}}\\[2em]

	\tikzmarknode{A4}{\begin{array}{c}
			\bullet\\
			\mathbf{n_k} + 2\mathbf{n_0}
	\end{array}}
	&
	\tikzmarknode{B4}{\begin{array}{c}
			\bullet\\
			-\mathbf{n_k} + 2\mathbf{n_0}
	\end{array}}\\[2em]

	\tikzmarknode{A5}{\begin{array}{c}
			\bullet\\
			\mathbf{n_k} + 4\mathbf{n_0}
	\end{array}}
	&
	\tikzmarknode{B5}{\begin{array}{c}
			\bullet\\
			-\mathbf{n_k} + 4\mathbf{n_0}
	\end{array}}
    \\[0.5em]

	\vdots & \vdots
\end{array}
\]
\caption{\small The interactions described by the Hamiltonian \eqref{eq:H2nondiag} only occur within chains like the one depicted in this diagram, here constructed denoting $\mathbf{n_k}$ a generic representative mode. The whole space of GFT modes is then conveniently split into infinitely many disconnected diagrams. See appendix \ref{app:AppDiag} for details regarding their proper definition and construction.}
\label{fig:bi-chainMAIN}
\end{figure}

Diagonalising such a system  requires that the kinetic couplings $K^{(0)}_{\mathbf{n}}$ and $K^{(2)}_{\mathbf{n}}$ are invariant within each chain (see Appendix \ref{app:AppDiag}). {This is a simplifying approximation adopted for analytical tractability, as it} effectively reduces the problem to diagonalising the chains independently. There is a well-known procedure to diagonalise quadratic Hamiltonians with such coupling structure, involving Fourier (or Bloch) transforms along each chain, which introduce a continuous angle $\theta \in [0, 2\pi)$ in place of the discrete position along the discrete ``lattice'' of modes. {In more realistic GFT models these kernels typically depend on the quantum numbers and therefore vary along the chains; while treating them would require more involved (or numerical) methods, this is not expected to spoil the central physical result that interactions generically give rise to collective excitations.}

After performing all these steps---explicitly done in appendix \ref{app:AppDiag}---one finds that the normal modes of $\hat{H}_2$ are labelled by a new set of quantum numbers $\pm{\boldsymbol{\nu}} := \{\mathbf{s}, \theta, \pm \varepsilon\}$. $\mathbf{s}$ is the coset representative identifying each chain, $\theta$ is the continuous Bloch angle parametrising each (Fourier transformed) chain, and  $\varepsilon \in \{1, -1\}$ is a discrete index distinguishing the two components of the coupled pair $\{\boldsymbol{\nu}\,,-\boldsymbol{\nu}\}$.\footnote{We point out that the label $\boldsymbol{\nu}$ reduces to the old out-of-condensate label $\mathbf{n}$ for $\lambda=0$, namely $\boldsymbol{\nu}\;\overset{\lambda\rightarrow0}{\longrightarrow}\; \mathbf{n}\neq \pm\mathbf{n_0}$.} In these labels, the canonical commutation relations \eqref{eq:originalCCR} and the reality condition for the field mode operators read
\begin{equation}
    [\hat{\varphi}_{\boldsymbol{\nu}},\hat{\pi}_{\boldsymbol{\nu'}}]
	={\rm i}\,\delta_{\boldsymbol{\nu},\boldsymbol{\nu'}}\,,
    \qquad
    \hat{\varphi}_{\boldsymbol{\nu}}^\dagger=\hat{\varphi}_{-\boldsymbol{\nu}}\,.
\end{equation}
This structure is strictly analogous to the free theory, where the pair $\{\mathbf{n}\,,-\mathbf{n}\}$ described the same physical degree of freedom. In other words, the Fourier transform over the chain index effectively decouples the physically distinct modes; the remaining algebraic structure couples only $\boldsymbol{\nu}$ and $-\boldsymbol{\nu}$ (which differ only by $\varepsilon$), and since these refer to the same underlying degree of freedom, the Hamiltonian can effectively be written in a diagonal form, with the normal modes being Bloch waves on each chain. 

A last remark about the diagonalisation concerns the modes in the special chain that contains the condensate modes $\pm\bn_0$. These constitute a \textit{severed} set of odd multiples of the condensates, where the labels $\pm \bn_0$ themselves have been removed. The procedure to diagonalise this family of modes is parallel to the one outlined for the other chains, except for subtleties regarding open boundaries at $\pm3\bn_0$, which change the form of the normal modes to standing waves (rather than Bloch waves). We refer to Appendix \ref{app:AppDiag} for all the technical details about the diagonalisation procedure, the definition of the normal modes and the corresponding labels.

As usual, in terms of the normal modes $\boldsymbol{\nu}$, the Hamiltonian \eqref{eq:H2nondiag} takes a diagonal form
\begin{equation}\label{eq:HamiltonianNU}
\begin{aligned}
			&\hat{H}_2 =  \intsum\,
			\hat{H}_{\boldsymbol{\nu}} \,, \\ 
            &\hat{H}_{\boldsymbol{\nu}}  = -\frac{1}{2}  \left(
		\frac{\hat{\pi}_{\boldsymbol{\nu}}\hat{\pi}_{-\boldsymbol{\nu}}}{K_{\boldsymbol{\nu}} ^{(2)}}  + \left( K_{\boldsymbol{\nu}} ^{(0)}
		+{\lambda}\Phi^2  f_{\boldsymbol{\nu}}  \right) \hat{\varphi}_{\boldsymbol{\nu}}\hat{\varphi}_{-\boldsymbol{\nu}}  \right)\,,
        \end{aligned}
\end{equation}
where we adopt the ``sum-integral'' notation $\intsumTEXT_{\boldsymbol{\nu}} := \sum_{\mathbf{s},\varepsilon} \int \dd \theta $ for convenience. Here $f_{\boldsymbol{\nu}} \equiv f(\theta)= 1+\cos\theta$ is known in the crystal literature as the ``Bloch factor'', and it appears as a direct result of the nearest-neighbour-like interactions along the abstract GFT chains. Notice that the value $\theta=\pi$ effectively describes modes that do not feel the interaction (they behave as if $\lambda=0$, and correspond to the edge of a Brillouin zone in solid-state  physics terminology).

We now define the creation and annihilation operators for normal modes
\begin{equation}\label{eq:bbdag}
	\begin{aligned}
		\hat{b}_{\boldsymbol{\nu}} &:= \frac{1}{\sqrt{2 \Omega^\lambda_{\boldsymbol{\nu}}}} \left(\Omega^\lambda_{\boldsymbol{\nu}} \hat{\varphi}_{\boldsymbol{\nu}}  + {\rm i} \hat{\pi}_{-\boldsymbol{\nu}} \right) \,, \\ 
		\hat{b}^\dagger_{\boldsymbol{\nu}} &: =  \frac{1}{\sqrt{2 \Omega^\lambda_{\boldsymbol{\nu}}}} \left(\Omega^\lambda_{\boldsymbol{\nu}} \hat{\varphi}_{-\boldsymbol{\nu}}  - {\rm i} \hat{\pi}_{\boldsymbol{\nu}} \right) \,,
	\end{aligned}
\end{equation}
where
\begin{equation}\label{eq:Omegalambda}
\Omega^\lambda_{\boldsymbol{\nu}} := \sqrt{\left| {K^{(2)}_{\boldsymbol{\nu}}} \left(K^{(0)}_{\boldsymbol{\nu}} +\lambda \Phi^2 f_{\boldsymbol{\nu}} \right) \right|} \,.
\end{equation}
These satisfy $[\hat{b}_{\boldsymbol{\nu}}, \hat{b}^\dagger_{\boldsymbol{\nu'}}]= \delta_{\boldsymbol{\nu},\boldsymbol{\nu'}}$.
In terms of these, \eqref{eq:HamiltonianNU} reads 
\begin{equation}\label{eq:BogoHamiltonian}
	\hat{H}_{\boldsymbol{\nu}}= 
    \begin{cases}
        \omega^\lambda_{\boldsymbol{\nu}} \left( \hat{b}^\dagger_{\boldsymbol{\nu}} \hat{b}_{\boldsymbol{\nu}} +\frac{1}{2} \right)\,,  &\text{if }{\boldsymbol{\nu}}\in \mathfrak{S}^\text{HO}\,,
        \\
    \frac{1}{2}\omega^\lambda_{\boldsymbol{\nu}}\left( \hat{b}^{\dagger}_{\boldsymbol{\nu}}\hat{b}^{\dagger}_{-\boldsymbol{\nu}} + \hat{b}_{\boldsymbol{\nu}}\hat{b}_{-\boldsymbol{\nu}} \right)\,, \quad\qquad &\text{if }{\boldsymbol{\nu}}\in \mathfrak{S}^\text{SQ}\,,
    \end{cases}
\end{equation}
where we have introduced the time-dependent parameter
\begin{equation}\label{eq:omegalambda}
	\begin{aligned}
	 \omega^\lambda_{\boldsymbol{\nu}} &= - \sgn \big(K^{(0)}_{\boldsymbol{\nu}}+ \lambda \Phi^2 f_{\boldsymbol{\nu}}\big) \sqrt{\frac{\left|K^{(0)}_{\boldsymbol{\nu}} + \lambda \Phi^2 f_{\boldsymbol{\nu}} \right|}{\left|K^{(2)}_{\boldsymbol{\nu}}\right|} }\,,
	 \end{aligned}
\end{equation}
and we have split the space of labels into two disjoint sets $\mathfrak{S}^{\text{HO}}$ and $\mathfrak{S}^{\text{SQ}}$ defined by 
\begin{equation}\label{eq:setsinteractingtheory}
	\begin{aligned}
		{\boldsymbol{\nu}}\in \mathfrak{S}^\text{HO} \; &\Leftrightarrow \; \sgn \big(K^{(2)}_{\boldsymbol{\nu}}\big) = \sgn \big(K^{(0)}_{\boldsymbol{\nu}}+ \lambda \Phi^2 f_{\boldsymbol{\nu}}\big)\,,\\
		{\boldsymbol{\nu}} \in \mathfrak{S}^\text{SQ} \; &\Leftrightarrow \; \sgn \big(K^{(2)}_{\boldsymbol{\nu}}\big) = - \sgn \big(K^{(0)}_{\boldsymbol{\nu}}+ \lambda \Phi^2 f_{\boldsymbol{\nu}}\big)\,.
	\end{aligned}
\end{equation}
Note that all these expressions are a direct generalisation of the free-theory scenario, see \eqref{eq:omegafree}--\eqref{eq:setsfreetheory}. We have the harmonic oscillator modes ${\boldsymbol{\nu}}\in \mathfrak{S}^\text{HO}$ (HO modes), governed by a standard harmonic oscillator Hamiltonian, and squeezing modes ${\boldsymbol{\nu}}\in \mathfrak{S}^\text{SQ}$ (SQ modes), governed by a two-mode squeezing Hamiltonian. We note that there exist laboratory condensates which exhibit this behaviour \cite{Fallani_2004,Klempt:2009dt,Bookjans_2011,Steinhauer:2014dra}. As a last remark, we note that while $K^{(0)}_{\bnu}+\lambda \Phi^2 f_{\bnu}$, unlike $K^{(2)}_{\bnu}$, in general depends on relational time, the relative sign, and therefore also the dynamical nature of the mode (whether HO or SQ), will not change within the regime of validity of the mean field expansion, since in this regime the correction proportional to $\lambda$ is small and cannot change the sign of $K^{(0)}_{\bnu}$.

We now discuss how these collective modes are related to the microscopic GFT degrees of freedom. To that end we find the specific Bogolyubov transformation \eqref{eq:general_Bogolyubov_transformation} bringing the Hamiltonian to its normal form.

First, we define the atomic creation and annihilation operators in the new basis of modes labelled by $\bnu$
\begin{equation}
	\begin{aligned}
			\hat{a}_{\boldsymbol{\nu}} &= \frac{1}{\sqrt{2 \Omega_{\boldsymbol{\nu}}}} \big(\Omega_{\boldsymbol{\nu}} \hat{\varphi}_{\boldsymbol{\nu}} + {\rm i} \hat{\pi}_{-\boldsymbol{\nu}}\big)   \,,   \\
		\hat{a}^\dagger_{\boldsymbol{\nu}} &= \frac{1}{\sqrt{2 \Omega_{\boldsymbol{\nu}}}} \big(
	\Omega_{\boldsymbol{\nu}} \hat{\varphi}_{-\boldsymbol{\nu}} - {\rm i} \hat{\pi}_{\boldsymbol{\nu}}
	\big) \,,
	\end{aligned}
\end{equation}
with $\Omega_{\boldsymbol{\nu}} = \sqrt{\big|K_{\boldsymbol{\nu}}^{(0)} K_{\boldsymbol{\nu}}^{(2)}\big|}$. Since the operators $\hat{a}_{\bnu}$, $\hat{a}_{\bnu}^\dagger$ are related to the original $\hat{a}_\bn$, $\hat{a}_\bn^\dagger$ by a canonical transformation {(essentially the Fourier transform\footnote{{Explicitly, one can write $\hat{a}_{\bnu} \equiv \hat{a}_{\mathbf{s},\varepsilon}(\theta) =\frac{1}{\sqrt{2\pi} }\sum_m e^{-{\rm i} m \theta}\hat{a}_{\mathbf{s},m,\varepsilon}$ where $\hat{a}_{\mathbf{s},m,\varepsilon}\equiv \hat{a}_\bn$, and the Fourier transform \eqref{eq:Fourier} can be cast as $\hat{\varphi}_{\bnu}\equiv \hat{\varphi}_{\mathbf{s},\varepsilon}(\theta)= \frac{1}{\sqrt{2 \Omega_\mathbf{s}}}\big(\hat{a}^\dagger_{\mathbf{s},-\varepsilon}(\theta)+ \hat{a}_{\mathbf{s},\varepsilon}(\theta)\big) \equiv \frac{1}{\sqrt{2 \Omega_{\boldsymbol{\nu}}}}\big(\hat{a}^\dagger_{-\bnu}+ \hat{a}_{\bnu}\big)$ and similarly for $\hat{\pi}_{\bnu}$, see appendix \ref{app:AppDiag} for details.}} of appendix \ref{app:AppDiag})} that does not mix creation and annihilation operators, preserving the vacuum and the corresponding Fock space. Hence $\hat{a}_{\bnu}$ and $\hat{a}_{\bnu}^\dagger$ also create and destroy microscopic GFT quanta.

Then, the corresponding Bogolyubov transformation becomes a block-diagonal two-mode squeezing transformation mixing only $\pm\bnu$ modes, given by
\begin{equation}
\label{eq:bogotransGFT}
	\begin{aligned}
		\hat{b}_{\boldsymbol{\nu}}  
	= u_{\boldsymbol{\nu}}  \, \hat{a}_{\boldsymbol{\nu}} + v_{\boldsymbol{\nu}} \, \hat{a}^\dagger_{-\boldsymbol{\nu}}\,,\qquad
		\hat{b}^\dagger_{\boldsymbol{\nu}}
	= v_{\boldsymbol{\nu}}  \, \hat{a}_{-\boldsymbol{\nu}} + u_{\boldsymbol{\nu}} \, \hat{a}^\dagger_{\boldsymbol{\nu}}\,,
	\end{aligned}
\end{equation}
where
\begin{equation}\label{eq:uvomega}
    u_{\boldsymbol{\nu}}
    =
    \frac{|\omega^\lambda_{\boldsymbol{\nu}}| + |\omega^0_{\boldsymbol{\nu}}|}{2\sqrt{|\omega^\lambda_{\boldsymbol{\nu}} \,\omega^0_{\boldsymbol{\nu}}|}} \,, \qquad     v_{\boldsymbol{\nu}} 
    =
    \frac{|\omega^\lambda_{\boldsymbol{\nu}}| - |\omega^0_{\boldsymbol{\nu}}|}{2\sqrt{|\omega^\lambda_{\boldsymbol{\nu}} \,\omega^0_{\boldsymbol{\nu}}|}}\,,
\end{equation}
are the real (time-dependent) Bogolyubov coefficients. These can be seen to satisfy  $u^2_{\boldsymbol{\nu}} - v^2_{\boldsymbol{\nu}} =1$ at all relational times which, because they are real, and block diagonal, is enough to guarantee that they satisfy all conditions imposed on Bogolyubov coefficients for the transformation to be canonical.

Because this transformation involves two-mode squeezing, the Bogolyubov vacuum $|0\rangle_b$, defined by $\hat{b}_{\bnu}|0\rangle_b=0\;\forall\;\bnu$, is not the vacuum for atomic GFT excitations. Correspondingly, normal mode operators define a different Fock space, built by acting with the creation operators $\hat{b}_{\bnu}^\dagger$ onto $|0\rangle_b$.

Physically, the Bogolyubov vacuum represents absence of collective excitations on top of the condensate. The new operators do not excite individual GFT atoms, but rather collective excitations of the whole system that arise due to the presence of microscopic interactions, in full analogy with phonons in a laboratory BEC. 

In fact, since $u_{\boldsymbol{\nu}}\overset{\lambda\rightarrow0}{\longrightarrow}1$ and $v_{\boldsymbol{\nu}}\overset{\lambda\rightarrow0}{\longrightarrow}0$, we note that $|0\rangle_b$ becomes the original vacuum in the non-interacting limit, exactly as occurs for atomic BECs, which do not exhibit phononic excitations in the idealised non-interacting case. However, for any non-zero coupling between microscopic quanta, collective modes of vibration of the full system describe its leading order dynamics beyond exact mean-field configurations.

This analogy between phonons in BECs and collective excitations in GFT can be pushed further by looking at the depletion of the condensate induced by the presence of interactions. As explained in section \ref{sec:BogoGFT}, the depletion is characterised by the mean number of out-of-condensate GFT quanta in the Bogolyubov vacuum. Inverting \eqref{eq:bogotransGFT} we find the GFT version of the general formula \eqref{eq:general_depletion_number},
\begin{equation}\label{eq:DEP}
       {}_b\langle 0| \hat{a}_{\boldsymbol{\nu}}^\dagger\hat{a}_{\boldsymbol{\nu}}|0\rangle_b = v_{\boldsymbol{\nu}}^2=\lr{\frac{1}{4}\lambda k_{\bnu}\Phi^2 (\chi)}^2 
    +{O}(\lambda^3)\,, 
\end{equation}
where we have defined 
\begin{equation}
\label{eq:knu}
    k_{\bnu} := \frac{f_{\boldsymbol{\nu}}}{K^{(0)}_{\boldsymbol{\nu}}}\,,
\end{equation}
in terms of which \eqref{eq:omegalambda} and \eqref{eq:uvomega} expand as $\omega^\lambda_{\bnu}=\omega^0_{\bnu}\big(1+\frac12\lambda k_{\bnu}\Phi^2\big)+O(\lambda^2)$, $u_{\bnu}=1+O(\lambda^2)$ and $v_{\bnu}=\frac14\lambda k_{\bnu}\Phi^2+O(\lambda^2)$. Equation~\eqref{eq:DEP} shows that depletion is non-vanishing in presence of interactions, as occurs for laboratory condensates. Note that though the leading order contribution to the effect is $O(\lambda^2)$, it can be fully computed from the $O(\lambda)$ contribution to $v_{\bnu}$. 
    
To conclude we remark that, as expected from the general setting discussed above, the leading corrections to the mean-field dynamics of GFT condensates are given by depletion and Bogolyubov collective excitations. Because the condensate is time-dependent, analysing the dynamics of these corrections, and the corresponding imprints in macroscopic observables, is not straightforward. In the following sections we will address these issues.

\section{Dynamics of GFT collective excitations}\label{sec:EvolutionofBogolons}

In this section we will discuss the dynamics of collective excitations in GFT, where the condensate is time-dependent. We will also characterise the regime in which the corrections that they imprint into the mean field remain controlled. 

To that end, we start by studying the relational evolution of the collective creation and annihilation operators. Since collective modes are normal, their evolution can be studied independently. According to \eqref{eq:BogoHamiltonian}, their relational Hamiltonian is either that of a standard harmonic oscillator or an inverted one, in both cases with a time-dependent frequency $\omega^\lambda_{\boldsymbol{\nu}}(\chi)$ due to the dynamical nature of the condensate. It will prove useful to introduce the integrated parameter
\begin{equation}\label{eq:Theta}
\Theta^\lambda_{\boldsymbol{\nu}}(\chi) := \int_0^\chi \dd\chi'\, \omega^\lambda_{\boldsymbol{\nu}}(\chi') \,,
\end{equation}
which, expanding~\eqref{eq:omegalambda} to order $\lambda$, boils down to
\begin{equation}\label{eq:thetaorderl}
\begin{aligned}
&\Theta^\lambda_{\boldsymbol{\nu}}(\chi) = \omega^0_{\boldsymbol{\nu}}\chi \Big({1+\lambda\Delta \Theta_{\bnu}(\chi)\Big)}+O(\lambda^2)\,, 
\\
&\Delta \Theta_{\bnu}(\chi)\coloneqq \frac{ 1}{4}k_{\bnu}\Phi_0^2\left(1+\frac{\sinh(2 \omega_{\bn_0}\chi)}{2\omega_{\bn_0}\chi} \right)\,,
\end{aligned}
 \end{equation}
where $k_{\bnu}$ is defined in \eqref{eq:knu}. The Heisenberg dynamics for the collective operators is given by
\begin{equation}
\label{eq:HeisEqb}
    {\rm i} \frac{\dd}{\dd \chi} \hat{b}_{\bnu} = \lrsq{\hat{b}_{\bnu},\hat{H}_2}+{\rm i} \frac{\partial}{\partial \chi} \hat{b}_{\bnu}\,,
\end{equation}
where the last term is present due to the (relational) time dependence of the background condensate, which manifests as time dependence of the Bogolyubov coefficients that relate collective excitations to GFT atoms. More precisely, expanding \eqref{eq:uvomega} up to order $\lambda$, we write \eqref{eq:HeisEqb} as
\begin{equation}\label{eq:newHeis}
     {\rm i} \frac{\dd}{\dd \chi} \hat{b}_{\bnu} = \lrsq{\hat{b}_{\bnu},\hat{H}_2} +{\rm i} v'_{\bnu}(\chi)\hat{b}^\dagger_{-\bnu}+O(\lambda^2)\,,
\end{equation}
where the last term would vanish if the background condensate were stationary. For both HO and SQ collective modes, the solution of \eqref{eq:newHeis} takes the general form
\begin{equation}\label{eq:balphabeta}
    \hat{b}_{\bnu} (\chi)= \alpha_{\bnu}(\chi) \hat{b}_{\bnu}(0)+\beta_{\bnu}(\chi)\hat{b}_{-\bnu}^\dagger(0)\,,
\end{equation}
with $\alpha_{\bnu}(0)=1$ and $\beta_{\bnu}(0)=0$. 

The Heisenberg dynamics of the collective operators can then be encoded in a Bogolyubov transformation. Before solving for these coefficients, there are important aspects of the dynamics related to the evolving condensate that we must discuss. 

The relevance of the partial derivative term in \eqref{eq:HeisEqb} can be assessed via an adiabaticity parameter such as \cite{Parker_Toms_2009,mukhanovBOOK}
\begin{equation}\label{eq:epsilon}
    \epsilon_{\bnu} := \left|\frac{\omega_{\bnu}'(\chi)}{\omega_{\bnu}^2(\chi)}\right| = \left|\frac{k_{\bnu}}{\omega_{\bnu}^0} \lambda \Phi(\chi)\Phi'(\chi)\right|
    \,,
\end{equation}
which essentially compares the timescale at which normal modes evolve against the scale characterising the background expansion. If this parameter is small, the physics of the system is well encoded by the adiabatic approximation, in which the state of the system essentially adjusts to the background as the condensate evolves. A sufficient condition to keep $\epsilon_{\bnu}$ small up to the time at which the Bogolyubov approximation breaks down is {that $|k_{\bnu} \,\omega_{\bn_0}/\omega^0_{\bnu}|\lesssim O(1)$, since $\Phi'/\Phi=\omega_{\bn_0}\tanh(\omega_{\bn_0}\chi)$ implies $|\Phi\Phi'|\leq|\omega_{\bn_0}|\Phi^2$, hence $\epsilon_{\bnu}\leq |k_{\bnu}\omega_{\bn_0}/\omega^0_{\bnu}|\,|\lambda|\Phi^2\ll1$.}

In this respect, note that the non-adiabatic term in the Heisenberg equation \eqref{eq:newHeis} enters at the same order $O(\lambda \Phi^2)$ as the adiabatic corrections encoded in the commutator $\big[{\hat{b}_{\bnu},\hat{H}_2}\big]$ for all types of modes. Therefore, we will need to retain this non-adiabatic contribution to ensure a consistent leading-order description. This in turn implies that the normal modes we defined are not dynamically decoupled, as they are not eigenstates of the corresponding evolution operator---except in the adiabatic limit. In other words, our definition of collective excitations entails the choice of diagonalising the Hamiltonian at each instant of time. In the literature of QFT in curved spacetimes, this is known as the \textit{instantaneous lowest-energy state} prescription \cite{mukhanovBOOK,Birrell:1982ix}. The mathematical ``price to pay'' for continuously redefining the vacuum at every moment is the introduction of the non-adiabatic term ${\rm i} v'_{\boldsymbol{\nu}}(\chi)\hat{b}^\dagger_{-\boldsymbol{\nu}}$ in the Heisenberg equation \eqref{eq:newHeis}. We note that since we will apply our framework to GFT cosmology, where the condensate is continuously growing, we lack the other typical options such as the ``in-out'' formalism, which relies on asymptotically static regions \cite{Birrell:1982ix}. While this instantaneous prescription is potentially pathological in highly non-adiabatic settings, the smallness parameter \eqref{eq:epsilon} ensures that the continuous family of vacua remains well-behaved \cite{Birrell:1982ix,mukhanovBOOK}.

{\setlength{\parskip}{1pt}We} remark that these non-adiabatic effects can bring in interesting phenomenology, for instance, a pair-creation mechanism for collective excitations as the GFT condensate expands, quantified by $|\beta_{\bnu}|^2$ \cite{Parker,Parker2,Parker_Toms_2009}. We leave a detailed investigation of the relevance of this mechanism for future work. However, we ask whether the creation of collective excitations during evolution could precipitate the breakdown of the Bogolyubov regime before interaction domination takes place---see \eqref{eq:InteractionBound}. This question is relevant given that the Bogolyubov approximation requires the number of out-of-condensate quanta to remain microscopic compared to the number of condensed quanta. To assess the validity of the Bogolyubov approximation, we define the quantity
\begin{equation}\label{eq:ratio}
R_{\boldsymbol{\nu}}(\chi) := \frac{\langle\hat{{N}}_{\bnu} (\chi)\rangle}{N_\text{c} (\chi)}\,,
\end{equation}
where $\hat{{N}}_{\bnu} (\chi):=\big(\hat{a}^\dagger_{\bnu}(\chi)\hat{a}_{\bnu}(\chi)+\hat{a}^\dagger_{-\bnu}(\chi)\hat{a}_{-\bnu}(\chi)\big)/2$ {and the expectation value is taken in a generic state belonging to the family defined by \eqref{eq:state}}. Independently of whether the modes are of HO or SQ type, one can write this quantity up to $O(\lambda)$ in terms of expectation values for collective operators as
\begin{widetext}
\begin{equation}\label{eq:ratioApprox}
R_{\boldsymbol{\nu}}(\chi) = \frac{\langle\hat{{N}}_{\bnu}^\text{b}(\chi)\rangle}{N_\text{c} (\chi)} - \frac{1}{4}\lambda k_{\bnu}  \frac{\Phi^2(\chi)}{N_\text{c} (\chi)}\langle \hat{b}_{\boldsymbol{\nu}}(\chi)\hat{b}_{-\boldsymbol{\nu}}(\chi)+\hat{b}^{\dagger}_{\boldsymbol{\nu}}(\chi)\hat{b}^{\dagger}_{-\boldsymbol{\nu}}(\chi)\rangle
+    O(\lambda^2)\,,
\end{equation}
\end{widetext}
where we have defined $\hat{N}^\text{b}_{\boldsymbol{\nu}}(\chi):=
	\big(\hat{b}^\dagger_{\boldsymbol{\nu}}(\chi)\hat{b}_{\boldsymbol{\nu}}(\chi)+\hat{b}^\dagger_{-\boldsymbol{\nu}}(\chi)\hat{b}_{-\boldsymbol{\nu}}(\chi)\big)/2$ and we have used the identity
\begin{equation}\label{eq:Nstheta}
\begin{aligned}	\hat{N}_{\boldsymbol{\nu}}(\chi)&=	\hat{N}^\text{b}_{\boldsymbol{\nu}}(\chi)+v_{\boldsymbol{\nu}}^2\left(2	\hat{N}^\text{b}_{\boldsymbol{\nu}}(\chi)+1\right) \\& - u_{\boldsymbol{\nu}}v_{\boldsymbol{\nu}} \left(\hat{b}_{\boldsymbol{\nu}}(\chi)\hat{b}_{-\boldsymbol{\nu}}(\chi)+\hat{b}^{\dagger}_{\boldsymbol{\nu}}(\chi)\hat{b}^{\dagger}_{-\boldsymbol{\nu}}(\chi)\right) \,,
\end{aligned}
\end{equation}
which will be useful in later sections. {Note that the coefficient $\Phi^2/N_\text{c}$ in \eqref{eq:ratioApprox} is bounded, see \eqref{eq:PhiNcRatio}.}

The above expressions make apparent that the specific behaviour of $R_{\boldsymbol{\nu}}(\chi)$ does depend on the type of collective mode considered. Hence, we now proceed to determine the coefficients $\alpha(\chi)$ and $\beta(\chi)$ defined in \eqref{eq:balphabeta} for HO and SQ collective modes. We will then use the results to discuss the behaviour of $R_{\bnu}(\chi)$ for each type of mode, assessing whether the regime of validity of the Bogolyubov approximation needs to be modified in the presence of stable or unstable Bogolyubov sectors.

\subsection{Collective HO modes}\label{sec:subsecHO}

For HO modes we have $\lrsq{\hat{b}_{\bnu},\hat{H}_2}=\omega_{\bnu}^\lambda(\chi) \hat{b}_{\bnu}$, so that the Heisenberg equation \eqref{eq:newHeis} up to $O(\lambda)$ reads 
\begin{equation}\label{eq:orders}
    {\rm i} \frac{\dd}{\dd \chi} \hat{b}_{\boldsymbol{\nu}} = \Big(\omega^0_{\boldsymbol{\nu}}  +   \frac{1}{2}\lambda k_{\boldsymbol{\nu}} {\omega^0_{\boldsymbol{\nu}} \Phi^2(\chi) \Big) \hat{b}_{\boldsymbol{\nu}}}  +  \frac{\rm i}{2}\lambda k_{\bnu}  \Phi(\chi)\Phi'(\chi)\,\hat{b}^\dagger_{-\boldsymbol{\nu}} \,,
\end{equation}
where we used the expansions below \eqref{eq:knu}. We clearly see that the microscopic interactions give rise to two types of contributions: a dynamical shift in the effective frequency, driven by the condensate amplitude, and a purely non-adiabatic correction due to the time evolution of the condensate. The solution of \eqref{eq:orders} is of the form \eqref{eq:balphabeta}, where the explicit expression for the coefficients $\alpha_{\bnu}(\chi)$ and $\beta_{\bnu}(\chi)$ is found by solving the coupled differential equations
\begin{equation}\label{eq:alphabeta}
    \begin{aligned}
        {\rm i} \alpha'_{\bnu}(\chi) &= \omega_{\bnu}^\lambda(\chi) \alpha_{\bnu}(\chi){+{\rm i} v'_{\bnu}(\chi) \beta^*_{-\bnu}(\chi)}\,,\\
     {\rm i} \beta'_{\bnu}(\chi) &= \omega_{\bnu}^\lambda(\chi) \beta_{\bnu}(\chi)+{\rm i} v'_{\bnu}(\chi) \alpha^*_{-\bnu}(\chi)\,,
    \end{aligned}
\end{equation}
perturbatively in $\lambda$, which leads to
\begin{equation}
    \label{eq:alpha}
    \alpha_{\boldsymbol{\nu}}(\chi)
    = \ee^{-{\rm i}\omega^0_{\boldsymbol{\nu}}\chi} \Big( 1 - {\rm i} \lambda \omega^0_{\boldsymbol{\nu}}\chi  \Delta\Theta_{\bnu}(\chi)\Big) +   O(\lambda^2) \,,
\end{equation}
and
\vspace{2cm}
\begin{widetext}
\begin{align}
    &\label{eq:beta}
    \beta_{\bnu } (\chi) = \frac{\lambda k_{\boldsymbol{\nu}}}{8\left( (\omega^0_{\boldsymbol{\nu}})^2 + \omega_{\bn_0}^2  \right)} 
    \left[ \Big( \big(\Phi'(\chi)\big)^2 +\omega_{\bn_0}^2 \Phi^2(\chi)- 2{\rm i}\omega^0_{\boldsymbol{\nu}} \Phi(\chi)\Phi'(\chi) \Big)\ee^{{\rm i}\omega^0_{\boldsymbol{\nu}}\chi}  - \omega_{\bn_0}^2 \Phi_0 ^2 \ee^{-{\rm i}\omega^0_{\boldsymbol{\nu}}\chi} \right]  +   O(\lambda^2) \,,
\end{align}
where $\Delta\Theta_{\bnu}(\chi)$ is given in \eqref{eq:thetaorderl} and we have used the mean-field equation $\Phi''(\chi) = \omega_{\bn_0}^2  \Phi(\chi)$. Note that these satisfy
$\alpha_{\boldsymbol{\nu}} = \alpha_{-\boldsymbol{\nu}}$, $\beta_{\boldsymbol{\nu}} = \beta_{-\boldsymbol{\nu}}$, and the condition $|\alpha_{\boldsymbol{\nu}}(\chi)|^2 - |\beta_{\boldsymbol{\nu}}(\chi)|^2 = 1$ to the considered order as expected, ensuring that the canonical commutation relations for the operators \eqref{eq:balphabeta} are  dynamically preserved.

We can now calculate $R_{\bnu}$ for HO modes at $O(\lambda)$. To that end, we first write explicitly the operators
\begin{align}
    \label{eq:Nb}
    &\hat{N}^\text{b}_{\boldsymbol{\nu}}(\chi) = {\hat{N}^\text{b}_{\boldsymbol{\nu}}}(0) +  \frac{1}{2}\lambda k_{\bnu}\lr{\tilde{\beta}_{\bnu}(\chi) \ee^{{\rm i}\omega_{\boldsymbol{\nu}}^0 \chi}  \hat{b}_{\boldsymbol{\nu}}^\dagger(0) \hat{b}_{-\boldsymbol{\nu}}^\dagger(0) + \tilde{\beta}_{\bnu}^*(\chi) \ee^{-{\rm i}\omega_{\boldsymbol{\nu}}^0 \chi}\hat{b}_{\boldsymbol{\nu}} (0)\hat{b}_{-\boldsymbol{\nu}}(0)}
    +   O(\lambda^2)\,,
    \\
    \label{eq:A}
    &\hat{b}_{\boldsymbol{\nu}}(\chi)\hat{b}_{-\boldsymbol{\nu}}(\chi)+\hat{b}^{\dagger}_{\boldsymbol{\nu}}(\chi)\hat{b}^{\dagger}_{-\boldsymbol{\nu}}(\chi) = \ee^{-2{\rm i}\omega_{\boldsymbol{\nu}}^0 \chi} 
    \hat{b}_{\boldsymbol{\nu}}(0)\hat{b}_{-\boldsymbol{\nu}}(0) \, + \ee^{2{\rm i}\omega_{\boldsymbol{\nu}}^0 \chi} \hat{b}^\dagger_{\boldsymbol{\nu}}(0)\hat{b}^\dagger_{-\boldsymbol{\nu}}  (0)  +  O(\lambda)\,,
\end{align}
\end{widetext}
where $\tilde{\beta}_{\bnu}(\chi)$ is defined by $\beta_{\bnu } (\chi) \coloneqq \frac{1}{2} \lambda k_{\bnu} \tilde{\beta}_{\bnu}(\chi)$, so it is independent of $\lambda$. While the expectation value of \eqref{eq:A} describes oscillations and hence does not contribute to the growth of $\langle\hat{N}_{\bnu}\rangle$, the expectation value of the number of collective excitations \eqref{eq:Nb} consists of a constant zeroth-order term (reflecting the conserved particle number typical of the HO-like dynamics) and a non-adiabatic correction due to the time-dependence of the background condensate. The latter imprints a growth in $ \langle\hat{N}^\text{b}_{\boldsymbol{\nu}}(\chi)\rangle$ at the same rate as the condensate---specifically, it imprints oscillations bounded by a growing envelope given by ${\beta}_{\bnu}(\chi)$, which grows as $\sim\lambda \Phi^2(\chi)$ in the Bogolyubov regime. Crucially, however, the growth of \eqref{eq:Nb} does not spoil the validity of the Bogolyubov regime, because the population of collective excitations relative to the condensate remains bounded. To see this, we write the expectation value of the total number of quanta \eqref{eq:Nstheta} in a generic state as
\begin{widetext}
    \begin{equation}\label{eq:NnuHO}
     \langle  \hat{N}_{\bnu}(\chi)\rangle =  \langle\hat{N}^\text{b}_{\boldsymbol{\nu}}(0)\rangle   +\frac{1}{4}\lambda k_{\bnu}\left[ 2\Big( \ee^{{\rm i}\omega_{\boldsymbol{\nu}}^0 \chi} \tilde{\beta}_{\boldsymbol{\nu}}(\chi) \langle \hat{b}_{\boldsymbol{\nu}}^\dagger(0) \hat{b}_{-\boldsymbol{\nu}}^\dagger (0)\rangle + \text{c.c.} \Big) 
     -\Phi^2(\chi)\Big( \ee^{-2{\rm i}\omega_{\boldsymbol{\nu}}^0 \chi} \langle \hat{b}_{\boldsymbol{\nu}}(0)\hat{b}_{-\boldsymbol{\nu}}(0) \rangle + \text{c.c.}     \Big)\right]  +   O(\lambda^2)\,,
\end{equation}
\end{widetext}
which allows to schematically write $R_{\bnu}(\chi)$ for HO modes:
\begin{equation}\label{eq:RHO}
\begin{aligned}
	R_{\text{HO}}(\chi) =& \frac{\text{constant}}{N_\text{c}(\chi)}+ \frac{1}{4} \lambda k_{\bnu} (\text{oscillations}) +   O(\lambda^2)\,,
    \end{aligned}
\end{equation}
where the oscillations collect two bounded, $\lambda$-independent contributions: the non-adiabatic term of \eqref{eq:Nb}, involving $\tilde\beta_{\bnu}/N_\text{c}$, {and the anomalous term of \eqref{eq:ratioApprox}, involving $\Phi^2/N_\text{c}$. Both ratios are bounded by $2/\Omega_{\bn_0}$ as a consequence of \eqref{eq:PhiNcRatio} (for the former, note that $|\tilde\beta_{\bnu}|<\Phi^2$ follows from \eqref{eq:beta} and $|\Phi'|\leq|\omega_{\bn_0}|\Phi$).} Hence, a sufficient condition to guarantee that $R_{\text{HO}}(\chi) \ll 1$ is that the initial state is such that $\langle\hat{N}^\text{b}_{\bnu}(0)\rangle\ll N_{\rm c}(\chi)$ and $|k_{\bnu}\langle \hat{b}_{\bnu}(0)\hat{b}_{-\bnu}(0) \rangle|/\Omega_{\bn_0} \lesssim O(1)$.

To conclude, we remark that non-adiabatic effects due to the condensate evolution, characterised by $\beta_{\boldsymbol{\nu}}\neq 0$, generically lead to the creation of GFT bogolon pairs with opposite labels via two-mode squeezing. Note that, as for depletion, the leading contribution to this effect, given by $|\beta_{\bnu}|^2$, appears only at $O(\lambda^2)$. In addition, when the initial state contains anomalous correlations, $\langle \hat{b}_{\boldsymbol{\nu}}(0)\hat{b}_{-\boldsymbol{\nu}}(0) \rangle \neq 0$, non-adiabatic effects show in \eqref{eq:NnuHO} at linear order in $\lambda$. Nonetheless, because these linear effects only induce bounded oscillations in the relative population, similarly to the adiabatic terms of \eqref{eq:A}, HO modes do not spoil the Bogolyubov approximation before interaction domination---in fact, one can show that $R_\text{HO}(\chi_*)\sim O(\lambda)$. Thus, for HO modes, the breakdown of the Bogolyubov approximation does not occur due to the growth of collective excitations, but rather to interactions becoming dominant, spoiling the validity of the mean-field expansion at times later than $\chi_*$, as discussed below \eqref{eq:InteractionBound}.

\subsection{Collective SQ modes}\label{sec:subsecSQ}

For SQ modes, on the other hand, we have $\lrsq{\hat{b}_{\bnu},\hat{H}_2}=\omega_{\bnu}^\lambda(\chi) \hat{b}^\dagger_{-\bnu}$, so the Heisenberg equation in this case reads
\begin{equation}
    {\rm i} \frac{\dd}{\dd \chi} \hat{b}_{\boldsymbol{\nu}} = \Big( \omega_{\boldsymbol{\nu}}^0+\frac{1}{2}\lambda k_{\bnu}\omega_{\boldsymbol{\nu}}^0 \Phi^2(\chi)   +    \frac{\rm i}{2}\lambda k_{\bnu}  \Phi(\chi) \Phi'(\chi) \Big) \hat{b}^\dagger_{-\boldsymbol{\nu}}\,.
\end{equation}
We see two contributions: the correction to the squeezing parameter proportional to $\Phi^2$, and the non-adiabatic term proportional to $\Phi\Phi'$. In contrast to the HO case, however, both contributions now source the creation operator $\hat{b}^\dagger_{-\boldsymbol{\nu}}$, and hence drive squeezing. Similarly to the previous subsection, the above equation is solved by \eqref{eq:balphabeta} where the Bogolyubov coefficients must satisfy
\begin{widetext}
\begin{equation}
\begin{aligned}
{\rm i} \alpha'_{\boldsymbol{\nu}}(\chi) &= \Big( \omega_{\boldsymbol{\nu}}^\lambda(\chi) + {\rm i} v'_{\boldsymbol{\nu}}(\chi) \Big) \beta^*_{-\boldsymbol{\nu}}(\chi)\,,\\
{\rm i} \beta'_{\boldsymbol{\nu}}(\chi) &= \Big( \omega_{\boldsymbol{\nu}}^\lambda(\chi) + {\rm i} v'_{\boldsymbol{\nu}}(\chi) \Big) \alpha^*_{-\boldsymbol{\nu}}(\chi)\,.
\end{aligned}
\end{equation}
The solution up to $O(\lambda)$ reads
\begin{align}
\begin{split}
\alpha_{\boldsymbol{\nu}}(\chi) =& \cosh\big(\omega^0_{\boldsymbol{\nu}}\chi\big) 
+ \lambda\omega^0_{\boldsymbol{\nu}}\chi \Delta\Theta_{\boldsymbol{\nu}}(\chi)\sinh \big(\omega^0_{\boldsymbol{\nu}}\chi\big)\\&- \frac{{\rm i}\lambda k_{\boldsymbol{\nu}} \Phi(\chi)}{4\big((\omega^0_{\boldsymbol{\nu}})^2 - \omega_{\bn_0}^2\big)}
\Big[
\omega_{\bn_0}^2\Phi(\chi)\sinh\big(\omega^0_{\boldsymbol{\nu}}\chi\big)
- \omega^0_{\boldsymbol{\nu}}\Phi'(\chi)\cosh\big(\omega^0_{\boldsymbol{\nu}}\chi\big)
\Big] +   O(\lambda^2)
\,,
\end{split}
\\
\begin{split}
\beta_{\boldsymbol{\nu}}(\chi) =&
-{\rm i}
\sinh\big(\omega^0_{\boldsymbol{\nu}}\chi\big) 
-{\rm i} \lambda\omega^0_{\boldsymbol{\nu}}\chi \Delta\Theta_{\boldsymbol{\nu}}(\chi)\cosh\big(\omega^0_{\boldsymbol{\nu}}\chi\big)\\& - \frac{\lambda k_{\boldsymbol{\nu}} \Phi'(\chi)}{4\big((\omega^0_{\boldsymbol{\nu}})^2 - \omega_{\bn_0}^2\big)}
\Big[
 \Phi'(\chi)\cosh\big(\omega^0_{\boldsymbol{\nu}}\chi\big)
-\omega^0_{\boldsymbol{\nu}}\Phi(\chi)\sinh\big(\omega^0_{\boldsymbol{\nu}}\chi\big)
\Big]+   O(\lambda^2)
\,,
\end{split}
\label{eq:BogoCoefSQmodes}
\end{align}
\end{widetext}
which again satisfy $\alpha_{\bnu}=\alpha_{-\bnu}$, $\beta_{\bnu}=\beta_{-\bnu}$, and $|\alpha_{\boldsymbol{\nu}}(\chi)|^2 - |\beta_{\boldsymbol{\nu}}(\chi)|^2 = 1$ as expected.

These solutions show a crucial physical difference between these SQ modes and the HO modes. In the HO case, the corresponding Bogolyubov sector is adiabatically stable, with any squeezing-like behaviour exclusively driven by the dynamics of the condensate background and the presence of interactions. Mathematically, this is why $\beta_{\boldsymbol{\nu}}(\chi)$ in \eqref{eq:beta} vanishes if $\Phi'(\chi)=0\;\forall\chi$ or $\lambda=0$. For the SQ modes, on the other hand, the free theory is already intrinsically unstable (see discussion below \eqref{eq:Hsumsinglemode} in section~\ref{sec:QGcosmo}). This is why $\beta_{\bnu}(\chi)$ does not vanish even if $\Phi'(\chi)=0\;\forall \chi$. Moreover, the non-interacting limit of \eqref{eq:BogoCoefSQmodes} shows that both Bogolyubov coefficients retain $\lambda$-independent contributions, so that the evolution of these modes is still governed by squeezing. This result should not be surprising, given that in the non-interacting limit collective excitation operators reduce to the original GFT atomic operators---see section \ref{sec:bogotransGFT}. In turn, this shows that the unstable Bogolyubov sector has origin in the fundamental unstable microscopic sector of the GFT. 

Following similar steps as for HO modes, we now compute $R_{\bnu}(\chi)$ for SQ modes, to assess whether their growth could break down the validity of the Bogolyubov regime before interactions start dominating the dynamics. To that end, we compute the relational evolution of the relevant operators up to $O(\lambda)$. These are
\begin{equation}
\hat{b}_{\bnu}(\chi)\hat{b}_{-\bnu}(\chi)+\text{h.c.}=\hat{b}_{\bnu}(0)\hat{b}_{-\bnu}(0)+\text{h.c.}+O(\lambda)\,,
\end{equation}
which is constant in relational time, and
\begin{widetext}
\begin{equation}
\begin{aligned}
\label{eq:NbSQ}
\hat{{N}}_{\bnu}^\text{b}(\chi) =& -\frac{1}{2} + \left(\hat{{N}}_{\bnu}^\text{b}(0)+\frac{1}{2}
\right)\Big[\cosh\big(2\omega^0_{\boldsymbol{\nu}}\chi\big) 
+ 2\lambda\omega^0_{\boldsymbol{\nu}}\chi \Delta\Theta_{\boldsymbol{\nu}}(\chi)\sinh \big(2\omega^0_{\boldsymbol{\nu}}\chi\big)\Big] \\ &+\frac{{\rm i }   }{2}\left(\hat{b}_{\boldsymbol{\nu}}(0)\hat{b}_{-\boldsymbol{\nu}}(0)-\hat{b}^{\dagger}_{\boldsymbol{\nu}}(0)\hat{b}^{\dagger}_{-\boldsymbol{\nu}}(0)\right) \Big[\sinh\left(2\omega^0_{\boldsymbol{\nu}}\chi\right) 
+2 \lambda\omega^0_{\boldsymbol{\nu}}\chi \Delta\Theta_{\boldsymbol{\nu}}(\chi)\cosh\big(2\omega^0_{\boldsymbol{\nu}}\chi\big)\Big] \\ 
&+ 
\frac{\lambda k_{\boldsymbol{\nu}} \omega_{\bn_0}^2\Phi_0^2}{8\big((\omega^0_{\boldsymbol{\nu}})^2-\omega_{\bn_0}^2\big)} \left( \hat{b}_{\boldsymbol{\nu}}^\dagger (0)\hat{b}_{-\boldsymbol{\nu}}^\dagger (0)+ \hat{b}_{\boldsymbol{\nu}}(0) \hat{b}_{-\boldsymbol{\nu}} (0)\right)\Big[ \cosh\big(2\omega^0_{\boldsymbol{\nu}}\chi\big)-\cosh\big(2\omega_{\bn_0}\chi\big) \Big]  +   O(\lambda^2) \,, 
\end{aligned}
\end{equation}
\end{widetext}
where we used $\Phi'^2=\omega_{\bn_0}^2\big(\Phi^2-\Phi_0^2\big)$, i.e.\ the conservation law stated just above \eqref{eq:PhiNc}. Again, we can clearly identify two beyond-mean-field contributions due to the interactions. In the first two lines of \eqref{eq:NbSQ}, we recognise the same evolution undergone by the condensate \eqref{eq:condensateevolution} but with a ${O}(\lambda)$ modified squeezing parameter as defined in \eqref{eq:thetaorderl}. In the third line, we see the purely non-adiabatic effect due to the time-dependence of the background condensate (as its prefactor $\omega_{\bn_0}^2\Phi_0^2=\Phi_0\Phi''(0)$ vanishes for a static condensate). With these ingredients, we can write $R_{\bnu}(\chi)$ for SQ modes schematically as
\begin{equation}\label{eq:RSQ}
R_\text{SQ}(\chi) = \frac{\langle\hat{{N}}_{\bnu}^\text{b}(\chi) \rangle}{N_\text{c}(\chi)} - \frac{1}{4}\lambda k_{\bnu} \frac{\Phi^2(\chi)}{N_\text{c}(\chi)}(\text{constant}) 
 +  O(\lambda^2)\,.
\end{equation}
{The second term is a bounded $O(\lambda)$ contribution, since $\Phi^2/N_\text{c}\leq2/\Omega_{\bn_0}$ by \eqref{eq:PhiNcRatio}.} Hence, to assess whether \eqref{eq:RSQ} remains small, we only need to check how the rate of growth of collective excitations compares with that of the condensate. Because the free squeezing modes are subdominant with respect to $\bn_0$ by assumption (i.e., $|\omega_{\boldsymbol{\nu}}^0| \equiv |\omega_\mathbf{n}| <|\omega_\mathbf{n_0}|$), the purely non-adiabatic terms in \eqref{eq:NbSQ} either decay relative to $N_\text{c}(\chi)$ or match its growth, contributing at most with a constant $  O(\lambda)$ offset to the ratio \eqref{eq:RSQ}. Therefore, the dynamics of $R_\text{SQ}$ are dominated by the adiabatic terms of $\hat{{N}}_{\bnu}^\text{b}(\chi)$---the first lines of \eqref{eq:NbSQ}---where again the perturbative effects coming from $\lambda \Delta \Theta_{\boldsymbol{\nu}}(\chi)$ will remain subleading corrections with respect to the zeroth-order squeezing behaviour. To test the relative importance of collective excitations more precisely, we compare the squeezing rate $|\Theta_{\boldsymbol{\nu}}^\lambda(\chi)|$ against $ |\omega_\mathbf{n_0} |\chi$. In practice, it suffices to examine the difference 
\begin{widetext}
\begin{equation}\label{eq:SQcomparison}
|\Theta_{\boldsymbol{\nu}}^\lambda(\chi)| -  |\omega_\mathbf{n_0}|\chi  = \left[|\omega_{\boldsymbol{\nu}}^0| - | \omega_\mathbf{n_0}| + \frac{1}{4}\lambda k_{\bnu}|\omega_{\boldsymbol{\nu}}^0 |\Phi_0^2  \left(  1+\frac{\sinh(2\omega_\mathbf{n_0} \chi)}{2\omega_\mathbf{n_0} \chi}\right)  \right]
\chi 
+   O(\lambda^2) \,,
\end{equation}
\end{widetext}
where we used the explicit expression \eqref{eq:thetaorderl}. Then, since $\sinh x/x\geq1$, the sign of the $\lambda$-dependent correction in \eqref{eq:SQcomparison} depends on the relative sign between $\lambda$ and $k_{\bnu}$ defined in \eqref{eq:knu}. Modes for which $\lambda k_{\bnu}<0$, squeeze even slower than the free SQ modes, and are therefore irrelevant. On the other hand, for modes such that $\lambda k_{\bnu}>0$ the first-order correction does increase the rate of growth of the collective excitations; however, such a small correction is insufficient to overcome the squeezing rate of the condensate before interaction domination breaks down the Bogolyubov regime at time $\chi_*$. 

This leads to the general conclusion that SQ normal modes are subdominant within the regime of validity of the Bogolyubov approximation regardless of the choice of GFT parameters, {and independently of any comparison with HO modes (since no meaningful statement can be made by extrapolating either sector beyond this regime, where the perturbative treatment itself ceases to be under control)}. For this reason, SQ normal modes will not be considered for the analytical calculations in the next section, where we discuss cosmological implications of the collective excitations on top of the GFT condensate. For completeness, we will nonetheless illustrate their behaviour in figure \ref{fig:GFTBogoCosmo}, showing that their contribution remains negligible in a cosmological context within the regime of validity of the approximation.

\section{Beyond mean field in GFT condensate cosmology}\label{sec:GFTbogocosmo}

In this section we will discuss the macroscopic imprints that collective excitations and depletion leave within the specific model of GFT condensate cosmology discussed in section \ref{sec:GFTcosmo}.

Following the definitions of section \ref{sec:QGcosmo} (see \eqref{eq:general_volume_operator}), we begin by introducing the expectation value of the volume
\begin{equation}
    V (\chi) = \mathfrak{v}_\text{c} N_\text{c}(\chi) + \intsum\intsumtwo \mathfrak{v}_{\bnu\bnu'}\langle \hat{a}_{\bnu}^\dagger(\chi) \hat{a}_{\bnu'}(\chi) \rangle\,,
\end{equation}
where we denote $\mathfrak{v}_\text{c}\equiv \mathfrak{v}_{\mathbf{n_0}}$ and $\mathfrak{v}_{\bnu\bnu'}$ are functions of the $\mathfrak{v}_{\bn\neq\bn_0}$.\footnote{The operator $\hat{V}_\text{out}:= \sum_{\mathbf n\neq \bn_0} \mathfrak{v}_{\mathbf n}\hat a^\dagger_{\mathbf n}\hat a_{\mathbf n}$ is non-diagonal in the $\boldsymbol\nu$–basis, and has the general form \eqref{eq:general_volume_operator} with $\mathfrak{v}_{\bnu\bnu'}=\frac{1}{2\pi} \sum_m \mathfrak{v}_{\mathbf{s},m,\varepsilon} \,\ee^{{\rm i}m (\theta-\theta')} $, see \eqref{eq:smlabels}. States for which $\langle \hat{a}^\dagger_{\bnu }\hat{a}_{\bnu'}\rangle\propto \delta_{\bnu,\bnu'} N_{\bnu}$ yield
    $\langle\hat{V}_\text{out}\rangle= \sum_{\mathbf{s},m,\varepsilon} \mathfrak{v}_{\mathbf{s},m,\varepsilon}  \langle\hat{a}^\dagger _{\mathbf{s},m,\varepsilon} \hat{a}_{\mathbf{s},m,\varepsilon}\rangle
    = \sum_{\mathbf{s},\varepsilon} \int \dd \theta \left[\frac{1}{2\pi} \sum_m \mathfrak{v}_{\mathbf{s},m,\varepsilon}\right] N_{\mathbf{s},\varepsilon}(\theta) = \intsumTEXT_{\bnu} \mathfrak{v}_{\bnu} N_{\bnu} 
$,
where $\mathfrak{v}_{\bnu} :=  \frac{1}{2\pi} \sum_m \mathfrak{v}_{\mathbf{s},m,\varepsilon}$ is the average over sites in the chains of fig.~\ref{fig:bi-chainMAIN}. {For concrete computations this is convergent because the chains can be thought as containing finitely many sites (see below \eqref{eq:VolumeBogo}).}\label{footnote:vnunu}} Just like in \eqref{eq:Friedmann}, $V(\chi)$ is the expectation value of \eqref{eq:volume} taken in any state defined by the properties discussed in \eqref{eq:state}. Within this class of states, we focus on those with mode-diagonal occupations for the non-condensate modes\footnote{Note that an initial state within this sub-class will remain there after relational evolution for the collective modes.} so that $\langle \hat{a}^\dagger_{\bnu }(\chi)\hat{a}_{\bnu'}(\chi)\rangle
=N_{\bnu}(\chi)\,  \delta_{\bnu,\bnu'}$ and we will use the simpler version of the above equation
\begin{equation}\label{eq:VolumeBogo}
    V (\chi) = \mathfrak{v}_\text{c} N_\text{c}(\chi) + \intsum \mathfrak{v}_{\bnu}N_{\bnu}(\chi)\,,
\end{equation}
which neglects coherences. This assumption is made to simplify the analytic calculations below, although similar results can be derived if it is lifted. The coefficients $\mathfrak{v}_{\boldsymbol\nu}$ in \eqref{eq:VolumeBogo} are interpreted as effective averaged volumes for out-of-condensate quanta arising from the microscopic volumes $\mathfrak{v}_{\mathbf n}$ (see footnote \ref{footnote:vnunu}). {In practice, as is customary in concrete GFT computations, one retains finitely many representation labels $\bn$ in the fundamental theory, although their number may be arbitrarily large. It follows that the sum in \eqref{eq:VolumeBogo} contains finitely many terms, as in the free theory, and that the chains of figure \ref{fig:bi-chainMAIN} are idealisations (each of them containing effectively only finitely many sites). We remark that out-of-condensate excitations remain perturbative in the regime of validity, so that higher modes are not significantly populated and thus contribute negligibly to the volume observable.}

With \eqref{eq:VolumeBogo}, we can explore the corrections to mean-field GFT cosmology \eqref{eq:Friedmann} due to GFT collective excitations. Restricting to HO collective modes, and using the results of section \ref{sec:subsecHO}, we have the single-mode expression
\begin{widetext}
    \begin{equation}\label{eq:NnuchiHO}
\hat{N}_{\boldsymbol{\nu}}(\chi) = \hat{N}^\text{b}_{\boldsymbol{\nu}} (0)+ \frac{1}{4}\lambda k_{\boldsymbol{\nu}} \left[ 2\Big( \ee^{{\rm i}\omega_{\boldsymbol{\nu}}^0 \chi} \tilde{\beta}_{\boldsymbol{\nu}}(\chi) \hat{b}_{\boldsymbol{\nu}}^\dagger(0) \hat{b}_{-\boldsymbol{\nu}}^\dagger (0)+ \text{h.c.} \Big) - \Phi^2(\chi)\Big( \ee^{-2{\rm i}\omega_{\boldsymbol{\nu}}^0 \chi} \hat{b}_{\boldsymbol{\nu}}(0)\hat{b}_{-\boldsymbol{\nu}}(0) + \text{h.c.} \Big)\right] +   O(\lambda^2)\,,
\end{equation}
\end{widetext}
which, in the non-interacting limit, reduces to the initial number of quanta in $\boldsymbol{\nu}$. Taking expectation values, we find
\begin{equation}
\begin{aligned}
N_{\boldsymbol{\nu}}(\chi) &= N^\text{b}_{\boldsymbol{\nu}} + \frac{1}{2}\lambda k_{\boldsymbol{\nu}} \Big[ 2\text{Re}\big(B_{\boldsymbol{\nu}}  \tilde{\beta}_{\boldsymbol{\nu}}^*(\chi)  \ee^{-{\rm i}\omega_{\boldsymbol{\nu}}^0 \chi}\big) \\&\quad- \Phi^2(\chi) \text{Re}\big( B_{\boldsymbol{\nu}} \ee^{-2{\rm i}\omega_{\boldsymbol{\nu}}^0 \chi} \big) \Big] +   O(\lambda^2)\,,\\
 N'_{\boldsymbol{\nu}}(\chi) &= -  \lambda k_{\boldsymbol{\nu}} \omega_{\boldsymbol{\nu}}^0 \Phi^2(\chi) \text{Im} \big( B_{\boldsymbol{\nu}}\ee^{-2{\rm i}\omega_{\boldsymbol{\nu}}^0 \chi}  \big) +   O(\lambda^2) \,,
\end{aligned}
\end{equation}
where we used $\tilde{\beta}_{\boldsymbol{\nu}}'(\chi) + {\rm i}\omega_{\boldsymbol{\nu}}^0 \tilde{\beta}_{\boldsymbol{\nu}}(\chi) = \Phi(\chi)\Phi'(\chi) \ee^{{\rm i}\omega^0_{\boldsymbol{\nu}}\chi}$ (which follows from \eqref{eq:alphabeta}) and we defined $N^\text{b}_{\bnu}:= \langle \hat{N}^\text{b}_{\bnu}(0)\rangle$ and $B_{\bnu}:= \langle\hat{b}_{\boldsymbol{\nu}}(0)\hat{b}_{-\boldsymbol{\nu}}(0)  \rangle$ in terms of initial conditions. Since we want to incorporate multiple out-of-condensate modes, we introduce the quantity
\begin{equation}\label{eq:DeltaV}
    \Delta V(\chi) := \intsum \mathfrak{v}_{\boldsymbol{\nu}} k_{\boldsymbol{\nu}} \text{Re}\Big[ B_{\boldsymbol{\nu}} \Big( \tilde{\beta}_{\boldsymbol{\nu}}^*(\chi) \ee^{-{\rm i}\omega_{\boldsymbol{\nu}}^0 \chi} - \frac{\Phi^2(\chi)}{2} \ee^{-2{\rm i}\omega_{\boldsymbol{\nu}}^0 \chi}\Big)\Big] \,,
\end{equation}
which is related to the generic definition in \eqref{eq:general_volume_split}. This allows to cast the effective Friedmann equation of GFT condensate cosmology as
\begin{equation}\label{eq:FrVdeltaV}
    \left( \frac{V'(\chi)}{V(\chi)} \right)^2 = \left( \frac{ V_\text{c}'(\chi) + \lambda \Delta V'(\chi) }{ V_\text{c}(\chi) + V_\text{b} + \lambda \Delta V(\chi) } \right)^2\,,
\end{equation}
where $V_\text{c}(\chi) =  \mathfrak{v}_\text{c} N_\text{c}(\chi)$ and $V_\text{b}:=\intsumTEXT_{\bnu} \mathfrak{v}_{\bnu} N^\text{b}_{\bnu}$ represents a \textit{constant} shift in the initial volume solely due to the number of collective excitations at $\chi=0$. In the free theory limit, \eqref{eq:FrVdeltaV} reduces to $[V_\text{c}'(\chi)/(V_\text{c}(\chi) + V_\text{b})]^2$ and the quantity $V_\text{b}$ reduces to $V_\text{HO}^\text{free}:=\sum_{\mathbf{n}} \mathfrak{v}_\bn N_\bn$, which corresponds to the initial number of quanta for all the modes $\mathbf{n}\in\mathfrak{N}^\text{HO}$ evolving with the HO-type free Hamiltonian in \eqref{eq:freeham}. Since the contribution of $V_\text{b}$ is subdominant, we will also show simplified expressions with $V_\text{b}=0$ later on.

Remarkably, we can obtain an analytic expansion in inverse volume powers for the effective Friedmann equation \eqref{eq:FrVdeltaV}, generalising the one obtained for the mean-field \eqref{eq:Friedmann}. To that end, we first need to reorganise the expression \eqref{eq:beta} for the non-adiabatic coefficient $\beta_{\bnu}(\chi)$ to isolate the contribution that scales with the number of condensed quanta $N_\text{c}(\chi)$. {Since \eqref{eq:beta} is built from the condensate mean field, this requires trading $\Phi^2$ for $N_\text{c}$ by means of \eqref{eq:PhiNc}.} We thus introduce
\begin{equation}\label{eq:JS}
    J_{n}(\chi) := \intsum  \, \frac{\mathfrak{v}_{\boldsymbol{\nu}} k_{\boldsymbol{\nu}}}{2\Omega_{\bn_0}}  \text{Re} \left[ B_{\boldsymbol{\nu}}^* \left( \frac{\omega_{\bn_0}}{\omega_{\bn_0} + {\rm i}\omega_{\boldsymbol{\nu}}^0} \right)^n \ee^{2{\rm i}\omega_{\boldsymbol{\nu}}^0 \chi} \right]\,,
\end{equation}
and
\begin{equation}\label{eq:SJ}
S(\chi) := - N_\text{c}(0)\Big(J_1(0) + J_0(\chi)+ \int_0^\chi \dd \tilde{\chi} \, J_1'(\tilde{\chi}) \ee^{-2\omega_{\bn_0}\tilde{\chi}} \Big)\,,
\end{equation}
in terms of which $\Delta V(\chi) = N_{\text{c}}(\chi)\big(J_1(\chi)-J_0(\chi)\big) +S(\chi)$ and $\Delta V'(\chi) = - \big(N_\text{c}(\chi)+N_\text{c}(0)\big) J_0'(\chi)$. The volume scaling of \eqref{eq:DeltaV}---implicitly hidden in $\tilde{\beta}_{\boldsymbol{\nu}}(\chi)$ and $\Phi^2(\chi)$---is now manifestly separated from the remainder $S(\chi)$. Crucially, the functions $J_n(\chi)$ and $S(\chi)$ have a strictly bounded behaviour, characterised by oscillatory and decaying contributions.\footnote{This holds for $\chi\geq0$. For $\chi<0$, where $N_\text{c}$ grows as $\ee^{-2\omega_{\bn_0}\chi}$, the same $\Delta V$ must be split differently to keep $S$ bounded: $\omega_{\bn_0}+{\rm i}\omega^0_{\bnu}\to\omega_{\bn_0}-{\rm i}\omega^0_{\bnu}$ in \eqref{eq:JS} and $\ee^{-2\omega_{\bn_0}\tilde{\chi}}\to\ee^{2\omega_{\bn_0}\tilde{\chi}}$ in \eqref{eq:SJ}.} By factoring out the growth of $N_\text{c}(\chi)$, we ensure that the following inverse-volume expansion is mathematically well-defined. We find:
\begin{equation}\label{eq:expansionEFE}
    \left(\frac{V'(\chi)}{V(\chi)}\right)^2  =         \mathcal{A}_\lambda + \frac{\mathcal{B}_\lambda}{V(\chi)} + \frac{\mathcal{C}_\lambda}{V(\chi)^2} + \frac{\mathcal{D}_\lambda}{V(\chi)^3}   + \cdots \,,
\end{equation}
where the coefficients of the expansion are given by
\begin{widetext}
\def\footnoterule{\kern-3pt\moveright\dimexpr(\textwidth+\columnsep)/2\relax\vbox{\hrule width .5in}\kern3pt}
\begin{equation}\label{eq:coefficients}
    \begin{aligned}
    \mathcal{A}_\lambda &=4\omega_{\bn_0}^2 \left( \frac{ 2\omega_{\bn_0} \mathfrak{v}_\text{c} \mp \lambda J_0' }{ 2\omega_{\bn_0} \big[ \mathfrak{v}_\text{c} + \lambda \big( J_1- J_0\big) \big] } \right)^2
    \,,\\
\mathcal{B}_\lambda & = 4\omega_{\bn_0}^2 \left( \mathfrak{v}_\text{c} \mp \frac{\lambda N_\text{c}(0) J_0'}{\omega_{\mathbf{n}_0}} \right) \left( \frac{ 2\omega_{\bn_0} \mathfrak{v}_\text{c} \mp \lambda J_0' }{ 2\omega_{\bn_0} \big[ \mathfrak{v}_\text{c} + \lambda \big( J_1 - J_0 \big) \big] } \right) - 2 \mathcal{A}_\lambda \big( V_\text{b} + \lambda S \big)
\,,\\
\mathcal{C}_\lambda &=-4\omega_{\bn_0}^2 \mathfrak{v}_\text{c}^2 \left[ \mathcal{I}_0 \mp \frac{\lambda J_0' \big( 4\mathcal{I}_0 + 1 - 4 N_\text{c}(0) \big)}{8\omega_{\bn_0}\mathfrak{v}_\text{c}} \right] + \big(\lambda N_\text{c}(0) J_0'\big)^2 - \big(V_\text{b} + \lambda S\big) \Big[ \mathcal{B}_\lambda + \mathcal{A}_\lambda \big(V_\text{b} + \lambda S\big) \Big]\,,\\
\mathcal{D}_\lambda &= 
\mp\frac{1}{4} \lambda \omega_{\bn_0} \mathfrak{v}_\text{c} J_0' (4 \mathcal{I}_0 + 1)\big(1 - 2N_\text{c}(0)\big) \big[ \mathfrak{v}_\text{c} + \lambda \big( J_1- J_0 \big) \big]
\,,
\end{aligned}
\end{equation}
\end{widetext}
and $\mathcal{I}_0$ is the same as in the free theory dynamics \eqref{eq:Friedmann}. {We remark that in the expansion we have used \eqref{eq:PhiNc} to express the mean field in terms of $N_\text{c}$.} 

{\setlength{\parskip}{1pt}Given} that the theory is in general asymmetric around $\chi=0$, the evolution follows two distinct branches. We thus adopt a branch-dependent notation for the coefficients in the inverse-volume expansion \eqref{eq:expansionEFE}, where the upper and lower signs ($\pm$, $\mp$) refer to the evolution in the rightward and leftward directions, respectively. Note that by keeping several terms in the expansion (in principle as many as desired), we can capture with sufficient fidelity nearly all relevant dynamics, and not just the asymptotic regions where $ \left({V'(\chi)}/{V(\chi)}\right)^2 \; \overset{\chi\rightarrow\pm \infty}{\longrightarrow}\;\mathcal{A}_\lambda$. The first four coefficients of \eqref{eq:expansionEFE} will be enough to illustrate the imprint of beyond mean-field corrections in GFT condensate cosmology---we note that by recursively computing more terms we find that the coefficients exhibit a hierarchical structure, with each being proportional to a combination of the preceding ones (e.g., $\mathcal{E}_\lambda$ is proportional to $\mathcal{D}_\lambda$, $\mathcal{F}_\lambda$ is expressed in terms of $\mathcal{E}_\lambda$ and $\mathcal{D}_\lambda$, and so on).

As a consistency check, we note that for $\lambda=0$ one finds $\mathcal{D}_0 = 0$, and the same is true for all the subsequent coefficients. This means that for the free theory, the expansion terminates exactly with the $1/V(\chi)^2$ correction with the coefficients $\mathcal{A}_{0} = 4\omega_{\bn_0}^2$, $\mathcal{B}_{0} =  4\omega_{\bn_0}^2 (\mathfrak{v}_\text{c}- 2 V_\text{b})$ and $\mathcal{C}_0 = -4 \omega_{\bn_0}^2\big(\mathfrak{v}_\text{c}^2 \mathcal{I}_0  + V_\text{b} (\mathfrak{v}_\text{c}-V_\text{b} )\big)$. These are precisely the coefficients one would obtain in \eqref{eq:Friedmann} with only one SQ mode which plays the role of the condensate and some out-of-condensate GFT quanta at the bounce, characterised by a nonvanishing $V_\text{b}$.\footnote{Recall that $V_\text{b}$ reduces to $V_{\text{HO}}^{\text{free}}$ in the non-interacting limit, as explained below \eqref{eq:FrVdeltaV}.} If we further assume that these vanish ($ V_\text{b}=0$) we recover exactly the coefficients of the mean-field equation \eqref{eq:Friedmann}, namely: $\mathcal{A}_{0} = 4\omega_{\bn_0}^2$, $\mathcal{B}_{0} = 4\omega_{\bn_0}^2 \mathfrak{v}_\text{c}$ and $\mathcal{C}_0 = -4 \omega_{\bn_0}^2  \mathfrak{v}_\text{c}^2 \mathcal{I}_0$.

\begin{figure*}[t]
	\begin{center}
		\includegraphics[width=.42\textwidth]{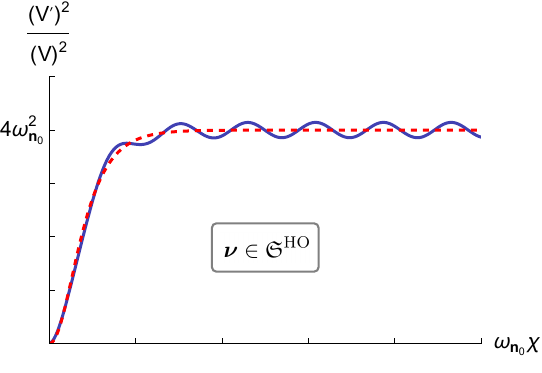}
				\hspace{0.3cm}
		\includegraphics[width=.42\textwidth]{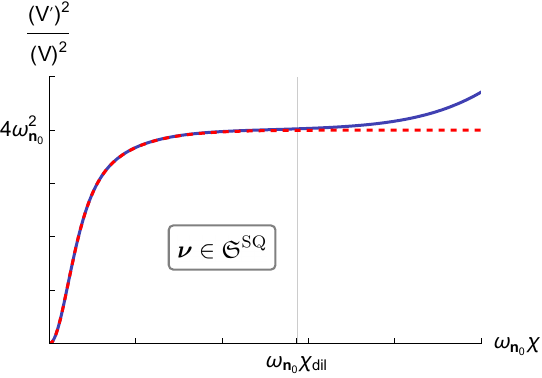}	
	\end{center}
	\caption{\small Effective Friedmann equation in the interacting theory (blue) and free theory (red), shown for a single collective mode $\bnu$ of HO type (left panel) and SQ type (right panel). The qualitative behaviour remains unchanged when multiple normal modes are included, as shown by the analytics. Left: $\omega_{\bnu}^0 = 2$,  $B_{\bnu} = -2{\rm i}$,  $k_{\bnu} = 5.5$. Right: $\omega_{\bnu}^0 = 0.4$,  $B_{\bnu} = 1$,  $k_{\bnu} = 0.2$. In both cases: $ \Omega_{\bn_0} = 1$, $\mathfrak{v}_\text{c} = 0.5$, $N_\text{c} = 5$, $C_0 = 0$, $N^\text{b}_{\bnu} = 2$, $\mathfrak{v}_{\bnu} = 1$, $\lambda = 0.001$ (blue) and $\lambda = 0$ (red). Note that the sign of $\lambda$ does not qualitatively affect the HO case, whereas it reverses the direction of the correction in the SQ case (as discussed below \eqref{eq:SQcomparison}). The vertical line in the right panel marks the time scale at which the diluteness condition breaks down, at $|\lambda| \Phi^2(\chi_\text{dil}) \sim 1$.
    }
	\label{fig:GFTBogoCosmo}
\end{figure*}

To finish the section, let us write equation \eqref{eq:FrVdeltaV} as an expansion in powers of $\lambda$, which reads
\begin{widetext}
    \begin{equation}\label{eq:EFEintOl}
\left( \frac{V'(\chi)}{V(\chi)} \right)^2 = \left( \frac{V_\text{c}'(\chi)}{V_\text{c}(\chi) + V_\text{b}} \right)^2 \left[ 1 + 2\lambda \left( \frac{\Delta V'(\chi)}{V_\text{c}'(\chi)} - \frac{\Delta V(\chi)}{V_\text{c}(\chi) + V_\text{b}} \right) \right]  +   O(\lambda^2)\,.
\end{equation}
This will allow us to identify more clearly the role of beyond-mean-field corrections. Neglecting the constant offset $V_\text{b}$, which is subdominant compared with the growing condensate volume $V_\text{c}(\chi)$, we obtain the simpler
expression
\begin{equation}
\begin{aligned}\label{eq:nobogo0}
\left( \frac{V'(\chi)}{V(\chi)} \right)^2 = \left( \frac{V_\text{c}'(\chi)}{V_\text{c}(\chi)} \right)^2& -  2\lambda\frac{ V_\text{c}'(\chi) \Phi^2(\chi)}{V_\text{c}^2(\chi)} \intsum \mathfrak{v}_{\boldsymbol{\nu}} k_{\boldsymbol{\nu}} \omega^0_{\boldsymbol{\nu}} \text{Im}\left(B_{\boldsymbol{\nu}} \ee^{-2{\rm i}\omega_{\boldsymbol{\nu}}^0 \chi} \right) \\&- {\lambda} \frac{V_\text{c}'(\chi)^2}{V_\text{c}(\chi)^3} \intsum \mathfrak{v}_{\boldsymbol{\nu}} k_{\boldsymbol{\nu}}\text{Re}\Big[ B_{\boldsymbol{\nu}} \Big( 2\tilde{\beta}_{\boldsymbol{\nu}}^*(\chi) \ee^{-{\rm i}\omega_{\boldsymbol{\nu}}^0 \chi} - \Phi^2(\chi) \ee^{-2{\rm i}\omega_{\boldsymbol{\nu}}^0 \chi}\Big)\Big]  +   O(\lambda^2) \,,
\end{aligned}
\end{equation}
\end{widetext}
where we can again clearly see the corrections to the free-theory equation \eqref{eq:Friedmann} as bounded oscillations {(since $V'_\text{c}/V_\text{c}$ is bounded by \eqref{eq:Friedmann}, $\Phi^2/V_\text{c}$ by \eqref{eq:PhiNcRatio} and $|\tilde{\beta}_{\bnu}|<\Phi^2$---see below \eqref{eq:RHO}). Unlike \eqref{eq:coefficients}, \eqref{eq:EFEintOl}--\eqref{eq:nobogo0} do not require \eqref{eq:PhiNc}.}

Physically, these are the leading corrections to the mean-field emergent cosmological evolution due to quantum fluctuations of the condensate. At the macroscopic level, they manifest as bounded oscillations in the effective Hubble rate $(V'/V)^2$, as seen in the left panel of figure \ref{fig:GFTBogoCosmo}. This means that the expansion of the universe exhibits small-amplitude modulations around the free-theory value, making the volume `wiggle' as it expands. 

Ultimately, this establishes a direct link between the microscopic excitation content of a quantum gravity condensate and the emergent cosmological dynamics, showing that the large-scale evolution retains a controlled and collective imprint of its underlying quantum degrees of freedom. The reason these perturbations are global oscillations in the spatial volume, and exhibit no spatial propagation (as opposed to phonons in BECs), is because we have not introduced a notion of spatial relational frame with respect to which we could localise the fluctuations. As a consequence, collective excitations are realised here as homogeneous `breathing modes' of the total volume. 

Upon inclusion of a spatial relational frame, these collective quantum-gravity excitations will be localisable inhomogeneities on top of the mean-field cosmological background. These will contribute to seed cosmological perturbations, with potential imprints in observables such as the cosmic microwave background. We leave the investigation of these ideas for future work.

\section{Outlook}

In this paper we have developed a beyond-mean-field framework for background-independent quantum gravity systems in which continuum spacetime arises as a coarse-grained condensate regime of many microscopic degrees of freedom. The central result is that interactions reorganise quantum fluctuations of the condensate into collective Bogolyubov modes, rather than independent excitations of individual atoms of geometry, in direct analogy to phonons in laboratory BECs. The construction identifies a new class of quantum-gravity excitations and provides the first controlled and systematic description of beyond mean-field dynamics in quantum gravity---a crucial step towards a more complete physical description of emergent spacetime.

We implemented this general construction in a tractable quantum-gravity model within group field theory. In this setting, the mean-field condensate reproduces nonsingular cosmological dynamics, while the collective excitations induce controlled corrections to macroscopic geometric observables. In the homogeneous sector considered here, these appear as bounded oscillatory modulations of the effective Friedmann equation, naturally interpreted as breathing modes of the total spatial volume.

The result establishes a direct bridge between microscopic quantum-gravity interactions, collective many-body excitations, and the effective dynamics of an emergent continuum universe. The mechanism we exploit is not speculative: it is grounded in the well-tested tradition of Bogolyubov theory and quantum many-body physics, drawing also upon QFT in curved spacetimes and adiabatic methods in time-dependent backgrounds. Consequently, it opens several compelling avenues for future research, ranging from theoretical refinements to cosmological phenomenology, even suggesting the possibility of analogue quantum simulations. We highlight the most promising directions below.

A central direction opened by this work is to reformulate the collective excitations directly in hydrodynamic variables. Here we have obtained them bottom-up, by applying the Bogolyubov construction to microscopic fluctuations around a quantum-gravity condensate. In ordinary BECs, collective excitations admit an equivalent hydrodynamic interpretation as quantised density--phase fluctuations of the condensate order parameter. The analogous description in quantum gravity would treat the emergent cosmological state as a `wavefunction' (or order parameter) $\Psi$ in minisuperspace. The collective excitations identified here should then arise as linearised fluctuations of the density around a cosmological background $\Psi_0$. Developing this hydrodynamic formulation would provide the complementary top-down approach, strengthening the idea of emergent cosmology as a quantum-gravity fluid on minisuperspace \cite{Oriti:2024qav}.

The most pressing cosmological extension is to go beyond the homogeneous sector. In relational quantum-gravity models, this requires introducing spatial reference fields so that fluctuations can be localised with respect to an emergent spatial frame \cite{Gielen:2018xph,Gerhardt:2018byq,Gielen_GFT_matterframes}. In such a setting, the collective excitations studied here would no longer describe only global breathing modes of the total volume, but genuine inhomogeneous perturbations propagating on the emergent Friedmann universe---i.e., the condensate background. A natural goal would then be to derive an effective analogue of the Mukhanov--Sasaki equation for quantum-gravity collective excitations, and to determine whether non-adiabatic production of Bogolyubov excitations can provide a quantum-gravitational origin of cosmological perturbations.

Another important direction is to test the robustness of the construction in more realistic microscopic models. In the GFT context, this includes expanding around solutions of the interacting mean-field equations rather than around the free condensate background, extending the analysis to non-Abelian $SU(2)$- or $SL(2,\mathbb C)$-based models, and incorporating the combinatorially non-local interactions that reproduce spin-foam amplitudes---see e.g., \cite{laghi2017excitations,bulakhov2018re,lewin2015bogoliubov,Ribeiro:2021zyz,Tang-Nonlocal} for collective excitations in BECs with non-local interactions. A conceptual extension would be to develop the present framework within the algebraic approach to canonical quantisation of GFT, allowing to investigate the interplay between the geometrical quantum fluctuations described by the collective modes and the quantum fluctuations of the matter clock itself \cite{LucaFluct}. {A further refinement would be to retain condensate-mode fluctuations: related issues have been studied in many-body condensates, both in equilibrium treatments of
condensate zero-mode fluctuations and in driven-dissipative condensates
\cite{LewensteinYou1996,Gardiner:1997yk,CastinDum1998,CarusottoWouters2007}.
Adapting these ideas to condensate cosmology, where the condensate occupation
grows exponentially in the Friedmann regime, requires a dedicated analysis,
which we leave to future work.}

The framework also opens the way to studying the imprint of many-body dynamics on emergent spacetime beyond the linear approximation. Higher-order terms in the fluctuation Hamiltonian would describe interactions among collective excitations and their backreaction on the condensate. In standard many-body systems, these are responsible for phenomena such as Beliaev decay and Landau damping \cite{RevModPhys.71.463}, and can also lead to transport properties, dissipation, and thermalisation \cite{Eisert:2014jea} (for thermal states in GFT, see \cite{IshaDaniele,Isha_thermalGFT,Isha_thermalGFT2}). They may also become essential when unstable squeezing sectors grow beyond the perturbative regime, potentially signalling transitions to new macroscopic phases in the underlying quantum-gravity phase diagram. Capturing these regimes will likely require non-perturbative techniques, such as phase-space methods \cite{Blakie:2008vka} or truncated-Wigner-type approximations \cite{Alice_Sinatra_2002}. In the cosmological interpretation, the same non-linearities would be a natural source of non-Gaussian signatures in the early universe \cite{Bartolo:2004if}, which could be used to constrain or rule out the framework \cite{Planck:2019kim}. The present Bogolyubov analysis should be understood as the leading term in a broader many-body expansion whose higher orders are both physically meaningful and technically accessible.

Finally, the collective-excitation language may provide a useful bridge towards quantum simulation in controlled many-body platforms \cite{Barcelo:2005fc,Oriti:2024qav}. Examples include cold atom systems \cite{Coutant:2009cu,Finazzi:2010nc,deNova:2023yyu,deNova:2024qeo,deNova:2025mwv}, polariton fluids \cite{Carusotto:2012vz,Bloch:2021qxj} and superconducting circuit-QED platforms (specifically TWPA/SNAIL devices) \cite{Marcos:2012zz,Nation:2011dka}, which can simulate growing and expanding background geometries that mimic an expanding condensate cosmology.

Ultimately, this work highlights a rare meeting point between background-independent quantum gravity, cosmology, QFT in time-dependent backgrounds, many-body physics, and quantum simulators. We expect future progress in quantum  gravity to rely on precisely this kind of cross-fertilisation, where the formal challenges of spacetime emergence are addressed leveraging robust tools from established disciplines as well as cutting edge technological developments.

\begin{acknowledgments}
The authors are grateful to Iacopo Carusotto, Luis J.~Garay, Steffen Gielen, Maxime J.~Jacquet, Mercedes Martín-Benito, Juan Ramón Muñoz de Nova, and Markus Oberthaler for fruitful discussions and their feedback on the manuscript. The authors acknowledge support to Grant PR28/23 ATR2023-145735 funded by MCIN/AEI/10.13039/501100011033. 
\end{acknowledgments}

\bibliography{Refs.bib}

\onecolumngrid
\appendix

\section{Hamiltonian framework for group field theory and its condensate cosmology}
\label{app:GFTAppendix}

This appendix provides a brief overview of the group field theory (GFT) formalism and the specific ingredients used to derive the results presented in section \ref{sec:GFTcosmo}. In particular, we present the construction of the underlying $U(1)$ simplicial model and its relational Hamiltonian framework, and explain the precise sense in which cosmology can be described in GFT as similar to condensation.

Group field theory (GFT) is a nonperturbative and background-independent approach to quantum gravity that in general refers to a class of quantum field theories defined on group manifolds rather than on spacetime, and that typically involve non-local interactions \cite{Oriti:2006se,FreidelGFT}. The framework describes candidate ``spacetime constituents'' by means of tensor fields, from which continuum geometry is reconstructed in a combinatorial and algebraic way. Originally developed as a generalisation of matrix and tensor models \cite{DiFrancesco:1993cyw,Gurau:2011xp}, GFT was later shown to arise naturally also in the context of spin foam models for quantum gravity \cite{DePietri:1999bx,Reisenberger_2000}. In the language of path-integral quantisation, the GFT partition function defines a non-perturbative sum over simplicial geometries and topologies (or `spacetime histories') with Feynman graphs that are associated to discrete gravity amplitudes. Specifically, when the field domain and interaction patterns are suitably tailored (e.g., choosing the local gauge group of general relativity and specific combinatorially non-local interaction kernels \cite{Perez:2012wv,Perez:2003vx,Oriti:2006se,Oriti:2011jm}) these Feynman amplitudes reproduce the discrete gravity amplitudes of first-order gravitational actions (e.g., Pleba\'nski- or Palatini-like actions).

While GFTs were introduced within this covariant context, several studies (in particular on applications to cosmology) have focused on a {\em canonical quantisation} of the theory \cite{Oriti_GFT2ndLQG,Oriti_GFTandLQG}. This perspective brings GFT closer to the language of many-body physics, enabling the use of second-quantisation methods and the interpretation of the fundamental excitations as ``quanta of geometry'' of an underlying quantum field theory of gravity. In this setting, the GFT Fock states can be related to the spin network (kinematical) states of canonical loop quantum gravity (LQG), yielding a clear interpretation in terms of quantised geometry \cite{Oriti_GFT2ndLQG,Oriti_GFTandLQG}. In concrete applications, the group manifold is typically chosen to include $SL(2,\mathbb{C})$ or $SU(2)$---namely the gauge group of general relativity or that of its Ashtekar--Barbero formulation \cite{barbero}---but models based on different groups can be defined and have been used (see, e.g., \cite{Perez:2003vx,Jercher:2021bie,GFTcosmoLONGpaper} for $Spin(4)$, and \cite{Carrozza:2012uv,BenGeloun:2011rc,Lahoche:2018oeo,Lahoche:2018vun,Lahoche:2018ggd,Lahoche:2018hou}
for Abelian $U(1)$ settings). 

An important development in this framework was enabled by the idea of relational dynamics \cite{Tambornino,Trinity,Goeller:2022rsx}, where internal degrees of freedom (such as massless scalar fields) are used as (quantum) reference frames, providing a {definition} of physical spacetime localisation. Two relational formulations are possible in GFT \cite{Oriti_2016,relham_Wilson_Ewing_2019,Marchetti2021,Gielen:2024sxs,AC_thesis}. An algebraic, timeless setting where no clock is selected \textit{a priori} and equations of motion are imposed at the quantum level à la Schwinger--Dyson \cite{GFTcosmoLONGpaper,Gielen_2016,Oriti_2016,BOriti_2017}, or a deparametrised approach where the clock variable is singled out at the classical level \cite{relham_Wilson_Ewing_2019,relhamadd,toy}. In the first scheme, one selects a physical clock only approximately, by focusing on specific observables \cite{Marchetti:2024nnk} and quantum states (e.g., coherent states that are peaked with respect to the particular internal clock variable \cite{Marchetti2021}). This approach is more ``agnostic'' in terms of clock choices and makes it possible to study the quantum aspect of the clock itself \cite{LucaFluct}, although it is formulated in a language that remains quite distant from standard canonical quantisation methods and many-body physics. Conversely, deparametrisation enables a standard canonical quantisation by describing the system as a traditional Hamiltonian quantum many-body theory (and has also been recently formulated so as to retain the quantum nature of the clock in a gauge-invariant way \cite{PWGFT}). In this work we follow the deparametrised route, which is particularly well suited for the analysis of emergent cosmological dynamics and Bogolyubov theory we develop in the main text.

Interestingly, regardless of the approach, canonically quantising GFTs and adopting a relational strategy gives rise to dynamical equations that can be contrasted to symmetry-reduced settings of general relativity, thus describing effective cosmology from a full quantum gravity standpoint \cite{Oriti_2016,BOriti_2017,relham_Wilson_Ewing_2019,relhamadd}. In the following, we provide an example of the paradigm by describing the details of the model used in this work.

\subsection{Specifics of our GFT model}

We here define the GFT model used in the main text (see section \ref{sec:GFTcosmo}). We focus on an Abelian GFT over $U(1)^4$ as a more tractable alternative to the standard models found in the literature (based on $SU(2)$ or $SL(2,\mathbb{C})$, see discussion above). Despite the Abelian nature of the group, the model remains physically relevant for cosmology, as it consistently yields bouncing FLRW dynamics by assuming the quanta have a geometrical interpretation. {What the model retains from realistic constructions are precisely the features this mechanism relies on: a field with four group arguments, so that the quanta have the same valence; a compact group, hence discrete representation labels; and a kinetic kernel with the same mode-dependent sign structure, which is what singles out a dominant mode. What it does not inherit is the geometric reading of the labels, which is why the interpretation of the quanta is here posited.}

We couple the theory to a (free) massless scalar field $\chi$, which we will use as relational clock. Working in the deparametrised perspective \cite{relham_Wilson_Ewing_2019}, we follow a conventional canonical quantisation using the scalar field label $\chi$ as a notion of time variable. Our GFT is then defined by a real-valued ``group field'' $\varphi (g_I,\chi)$,
\begin{equation}\label{eq:GFTfieldApp}
		\varphi\,:\, U(1)^4 \times \mathbb{R} \rightarrow \mathbb{R} \,,
\end{equation}
which takes five arguments: $g_I \equiv (g_1\,,\dots\,,g_4)\in U(1)^4$ and $\chi\in \mathbb{R}$, which is the massless scalar field that will be used as relational clock. We start from the action
\begin{equation}\label{eq:GFTactionApp}
	S [\varphi] = \frac{1}{2}\int \dd g_I \, \dd g_I' \, \dd \chi \; \varphi (g_I,\chi) K(g_I,g_I') \varphi (g_I',\chi) + V[\varphi] \,,
\end{equation}
where we integrate with respect to the normalised Haar measure. Following the literature of GFT cosmology \cite{Oriti_2016,BOriti_2017,Gielen_2016,relham_Wilson_Ewing_2019}, phase transitions \cite{Marchetti:2022igl,Marchetti:2022nrf}, and renormalisation \cite{BenGeloun:2011jnm,BenGeloun:2013mgx,Carrozza:2016vsq}, we take a standard kinetic term of the form
\begin{equation}\label{eq:genericKApp}
	K(g_I,g_I') = K^{(0)}(g_I,g_I') + K^{(2)} (g_I,g_I') \partial_\chi^2 \,,
\end{equation}
and we leave the interaction potential $V[\varphi]$ unspecified for now. While renormalisation analyses \cite{BenGeloun:2011jnm,BenGeloun:2013mgx,Carrozza:2016vsq} can further specify the form of $K^{(0)}(g_I,g_I')$ and $K^{(2)} (g_I,g_I')$, we will keep them generic and simply assume $K(g_I,g_I')$ satisfies $K(g_I,g_I')= K(g_I^{-1}  g_I')$, namely group ``translation'' invariance. 

We now expand the field in modes, as usual in quantum field theories (QFTs). The analogue of Fourier decomposition for GFTs stems from the Peter--Weyl theorem for Lie groups. Since $U(1)$ is compact, its irreducible unitary representations are labelled by integers $n\in \mathbb{Z}$. Hence, our mode expansion reads
\begin{equation}\label{eq:PWApp}
	\varphi(g_I, \chi) = \sum_{\mathbf{n}\in\mathbb Z^4}
	\varphi_{\mathbf{n}} (\chi)\,\ee^{{\rm i}\,\mathbf{n}\cdot\boldsymbol{\alpha}}\,,
\end{equation}
where $\mathbf{n}=(n_1,n_2,n_3,n_4)\in \mathbb{Z}^4$ and $\boldsymbol{\alpha}=(\alpha_1,\alpha_2,\alpha_3,\alpha_4)\in [0,2\pi)^4$. Since the field is real, the field modes satisfy ${\varphi}^*_\mathbf{n} = \varphi_{-\mathbf{n}}$, which implies that the labels $\pm\mathbf{n}$ are not independent, but they describe the same degree of freedom. 

The mode decomposition \eqref{eq:PWApp} shifts the focus from group to ``momentum'' labels, in which the action reads
\begin{equation}\label{eq:actionmodesApp}
	S[\varphi] = \frac{1}{2} \int \dd \chi \sum_\mathbf{n} \varphi_{- \mathbf{n}}  \left( K^{(0)}_\mathbf{n} + K^{(2)}_\mathbf{n} \partial_\chi^2 \right) \varphi_\mathbf{n} + V[\varphi]\,,
\end{equation}
where $K^{(0)}_\mathbf{n}$ and $K^{(2)}_\mathbf{n}$ can in general be positive or negative, and where the kinetic term $K_\mathbf{n}=K^{(0)}_\mathbf{n} + \partial_\chi^2 K^{(2)}_\mathbf{n}$ is assumed to satisfy $K_\mathbf{-n}= K_\mathbf{n}$. Similar to what happens in QFTs, the fact that $K_\mathbf{n}$ is diagonal follows from $K(g,h)=K(g^{-1} h)$ (see discussion below \eqref{eq:genericKApp}), and the property $K_\mathbf{-n}= K_\mathbf{n}$ is then a consequence of assuming invariance under group inversion $K(g)=K(g^{-1})$, the analogue of parity. After integration by parts one finds
\begin{equation}
\label{eq:GFTFreeActionModesApp}
	S[\varphi] = \frac{1}{2} \int \dd \chi\, \sum_\mathbf{n} \left( K^{(0)}_\mathbf{n} \varphi_\mathbf{-n}(\chi) \varphi_\mathbf{n}(\chi) - K^{(2)}_\mathbf{n} \partial_\chi \varphi_\mathbf{-n}(\chi) \partial_\chi \varphi_\mathbf{n}(\chi)\right)+ V[\varphi]\,,
\end{equation}
from which one defines the momentum of the group field $\pi_\mathbf{n} (\chi)= - K^{(2)}_\mathbf{n} \partial_\chi \varphi_\mathbf{-n} (\chi)$. Performing the Legendre transform of the Lagrangian with respect to $\chi$ gives a relational Hamiltonian \cite{relham_Wilson_Ewing_2019,relhamadd}
\begin{equation}\label{eq:freehamApp}
	H= -\frac{1}{2} \sum_\mathbf{n} \left( \frac{\pi_\mathbf{n}(\chi) \pi_\mathbf{-n}(\chi)}{K^{(2)}_\mathbf{n}} +K^{(0)}_\mathbf{n} \varphi_\mathbf{n}(\chi) \varphi_{-\mathbf{n}}(\chi)\right) - V[\varphi]\,,
\end{equation}
describing evolution with respect to the relational clock $\chi$. With a Hamiltonian at hand, canonical quantisation becomes straightforward. The classical Poisson structure in phase space is mapped onto commutators in the algebra of quantum observables
\begin{equation}\label{eq:originalCCRApp}
	[\hat{\varphi}_\mathbf{n}(\chi) \, ,\, \hat{\pi}_\mathbf{m}(\chi)] = \ci\delta_{\mathbf{n,m}} \,,
\end{equation}
and the Heisenberg equation dictates the $\chi$-evolution for any observable as ${\rm i} \frac{\dd \hat{ O}}{\dd \chi} = [\hat{ O}, \hat{H}]+ {\rm i} \partial_\chi \hat{O}$. As usual, we can define creation and annihilation operators satisfying $[\hat{a}_\mathbf{n}\,,\, \hat{a}_\mathbf{m}^\dagger]= \delta_\mathbf{n,m}$ for each mode as
\begin{equation}
\label{eq:aadagApp}
		\hat{a}_\mathbf{n} = \frac{1}{\sqrt{2 \Omega_\mathbf{n}}} (\Omega_\mathbf{n} \, \hat{\varphi}_\mathbf{n} + {\mathrm i} \hat{\pi}_{-\mathbf{n}}) \,,   \qquad 
		\hat{a}_\mathbf{n}^\dagger = \frac{1}{\sqrt{2 \Omega_\mathbf{n}}}( \Omega_\mathbf{n} \, \hat{\varphi}_{-\mathbf{n}} - {\mathrm i} \hat{\pi}_\mathbf{n}) \,,
\end{equation}
where 
\begin{equation}\label{eq:OmegafreeApp}
	\Omega_\mathbf{n} = \sqrt{\left|K_\mathbf{n}^{(0)} K_\mathbf{n}^{(2)}\right|} \,.
\end{equation}
These can be used to define a Fock vacuum as the state satisfying $\hat{a}_\mathbf{n}|0 \rangle=0$ for all $\mathbf{n}$, and to construct the full Fock space by acting with creation operators on the vacuum. To interpret the corresponding quanta physically, we generalise the natural interpretation of more realistic {(e.g., SU(2)-based)} GFTs to our Fock space. We thus \textit{assume} our quanta to be {three-simplices (i.e., quantum tetrahedra as for four-valent spin network nodes)} with a definite volume $\mathfrak{v}_\mathbf{n}$, given as a function of the four integers $\mathbf{n}$. In other words, for each mode, we assume the operator $\hat{V}_\mathbf{n}= \mathfrak{v}_\mathbf{n} \hat{a}^\dagger_\mathbf{n} \hat{a}_\mathbf{n}$ characterises the contribution of the mode $\bn$ to the volume observable, {and we note that no result depends on the specific form of $\mathfrak{v}_\mathbf{n}$}. Then, it is straightforward to define a total volume operator as
\begin{equation}
\label{eq:volumeApp}
	\hat{V}(\chi)= \sum_\mathbf{n} \hat{V}_\mathbf{n} (\chi)= \sum_\mathbf{n} \mathfrak{v}_\mathbf{n}  \hat{N}_{\mathbf{n}}(\chi)= \sum_\mathbf{n} \mathfrak{v}_\mathbf{n} \hat{a}^\dagger_\mathbf{n}(\chi) \hat{a}_\mathbf{n} (\chi) \,.
\end{equation}
The relational evolution of such observables is governed by $\hat{U}= \ee^{-{\rm i} \hat{H} \chi}$, where $\hat{H}$ is the quantum version of \eqref{eq:freehamApp}. 

The second-quantised language described above suggests to treat spacetime as a many-body system of GFT atoms. In this perspective, the transition from a discrete geometry to a continuous spacetime is understood as a collective phenomenon, akin to the formation of a macroscopic state in condensed matter physics. {Note that $\hat{V}$ is a one-body operator, additive over quanta and insensitive to their connectivity: the geometry it describes is extended in this collective sense, and not in the sense of a glued simplicial complex generated by the Feynman expansion of the interacting theory.} Crucially, the physical validity of a smooth geometric interpretation relies on the existence of a regime with macroscopic occupation numbers, namely with a large number of quanta occupying the same quantum state, effectively forming a GFT condensate \cite{ORITI2017235}. This leads to the idea that our large-scale spacetime corresponds to a particular geometric phase of an underlying field theory of quantum gravity---a\textit{ quantum fluid of geometry}---and that classical gravity should emerge as the hydrodynamic equations describing the collective behaviour of quantum gravity atoms \cite{GFTquantumST_Oriti,Oriti:2024qav}. In the following subsection we provide an instance of this paradigm in a simple context, with an explicit dynamical mechanism leading to macroscopic occupation and where such a many-body description yields the dynamics of an expanding universe.

\subsection{Extracting cosmological dynamics}

We here briefly discuss the emergence of relational Friedmann--Lema\^itre--Robertson--Walker (FLRW) cosmology in the deparametrised GFT formalism outlined above. Within this framework, cosmological dynamics are recovered from the relational evolution of the expectation value of the volume operator over suitable quantum states. We will not go into the details of the full derivation here, and only discuss the aspects that are essential to understand our work. The interested reader is referred to, e.g., \cite{Gielen_2020,relham_Wilson_Ewing_2019,AC_thesis} for further details (see also \cite{Oriti_2016,BOriti_2017} for earlier derivations).
 
As usual in the GFT cosmology literature, we start by assuming that the interaction term $V[\varphi]$ is negligible. Then, the free quantum Hamiltonian can easily be written in terms of the ladder operators defined in \eqref{eq:aadagApp}. Crucially, its specific form depends on the relative sign of $K^{(0)}_\mathbf{n}$ and $K^{(2)}_\mathbf{n}$ for the various modes $\mathbf{n}$, which can yield a standard harmonic oscillator Hamiltonian or a two-mode squeezing Hamiltonian. Specifically, assuming non-vanishing kinetic coefficients, one can split $\mathbb{Z}^4$ into disjoint sets $\mathfrak{N}^{\text{HO}}$ and $\mathfrak{N}^{\text{SQ}}$ such that 
\begin{equation}\label{eq:setsfreetheoryApp}
	\begin{aligned}
		&\mathbf{n}\in   \mathfrak{N}^{\text{HO}}\quad \Longleftrightarrow \quad\sgn \big(K^{(2)}_\mathbf{n}\big) = \sgn \big(K^{(0)}_\mathbf{n}\big)\,,\\
		&\mathbf{n}\in   \mathfrak{N}^{\text{SQ}} \quad \Longleftrightarrow \quad\sgn \big(K^{(2)}_\mathbf{n}\big) =- \sgn \big(K^{(0)}_\mathbf{n}\big)\,.
	\end{aligned}
\end{equation}
These allow to write the full Hamiltonian as
\begin{equation}\label{eq:HsumsinglemodeApp}
	\hat{{H}}  = \sum_{\mathbf{n} \in \mathfrak{N}^{\text{HO}}} \omega_\mathbf{n} \left(\hat{a}^\dagger_\mathbf{n} \hat{a}_\mathbf{n} + \frac{1}{2}\right) + \sum_{\mathbf{n} \in \mathfrak{N}^{\text{SQ}}} \frac{\omega_\mathbf{n}}{2} \left( \hat{a}^{\dagger}_\mathbf{n} \hat{a}^{\dagger}_\mathbf{-n}  + \hat{a}_\mathbf{n} \hat{a}_\mathbf{-n}  \right) \,,
\end{equation}
where
\begin{equation}\label{eq:omegafreeApp}
	\omega_\mathbf{n} = - \sgn\big(K^{(0)}_\mathbf{n}\big)\sqrt{\left|K^{(0)}_\mathbf{n}\big/K^{(2)}_\mathbf{n}\right|} \,
\end{equation}
plays the role of a frequency for the harmonic oscillator modes and of a squeezing intensity rate for the two-mode squeezing modes.

The time evolution of the two sets of modes is qualitatively different. While for the harmonic oscillator modes the number of quanta is conserved under time evolution, the number of quanta in the squeezing modes grows monotonically in time, approaching exponential growth for $\chi\gtrsim (2\omega_\mathbf{n})^{-1}$, as we explicitly show below. Standard choices for the kinetic term $K_\mathbf{n}$ (motivated by renormalisation studies \cite{BenGeloun:2011jnm,BenGeloun:2013mgx,Carrozza:2016vsq}, see \eqref{eq:genericKApp}), lead to the existence of a mode $\mathbf{n_0}\in\mathfrak{N}^{\text{SQ}}$ for which the squeezing intensity rate is maximised \cite{Gielen_lowspin}. In that case, the quantum dynamics of the theory is dominated by that mode at times $\chi\gtrsim(2\omega_{\mathbf{n}_0})^{-1}$.

Here we will not use a specific form for the kinetic term $K_\mathbf{n} = K_\mathbf{n}^{(0)} + \partial_\chi^2 K_\mathbf{n}^{(2)}$ but we simply assume that there is a dominating mode $\mathbf{n_0}$, such that $|\omega_\mathbf{n_0}| > |\omega_\mathbf{n}|$ for all $\mathbf{n}\neq \mathbf{n_0}$. This is enough to derive the standard results of GFT cosmology. To do that, we compute the relational Heisenberg evolution of the number operator for the dominant mode as $\hat{N}_{\pm\mathbf{n}_0} (\chi)= \ee^{{\rm i} \hat{H} \chi}\, \hat{N}_{\pm\mathbf{n}_0} (0)\,  \ee^{-{\rm i} \hat{H} \chi}$. Then, given that the field is real, the operator counting the number of physical quanta due to the squeezing in the dominant mode will be given by $\hat{N}_{\rm c}\coloneqq(\hat{N}_{\bn_0}+\hat{N}_{-\bn_0})/2$, whose Heisenberg evolution is given by
\begin{equation}\label{eq:condensateevolutionApp}
    \hat{N}_{\rm c} (\chi)=-\frac{1}{2}+ \lr{\hat{N}_{\rm c}(0)+\frac{1}{2}}\cosh(2\omega_{\bn_0}\chi)+ \frac{{\rm i}}{2}  \left(\hat{a}_{\mathbf{n}_0}(0)\hat{a}_{-\mathbf{n}_0}(0)-\hat{a}^{\dagger}_{\mathbf{n}_0}(0)\hat{a}^{\dagger}_{-\mathbf{n}_0}(0)\right) \sinh(2\omega_{\mathbf{n}_0} \chi)\,.
\end{equation}
As explained below \eqref{eq:OmegafreeApp}, the physical interpretation of these quanta comes from assuming that they carry a unit of volume $\mathfrak{v}_{\bn_0}$, and cosmological evolution is obtained from the expectation value of $\hat{V}_{\rm c}=\mathfrak{v}_{\bn_0}\hat{N}_{\rm c}$ as a function of relational time, which for generic states takes the form \cite{Oriti_2016,BOriti_2017,relham_Wilson_Ewing_2019}
\begin{equation}\label{eq:FriedmannApp}
	\left(\frac{1}{\langle\hat{V}_\text{c}(\chi)\rangle} \frac{{\mathrm d} \langle\hat{V}_\text{c}(\chi)\rangle}{{\mathrm d} \chi}\right)^2 = 4\omega_\mathbf{n_0}^2 \left( 1+\frac{\mathfrak{v}_\mathbf{n_0}}{ \langle\hat{V}_\text{c}(\chi)\rangle} -\frac{\mathfrak{v}_\mathbf{n_0}^2}{ \langle\hat{V}_\text{c}(\chi)\rangle^2} \mathcal{I}_0\right)\,,
\end{equation}
where all the dependence on the choice of initial state is encoded in $\mathcal{I}_0:=\langle \hat{N}_{\rm c}(0)\rangle\big(\langle \hat{N}_{\rm c}(0)\rangle+1\big)-C_0^2$, with $C_0:=\frac{\rm i}{2}\langle \hat{a}_{\mathbf{n}_0}(0)\hat{a}_{-\mathbf{n}_0}(0)-\hat{a}^{\dagger}_{\mathbf{n}_0}(0)\hat{a}^{\dagger}_{-\mathbf{n}_0}(0)\rangle$. Note that $\mathcal{I}_0\geq 0$ follows from the positivity of the bosonic covariance matrix, which is required for consistency with a valid quantum state. While technically \eqref{eq:FriedmannApp} holds for \textit{any} state, this can be a good description of the dynamics of cosmological observables only adopting states which have small quantum fluctuations; the rich family of Gaussian states was shown to have semiclassical properties in this respect, see \cite{Gauss,AC_thesis}. 

Notice that while the GFT literature usually adopts a basis in which the Hamiltonian is a single-mode squeezing operator, we retain the two-mode nature of squeezing as this will make the comparison with the interacting theory of later sections more natural. Instead of creating pairs in a single mode label, the Hamiltonian here distributes the pair across $\pm\bn_0$. As is well known, single- and two-mode squeezers are unitarily related by a change of basis, namely a $\pi/4$ rotation defining symmetric and anti-symmetric modes, also known as Hadamard transform in quantum information and 50:50 beam splitter in quantum optics \cite{TextSerafini}. This correspondence was formalised for GFT in \cite{AC_thesis,PWGFT}.

We now compare \eqref{eq:FriedmannApp} with the relational evolution for the volume of the universe in standard FLRW cosmology coupled to a massless scalar field $\chi$ (used as relational time), which is given by $\lr{V'/V}^2=12\pi G $ (see \eqref{eq:RelationalFLRW}). We see that the relational FLRW dynamics are recovered for large values of the expectation value of the volume of the universe (i.e., at large enough relational time) provided that the GFT parameters are identified with Newton's constant as $\omega_\mathbf{n_0} ^2 =3\pi G$. Importantly, the inverse volume corrections encode deviations from general relativity. In particular, the $1/\langle \hat{V}\rangle^2$ term describes a nonsingular bounce replacing the initial Big Bang singularity, reproducing the results of loop quantum cosmology in this regard \cite{Ashtekar:2006uz,Ashtekar:2007em,Bojowald:2001xe}. This is a very robust finding that can be obtained within the GFT framework using different quantisation methods and for different GFT models (see, e.g., \cite{Oriti_2016,BOriti_2017,toy,relham_Wilson_Ewing_2019,relhamadd,Gielen_2020,Marchetti2021}), including extensions in which additional matter content is included \cite{Li:2017uao,Ladstatter:2025kgu}.

We first note that the growth of the expectation value of the volume operator $\langle \hat{V}_\text{c} \rangle$ is entirely driven by the growth in the number of GFT quanta $\langle\hat{N}_\text{c}\rangle$, cf.~\eqref{eq:condensateevolutionApp}. FLRW cosmology is therefore recovered in a regime where the occupation number of these quanta becomes sufficiently large, suggesting that cosmological spacetime arises as the mean-field description of our underlying quantum gravity system in the large-$N$ limit. In particular, at large relational times, a single mode $\mathbf{n}_0$ becomes macroscopically occupied, in close analogy with Bose--Einstein condensation. This mode constitutes a \textit{condensate} in the precise sense that its occupation number is parametrically larger than that of all other modes, thereby justifying a mean-field treatment as the leading-order description of the system. In this way, cosmology emerges in GFT as a condensate regime of fundamental quantum gravity degrees of freedom.

Importantly, our system bears a striking difference with the standard BECs arising in condensed matter. In laboratory settings, condensation is reached by lowering the temperature of a system of a large number of bosonic microscopic constituents such that most of them are forced to fall in the ground state. In such systems, the number of quanta is typically conserved as a consequence of a global symmetry, and the condensate is a phase of thermodynamic equilibrium at low enough temperatures, in which the global symmetry is spontaneously broken \cite{pitaevskii2016bose}. In the present GFT setting, by contrast, macroscopic occupation is not driven by thermodynamic equilibrium \cite{Isha_thermalGFT,Isha_thermalGFT2,IshaDaniele}, nor does it require the breaking of a global symmetry. Instead, it is due to a dynamical mechanism arising naturally in GFTs (both in the deparametrised setting used here and the general timeless approach once relational dynamics is implemented \cite{Oriti_2016,Gielen_lowspin,Marchetti2021,Marchetti:2024nnk}). This mechanism leads to the production of GFT quanta in the dominant mode at a rate that is \textit{exponentially larger} than in the rest of the modes. More precisely, for any of the squeezing modes, the ratio $\langle\hat{N}_{\mathbf{n}\neq\mathbf{n}_0}\rangle/\langle\hat{N}_{\mathbf{n}_0}\rangle$ is exponentially suppressed as $\ee^{2(\omega_{\mathbf{n}}-\omega_{\mathbf{n}_0})\chi}$ at large enough relational time (cf.~\eqref{eq:condensateevolutionApp}).

We conclude this appendix by commenting on the role of interactions. Throughout the discussion above we have neglected the interaction term $V[\varphi]$ in \eqref{eq:GFTactionApp}, since the free theory already captures the emergence of an expanding FLRW universe from the relational dynamics of a GFT condensate. At the same time, interacting GFT models have been shown to describe a richer set of cosmological regimes and phenomena; see, e.g., \cite{deCesare:2016rsf,OritiPang,Gielen:2023han,Pang:2025jtk,Ladstatter:2025kgu,Marchetti:2025jze}. The free approximation is expected to be valid in a regime where the number of GFT quanta is large enough to admit a macroscopic geometric interpretation, but still sufficiently small that interactions remain subdominant, so that the system behaves as a dilute, weakly interacting gas \cite{Oriti_2016,BOriti_2017}. Beyond this regime, interactions can no longer be neglected. In particular, insights from Bogolyubov theory suggest the existence of an intermediate, mesoscopic regime in which the condensate remains well described at leading order by the free dynamics, while interaction effects produce perturbative corrections to the mean-field picture. In fact, an open question in GFT cosmology is how to describe and understand beyond mean-field effects at both the mathematical and physical levels \cite{GFTcosmoLONGpaper,Gielen:2017eco,Gielen_GFT_matterframes}. This is precisely the regime studied in the main text. Motivated by the Bogolyubov theory of weakly interacting Bose gases, in section \ref{sec:GFTbogolons} we treat $V[\varphi]$ perturbatively and investigate the leading beyond-mean-field effects induced by interactions.

\section{Diagonalisation procedure} \label{app:AppDiag}

We begin the appendix by briefly recalling that the free GFT Hamiltonian is diagonal with respect to the momentum labels $\mathbf{n}$. Indeed, using the reality condition $\hat{\varphi}_\mathbf{n}^\dagger = \hat{\varphi}_{-\mathbf{n}}$, one writes
\begin{equation}\label{eq:HfreeAppendix}
    \hat{H}^\text{free}= -\frac{1}{2} \sum_\mathbf{n} \left( \frac{\hat{\pi}_\mathbf{n} \hat{\pi}_\mathbf{-n}}{K^{(2)}_\mathbf{n}} +K^{(0)}_\mathbf{n} \hat{\varphi}_\mathbf{n} \hat{\varphi}_{-\mathbf{n}}\right) = -\frac{1}{2} \sum_\mathbf{n} \left( \frac{|\hat{\pi}_\mathbf{n}|^2}{K^{(2)}_\mathbf{n}} +K^{(0)}_\mathbf{n} |\hat{\varphi}_\mathbf{n} |^2\right) \,.
\end{equation}
In other words, since the pair $\{\mathbf{n}\,, -\mathbf{n}\}$ describes the same physical degree of freedom, no further diagonalisation is required at the free level.

The Hamiltonian of the interacting theory \eqref{eq:H2nondiag}, on the other hand, shows a more intricate structure:
\begin{equation}\label{eq:H2Appendix}
	\hat{H}_2 = 	-\frac{1}{2}\sum_{\mathbf{n\neq \pm n_0}} \left[ 
	\left( \frac{\hat{\pi}_\mathbf{n} \hat{\pi}_\mathbf{-n}}{K^{(2)}_\mathbf{n}} +K^{(0)}_\mathbf{n} \hat{\varphi}_\mathbf{n} \hat{\varphi}_{-\mathbf{n}}\right) + \frac{\lambda}{2} \Phi^2 \left(2 \hat{\varphi}_\mathbf{n}\hat{\varphi}_{\mathbf{-n}} + \hat{\varphi}_\mathbf{n}\hat{\varphi}_{\mathbf{-n-2\mathbf{n_0}}}+ \hat{\varphi}_\mathbf{n}\hat{\varphi}_{\mathbf{-n+ 2 \mathbf{n_0}}}\right)
	\right] \,,
\end{equation}
which in particular describes coupling between different modes due to the last two terms. We now describe the procedure to diagonalise such Hamiltonian in detail.

We begin by pointing out that unlike in traditional field theories, GFT models have in general mode-dependent kinetic couplings: the ``mass-like'' term $K_\mathbf{n}^{(0)}$ and the coefficient $K_\mathbf{n}^{(2)}$ of the second $\chi$-derivative in \eqref{eq:actionmodesApp}. This is an important feature which in particular can explain how to single out a condensate mode (i.e., the mode $\mathbf{n_0}$ for which squeezing is maximised, see discussion below \eqref{eq:omegafree}) yielding the GFT cosmological results shown in section \ref{sec:QGcosmo}. Without spoiling such GFT condensation mechanism, we will now assume that the kinetic couplings {in \eqref{eq:H2Appendix} (we use $K_\mathbf{n}$ to denote both at once)} are invariant under {all} the ``symmetries of the labels $\mathbf{n}$'' (i.e., the various types of couplings) of $\hat{H}_2$: reflections $\mathbf{n}\rightarrow\mathbf{-n}$, as well as shifts $\mathbf{n}\rightarrow\mathbf{-n} \pm 2 \mathbf{n_0}$ (which is the real novelty). In other words, we will impose \begin{equation}\label{eq:newassumption}
K_\mathbf{n} = K_{T_\mathbf{a}(\mathbf{n})}\,,
\end{equation}
where we defined the three involutions
 \begin{equation}\label{eq:Tas}
     T_\mathbf{a}(\mathbf{n}):= 2\mathbf{a} -\mathbf{n}\,,
 \end{equation}
with centre of reflection $\mathbf{a}=\{\mathbf{0}\,, \mathbf{n_0}\,, -\mathbf{n_0}\}$. We point out that, since the condensate coupling $K_{\mathbf{n}_0}$ is assumed to be unique (in that it provides maximum squeezing rate \eqref{eq:omegafree}), there is a special family of modes -- a zero-measure subset of all the modes -- for which property \eqref{eq:newassumption} needs to be handled with care. Specifically, since the maps $T_\mathbf{a}(\mathbf{n}_0)$ only relate odd multiples of the condensate labels, we assume that the kinetic couplings are uniform for the set $\mathcal{M}:=\{\pm 3 \bn_0, \pm 5 \bn_0,\pm 7 \bn_0,\dots\}$; we will denote $K_{\mathbf{n}} = K_{\mathcal{M}}$ for all $\mathbf{n} \in \mathcal{M}$, where $K_\mathcal{M} \neq K_{\mathbf{n}_0}$. This ensures that while the odd multiples of the condensate share the same kinetic term, their evolution remains distinct from that of the condensate. The modes in $\mathcal{M}$ require a separate treatment in the diagonalisation procedure, as we comment below.

\begin{figure}[h]
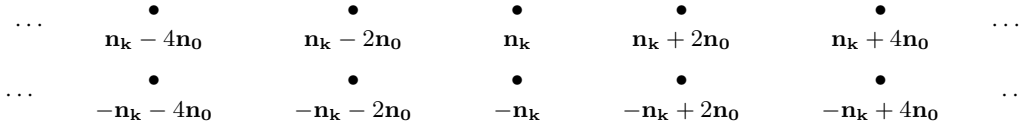

	\centering
\[
\begin{array}{ccccccc}
	\cdots\qquad
	\tikzmarknode{A1}{\begin{array}{c}
			\bullet\\
			\mathbf{n_k} - 4\mathbf{n_0}
	\end{array}}
	& \qquad
	\tikzmarknode{A2}{\begin{array}{c}
			\bullet\\
			\mathbf{n_k} - 2\mathbf{n_0}
	\end{array}}
	& \qquad
	\tikzmarknode{A3}{\begin{array}{c}
			\bullet\\
			\mathbf{n_k}
	\end{array}}
	& \qquad
	\tikzmarknode{A4}{\begin{array}{c}
			\bullet\\
			\mathbf{n_k} + 2\mathbf{n_0}
	\end{array}}
	& \qquad
	\tikzmarknode{A5}{\begin{array}{c}
			\bullet\\
			\mathbf{n_k} + 4\mathbf{n_0}
	\end{array}}\qquad\cdots
	\\[1em]
	\cdots\qquad
	\tikzmarknode{B1}{\begin{array}{c}
			\bullet\\
			-\mathbf{n_k} - 4\mathbf{n_0}
	\end{array}}
	& \qquad
	\tikzmarknode{B2}{\begin{array}{c}
			\bullet\\
			-\mathbf{n_k} - 2\mathbf{n_0}
	\end{array}}
	& \qquad
	\tikzmarknode{B3}{\begin{array}{c}
			\bullet\\
			-\mathbf{n_k}
	\end{array}}
	& \qquad
	\tikzmarknode{B4}{\begin{array}{c}
			\bullet\\
			-\mathbf{n_k} + 2\mathbf{n_0}
	\end{array}}
	& \qquad
	\tikzmarknode{B5}{\begin{array}{c}
			\bullet\\
			-\mathbf{n_k} + 4\mathbf{n_0}
	\end{array}}\qquad\cdots
\end{array}
\]\caption{\small A bi-chain represents all the interactions between different modes, and is constructed by selecting a generic mode $\mathbf{n_k}$ and applying the involutions $T_{\mathbf{a}}$. Of course, a single bi-chain does not represent all modes: one can now pick a different $\mathbf{n_j}$ (which cannot be reached from $\mathbf{n_k}$ using the maps \eqref{eq:Tas}) and find another \textit{independent} bi-chain. Indeed, in order to describe all the GFT modes $\mathbf{n}\in \mathbb{Z}^4$, one needs an infinite number of such disconnected diagrams.}
\label{fig:bi-chain}
\end{figure}

Note that the property \eqref{eq:newassumption} for $\mathbf{a=0}$ is usually already assumed in the setup for GFT cosmology, the reason being that $V[\varphi]$ in \eqref{eq:actionmodesApp} is usually neglected and hence the property $K_\mathbf{n}=K_\mathbf{-n}$ suffices for the free theory \eqref{eq:HfreeAppendix}. In the same spirit, one here requires the coupled modes to have the same kinetic terms when retaining $V[\varphi]$ and treating it \`a la Bogolyubov. {This is a strong requirement, adopted for analytical tractability rather than on physical grounds, and we do not expect it to hold in realistic models where the kinetic coefficients typically vary with the representation labels.} At the same time, we stress that property \eqref{eq:newassumption} implies that only special subsets of out-of-condensate modes have the same kinetic terms; in particular, there still are infinitely many classes of modes that have different kinetic couplings. If we represent abstract mode labels as points in an infinite ``$\mathbb{Z}^4$-grid'', then interactions between modes only happen via the involutions \eqref{eq:Tas} and only relate the modes that belong to the graph depicted in figure \ref{fig:bi-chain}, which we dub ``bi-chain''. To understand this in detail, first note that the repeated action of multiple $T_\mathbf{a}$ maps yields
\begin{equation}\label{eq:compositionTi}
	\begin{aligned}
	T_{\mathbf{a_1}}\circ T_{\mathbf{a_2}}\circ \dots \circ T_{\mathbf{a_k}} (\mathbf{n}) &= (-1)^k \mathbf{n} + 2 \sum_{i=1}^k (-1)^{i-1} \mathbf{a_i}\\ & = \begin{cases}
		\mathbf{n }+ 2 (\mathbf{a_1}-\mathbf{a_2}+\mathbf{a_3}- \cdots -\mathbf{a_k}) \quad &k\;\text{even}\,,\\
		-\mathbf{n}+2(\mathbf{a_1}-\mathbf{a_2}+\mathbf{a_3}-\cdots +\mathbf{a_k}) \quad &k\;\text{odd}\,.
	\end{cases}
		\end{aligned}
\end{equation}
Thus, on the one hand, one has infinitely long bi-chains of modes, where a given mode {$\mathbf{n}$ is coupled {directly} only to the shifted and reflected labels $-\mathbf{n}\pm2\mathbf{n_0}$ and $-\mathbf{n}$}. On the other hand, one has infinitely many such bi-chains that, crucially, are independent from each another. This motivates us to visualise $\mathbb{Z}^4$ as disjoint pieces, as we explain below.

More formally, let us take $G\coloneqq\langle T_\mathbf{0}, T_\mathbf{n_0}, T_\mathbf{-n_0}\rangle$ to be the subgroup of the permutation group of $\mathbb{Z}^4$ generated by the three involutions in \eqref{eq:Tas}.\footnote{\label{footnote:IDG}{This} is the so-called \textit{infinite dihedral group}, which is usually presented as $D^{xy}_{\infty} : =\langle x,y | x^2 = y^2 =1\rangle$ or as $D^{rs}_{\infty} : = \langle r,s | s^2=1, srs=r^{-1}\rangle$. $D_\infty$ is a two-dimensional group; indeed one of the $T_\mathbf{\pm n_0}$ maps in \eqref{eq:Tas} is redundant as $T_\mathbf{\pm n_0}= T_\mathbf{0} T_\mathbf{\mp n_0} T_\mathbf{0}$. The group $G$ can then be cast in those specific (``$xy$'' and ``$rs$'') presentations by defining translation elements as $\tau_\mathbf{\pm n_0} = T_\mathbf{0} \circ T_\mathbf{\pm n_0}$ (such that $\tau_\mathbf{\pm n_0}(\mathbf{n}) = \mathbf{n} \pm 2 \mathbf{n_0}$). Note that these define a (normal) translation subgroup $\langle \tau\rangle \cong  \mathbb{Z}$ of $D_\infty$. Since $T_\mathbf{a}^2=1$ and one also checks that $\tau_\mathbf{n_0} \tau_\mathbf{-n_0} =1$ and $T_\mathbf{0} \tau_\mathbf{\pm n_0} T_\mathbf{0} = \tau_\mathbf{\mp n_0} = \tau_\mathbf{\pm n_0}^{-1}$, then $ \langle T_\mathbf{0}, T_\mathbf{ n_0} \rangle$ and $\langle  \tau_\mathbf{ n_0}, T_\mathbf{0} \rangle$ provide precisely the standard presentations of the infinite dihedral group $D_\infty$.} Then for any $\mathbf{n}\in\mathbb{Z}^4$ one defines the orbit
\begin{equation}\label{eq:orbit}
 O (\mathbf{n}) := \{W(\mathbf{n}) \, |\, W\in G\} = \{ \pm \mathbf{n} +  2k \mathbf{n_0}  \, |\, k \in \mathbb{Z}\} \,,
\end{equation}
which is the set of all points in $\mathbb{Z}^4$ one can reach by applying any sequence and any number of $T_\mathbf{a}$'s. This provides an equivalence relation
\begin{equation}\label{eq:equivalencerelation}
	\mathbf{n} \sim \mathbf{n}' \qquad \Longleftrightarrow \qquad \mathbf{n'} \in   O(\mathbf{n})\,,
\end{equation}
where each equivalence class $  O(\mathbf{n})$ is exactly one of those bi-chains depicted in figure \ref{fig:bi-chain}. Proving that orbits are disjoint is fairly straightforward, as one simply needs to check that for a generic element, say $\mathbf{x}=\pm \mathbf{n} + 2 k \mathbf{n_0}$, the maps $T_\mathbf{a}$ preserve the orbit. Explicitly,
\begin{equation}
	T_\mathbf{a}( \mathbf{x}) = T_\mathbf{a} (\pm \mathbf{n} + 2 k \mathbf{n_0}) = \mp \mathbf{n} + 2(\mathbf{a}-k \mathbf{n_0}) \,,
\end{equation}
and since $\mathbf{a}=\{\mathbf{0}, \mathbf{n_0}, \mathbf{-n_0}\}$, one can again write the last term as some integer $k' \in \mathbb{Z}$ multiplying $\mathbf{n_0}$ as
\begin{equation}
	T_\mathbf{a}( \mathbf{x}) = \mp \mathbf{n} + 2 k' \mathbf{n_0}  \in   O(\mathbf{n})\,,
\end{equation}
where, specifically, $k'=\{-k\,,1-k\,,-1-k\}$ in the three cases respectively. Because all three generators $T_\mathbf{a}$ preserve the orbit, any composition \eqref{eq:compositionTi} does as well. This allows us to write
\begin{equation}\label{eq:partition1}
	\mathbb{Z}^4 = \bigsqcup_{[\mathbf{n}]\in \mathbb{Z}^4/\sim}   O(\mathbf{n})\,,
\end{equation}
where $[\mathbf{n}]$ denotes the equivalence class of $\mathbf{n}$ under the relation $\sim$ defined in \eqref{eq:equivalencerelation} and \eqref{eq:orbit}. Note that $\sim$ takes both the shifts and the reflections into account (technically, we mod out by $\mathbb{Z} \rtimes \mathbb{Z}_2$).

\begin{figure}[h]
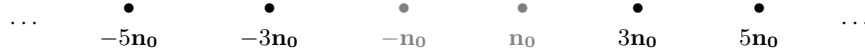

	\centering
\[
\begin{array}{ccccccc}
    \cdots\qquad
	\tikzmarknode{A1}{\begin{array}{c}
			\bullet\\
			 - 5\mathbf{n_0}
	\end{array}}
	& \qquad
	\tikzmarknode{A2}{\begin{array}{c}
			\bullet\\
			 - 3\mathbf{n_0}
	\end{array}}
	& \qquad
	\tikzmarknode{A3}{\begin{array}{c}
		{\color{gray}	\bullet}\\
		{\color{gray}	-\mathbf{n_0}}
	\end{array}}
	& \qquad
	\tikzmarknode{A4}{\begin{array}{c}
			{\color{gray}	\bullet}\\
			{\color{gray} \mathbf{n_0}}
	\end{array}}
	& \qquad
	\tikzmarknode{A5}{\begin{array}{c}
			\bullet\\
			3\mathbf{n_0}
	\end{array}}
	& \qquad
	\tikzmarknode{A6}{\begin{array}{c}
			\bullet\\
			5\mathbf{n_0}
	\end{array}}\qquad\cdots
\end{array}
\]\caption{\small The orbit $  O(\bn_0)$ can be represented as a single chain due to the simplification of the involutions $T_{\mathbf{a}}(\bn_0)$, which only relate odd multiples of the condensates amongst themselves. Removing the condensate labels $\pm \bn_0$ (grey dots) one is left with the severed set $\mathcal{M}$ defined in \eqref{eq:M}.}
\label{fig:condensatechain}
\end{figure}

A comment on the ``condensate orbit'' $  O(\mathbf{n}_0)$ is now in order. First, this is not a bi-chain with distinct top and bottom rows; rather, its structure collapses to a single chain containing all odd multiples of $\pm\mathbf{n}_0$ (see figure \ref{fig:condensatechain}). Importantly, in order to describe the out-of-condensate modes and diagonalise their dynamics, we need to exclude the condensate labels themselves from $  O(\mathbf{n}_0)$, as mentioned below \eqref{eq:Tas}. More precisely, we will consider the \textit{severed} set
\begin{equation}\label{eq:M}
\mathcal{M} :=   O(\mathbf{n}_0) \setminus \{\pm \mathbf{n}_0\} \,,    
\end{equation}
which together with all the other generic orbits $  O(\mathbf{n})$ describe all possible modes, excluding the condensate.

We have seen that a generic label $\mathbf{n'}$ that is not related to $\mathbf{n}$ via the maps \eqref{eq:Tas} forms a separate orbit $  O(\mathbf{n}')$. Indeed, \eqref{eq:partition1} is a way to partition the whole set of modes into disjoint \textit{cosets}. Inspired by this partition, we now abandon the notation $\mathbf{n}\in \mathbb{Z}^4$, and relabel the modes so as to reflect the splitting into disjoint pieces. We first introduce a coordinate system and discuss the diagonalisation procedure for all the generic bi-chains $  O(\mathbf{n})$ except for $\mathcal{M}$, and later comment on the remaining severed chain, which needs a separate (but fully parallel) diagonalisation procedure.

For any generic orbit $  O(\mathbf{n})$ such that $\bn \notin \mathcal{M}$, we define the label $\mathbf{s} $ as the \textit{coset representative} that uniquely identifies which bi-chain a given mode belongs to. Next, one identifies a binary parameter $\varepsilon$ and an integer $m$ that together ``coordinatise'' the position on a given bi-chain. Then, every point of the orbit can uniquely be written as 
\begin{equation}\label{eq:smlabels}
	\mathbf{n} =  \varepsilon \, \mathbf{s} + 2m\bn_0 \,, \qquad\qquad \mathbf{s} \in \mathcal{S}\setminus\{[\bn_0]\} \,, \quad  m \in \mathbb{Z} \,, \quad \varepsilon \in\{-1,1\}\,, 
\end{equation}
where $\mathcal{S}:=\mathbb{Z}^4/\sim$ is the set of classes which distinguish the bi-chains from each other (i.e., it contains only one representative for each one) and indeed the label $\mathbf{s}$ is ``blind'' to both reflections and shifts. Then, we can relabel the orbits \eqref{eq:orbit} as $  O(\textbf{s})$ and split the space of all modes \eqref{eq:partition1} as
\begin{equation}\label{eq:partition2}
	\mathbb{Z}^4 = \bigsqcup_{\mathbf{s}\in \mathcal{S}}   O(\mathbf{s})	\,.
\end{equation}

Clearly, \eqref{eq:smlabels} is just a convenient relabelling, and we are still considering all possible modes in $\mathbb{Z}^4$ (except for those in $\mathcal{M}$, which will be addressed below). From now on, every $\mathbf{n}$ gets identified by a triple $(\mathbf{s}, m,\varepsilon)$ where $\mathbf{s}$ is the canonical representative\footnote{This is just a matter of bookkeeping, standard jargon in set theory. Essentially, for every bi-chain one can identify a unique ``anchor'' or ``origin'', so that $\mathbf{s}$ represents that bi-chain only.} for each coset (and tells us on which bi-chain we are), $\varepsilon$ specifies whether we are on the ``top'' or ``bottom'' row of the bi-chain, and $m$ is an integer telling us where exactly we are ``along'' that bi-chain. For example the field and momentum operators now get new labels: $	\hat{\varphi}_\mathbf{n}  \rightarrow \hat{\varphi}_{\mathbf{s},m,\varepsilon}$ and $	\hat{\pi}_\mathbf{n}  \rightarrow \hat{\pi}_{\mathbf{s},m,\varepsilon}$. In the new labels, the reality condition $\hat{\varphi}_\bn^\dagger=\hat{\varphi}_{-\bn}$ reads $\hat{\varphi}_{\mathbf{s}, m,\varepsilon}^\dagger = \hat{\varphi}_{\mathbf{s},-m,-\varepsilon}$, and the canonical commutator \eqref{eq:originalCCR} becomes $[\hat{\varphi}_{\mathbf{s},m,\varepsilon}, \hat{\pi}_{\mathbf{s'},m',\varepsilon'}] = {\rm i} \delta_{\mathbf{s},\mathbf{s'}}\delta_{m,m'}\delta_{\varepsilon,\varepsilon'}$.

Importantly, all the terms in the Hamiltonian \eqref{eq:H2nondiag} now adopt the $(\mathbf{s}, m,\varepsilon)$ bookkeeping. Indeed using \eqref{eq:smlabels} and the property \eqref{eq:newassumption} one now rewrites the kinetic terms as
\begin{equation}\label{eq:Ks}
K^{(0)}_\mathbf{n} \; \rightarrow \; K^{(0)}_\mathbf{s} \,, \qquad K^{(2)}_\mathbf{n} \; \rightarrow \; K^{(2)}_\mathbf{s} \,,
\end{equation}
and, crucially, one finds that the field and momentum modes couple as (using the reality condition)
\begin{equation}\label{eq:couplingsm}
	\begin{aligned}
    |\hat{\pi}_\mathbf{n}|^2  &= \hat{\pi}_\mathbf{n} \hat{\pi}_{\mathbf{-n}}&&\; \rightarrow\; &  \hat{\pi}_{\mathbf{s},m,\varepsilon} \hat{\pi}_{\mathbf{s}, -m,-\varepsilon} &= |\hat{\pi}_{\mathbf{s},m,\varepsilon} |^2\,,\\
		|\hat{\varphi}_\mathbf{n}|^2&=\hat{\varphi}_\mathbf{n} \hat{\varphi}_{\mathbf{-n}}&&\; \rightarrow\;  & \hat{\varphi}_{\mathbf{s},m,\varepsilon} \hat{\varphi}_{\mathbf{s}, -m,-\varepsilon} &= |\hat{\varphi}_{\mathbf{s},m,\varepsilon}|^2\,, \\  	\hat{\varphi}_\mathbf{n} \hat{\varphi}_{\mathbf{n}\mp2\mathbf{n}_0}^\dagger&=\hat{\varphi}_\mathbf{n} \hat{\varphi}_{\mathbf{-n}\pm 2 \mathbf{n}_0}&&\; \rightarrow\; &\hat{\varphi}_{\mathbf{s},m,\varepsilon} \hat{\varphi}_{\mathbf{s}, -m\pm1,-\varepsilon} &=   \hat{\varphi}_{\mathbf{s},m,\varepsilon} \hat{\varphi}^\dagger_{\mathbf{s},m\mp 1,\varepsilon} \,.
	\end{aligned}
\end{equation}
Explicitly, the relevant contributions to the Hamiltonian \eqref{eq:H2Appendix} in these labels can be written as
\begin{equation}\label{eq:pluginhere}
\hat{H}_2^{\cancel{\mathcal{M}}} =	-\frac{1}{2}\sum_{\mathbf{s},m,\varepsilon} 
\left[ \left( \frac{|\hat{\pi}_{\mathbf{s},m,\varepsilon} |^2}{K_\mathbf{s}^{(2)}} +K_\mathbf{s}^{(0)} |\hat{\varphi}_{\mathbf{s},m,\varepsilon}|^2\right)+ \frac{\lambda}{2} \Phi^2  \big(2 |\hat{\varphi}_{\mathbf{s},m, \varepsilon}|^2 + \hat{\varphi}_{\mathbf{s},m,\varepsilon} \hat{\varphi}^\dagger_{\mathbf{s},m+ 1,\varepsilon}+ \hat{\varphi}_{\mathbf{s},m,\varepsilon} \hat{\varphi}^\dagger_{\mathbf{s},m- 1,\varepsilon}\big)\right]   \,,
\end{equation}
where the notation $\hat{H}_2^{\cancel{\mathcal{M}}}$ reminds us that this is the Hamiltonian for all non-condensate modes that are not in the special set $\mathcal{M}$ (i.e., encompassing all the other bi-chains). It is clear that the only non-diagonal features appearing in the last two terms have been isolated into the integer $m$ index. One can see that such labels, say $m$ and $m'$, are always coupled as (dropping the fixed labels $\mathbf{s}$ and $\varepsilon$)
\begin{equation}\label{eq:Kmm}
	\hat{\varphi}_{m} \delta_{m',f(m)} \hat{\varphi}_{m'} \,, \qquad f(m)=-m+a\,, \qquad a=\{0\,,1\,,-1\}\,,
\end{equation}
where $f(m)$ is an affine involution (essentially the one-dimensional leftover of \eqref{eq:Tas}), with $a=0$ corresponding to pure reflections. This structure naturally suggests to introduce a one-dimensional discrete {Fourier} transform {along each coset} (or Floquet--Bloch transform). Specifically, for each fixed $\mathbf{s}$ and $\varepsilon$, we now perform a ``chain-Fourier transform'' over the label $m\in\mathbb{Z}$ as 
\begin{equation}\label{eq:Fourier}
	\hat{\varphi}_{\mathbf{s},\varepsilon} (\theta) := \frac{1}{\sqrt{2 \pi}} \sum_{m \in \mathbb{Z}} \ee^{- {\rm i} m \theta} \hat{\varphi}_{\mathbf{s},m,\varepsilon}\,, \qquad  	\hat{\pi}_{\mathbf{s},\varepsilon}(\theta) := \frac{1}{\sqrt{2 \pi}} \sum_{m \in \mathbb{Z}} \ee^{ {\rm i} m \theta} \hat{\pi}_{\mathbf{s},m,\varepsilon}\,,
\end{equation}
which naturally introduces a continuous parameter $\theta \in [0\,, 2\pi)$ called ``Bloch angle'' in condensed matter literature. The inverse transforms read
\begin{equation}\label{inverseFT}
	\hat{\varphi}_{\mathbf{s},m,\varepsilon} = \frac{1}{\sqrt{2 \pi}} \int_0^{2\pi} \dd \theta \, \ee^{{\rm i} m \theta} \, \hat{\varphi}_{\mathbf{s},\varepsilon} (\theta) \,, \qquad \hat{\pi}_{\mathbf{s},m,\varepsilon} = \frac{1}{\sqrt{2 \pi}} \int_0^{2\pi} \dd \theta \, \ee^{- {\rm i} m \theta} \, \hat{\pi}_{\mathbf{s},\varepsilon} (\theta)\,.
\end{equation}
Since a Fourier transform is a canonical transformation, one quickly checks that the canonical commutation relations are preserved as $\left[\hat{\varphi}_{\mathbf{s},\varepsilon} (\theta) , \hat{\pi}_{\mathbf{s'},\varepsilon'}(\theta')\right] = {\rm i} \delta_{\mathbf{s},\mathbf{s'}} \delta(\theta-\theta')\delta_{\varepsilon,\varepsilon'}$; moreover, the reality condition becomes $\hat{\varphi}^\dagger_{\mathbf{s},\varepsilon}(\theta) = \hat{\varphi}_{\mathbf{s},-\varepsilon}(\theta)$. 

It is easy to see that in this basis the Hamiltonian becomes diagonal. Indeed, if we plug \eqref{inverseFT} into \eqref{eq:pluginhere} we find
\begin{equation}\label{magicofFourier}
	\begin{aligned}
	\sum_m \hat{\varphi}_{\mathbf{s},m,\varepsilon} \hat{\varphi}^\dagger_{\mathbf{s},m-a,\varepsilon} &=	\sum_m \hat{\varphi}_{\mathbf{s},m,\varepsilon} \hat{\varphi}_{\mathbf{s},-m+a,-\varepsilon}  =	\frac{1}{{2\pi}}\sum_m  \int \dd \theta  \dd \theta' \, \ee^{{\rm i} m \theta}\ee^{{\rm i}(a-m) \theta'} \hat{\varphi}_{\mathbf{s},\varepsilon}(\theta) \hat{\varphi}_{\mathbf{s},-\varepsilon}(\theta')  
	\\& = \int \dd \theta \, \ee^{{\rm i} a \theta}\, \hat{\varphi}_{\mathbf{s},\varepsilon}(\theta) \hat{\varphi}_{\mathbf{s},-\varepsilon}(\theta) = \int \dd \theta \, \ee^{{\rm i} a \theta}\, \hat{\varphi}_{\mathbf{s},\varepsilon}(\theta) \hat{\varphi}^\dagger_{\mathbf{s},\varepsilon}(\theta) 
    = \int \dd \theta \, \ee^{{\rm i} a \theta}\, |\hat{\varphi}_{\mathbf{s},\varepsilon}(\theta)|^2\,,
	\end{aligned}
\end{equation}
where $a=\{0\,,1\,,-1\}$, and we used the reality conditions as well as the property $ \sum_m \ee^{{\rm i} m(\theta-\theta')} = 2 \pi \delta (\theta-\theta')$.\footnote{This holds as a distribution for $2\pi$-periodic functions $f$ on $[0\,,2\pi)$ satisfying $\int \dd \theta' f(\theta') \sum \ee^{{\rm i} m(\theta-\theta')}= 2\pi f(\theta)$.} Then, assuming that kinetic operators are invariant along each orbit, i.e., using \eqref{eq:Ks},\footnote{Without property \eqref{eq:Ks} the Fourier transform of the kinetic part would read
\begin{equation*}
		\sum_m \frac{1}{K^{(2)}_{m}} \hat{\pi}_{\mathbf{s},m,\varepsilon} \hat{\pi}_{\mathbf{s},-m,-\varepsilon} =	 \frac{1}{{2\pi}} \int \dd \theta  \dd \theta' \,  \hat{\pi}_{\mathbf{s},\varepsilon}(\theta) \hat{\pi}_{\mathbf{s},-\varepsilon}(\theta')  \sum_m \frac{1}{K^{(2)}_{m}} \ee^{{\rm i} m(\theta-\theta')}\,,
\end{equation*}
and
\begin{equation*}
\sum_m K_{m}^{(0)} \hat{\varphi}_{\mathbf{s},m,\varepsilon} \hat{\varphi}_{\mathbf{s},-m,-\varepsilon} =	 \frac{1}{{2\pi}} \int \dd \theta  \dd \theta' \,  \hat{\varphi}_{\mathbf{s},\varepsilon}(\theta) \hat{\varphi}_{\mathbf{s},-\varepsilon}(\theta')  \sum_m K_{m}^{(0)} \ee^{{\rm i} m(\theta-\theta')} \,.
\end{equation*}
Only if the kinetic term does not depend on $m$, one can write the whole Hamiltonian in diagonal form using $ \sum_m \ee^{{\rm i} m(\theta-\theta')} = 2 \pi \delta (\theta-\theta')$. {The more general case of site-dependent coefficients would need to be treated numerically or perturbatively around the
uniform case. Importantly, the existence of the collective sector would not be affected, as it only requires mixing of creation and annihilation operators.}} the Hamiltonian becomes
\begin{equation}\label{eq:diagthetanotepsilon}
\hat{H}_2^{\cancel{\mathcal{M}}}  = -\frac{1}{2} \sum_{\mathbf{s}, \varepsilon}\int_0^{2\pi} \dd \theta \, \left[
\frac{|\hat{\pi}_{\mathbf{s},\varepsilon}(\theta) |^2 }{K_\mathbf{s}^{(2)}} +\left( K_\mathbf{s}^{(0)}  
 +{\lambda}\Phi^2 f(\theta) \right) |\hat{\varphi}_{\mathbf{s},\varepsilon}(\theta) |^2\right]\,,
\end{equation}
where $f(\theta)\coloneqq 1+\cos\theta$ is the analogue of a ``Bloch factor'' which appears due to the chain-Fourier transform.

We finally return to the only remaining chain of modes, the severed set $\mathcal{M}$ defined in \eqref{eq:M}, which has been excluded from the discussion so far. Since the full quadratic Hamiltonian \eqref{eq:H2Appendix} naturally decomposes into independent chain-sectors, we write $\hat{H}_2 =\hat{H}_2^\mathcal{M}+\hat{H}_2^{\cancel{\mathcal{M}}}$, where $\hat{H}_2^{\cancel{\mathcal{M}}}$ describes all possible bi-chains except for the set $\mathcal{M}$, as previously noted. To complete the description, we only need to address the last sector $\hat{H}_2^\mathcal{M}$, for $\bn \in \mathcal{M}$. While the procedure largely mirrors our previous treatment, the removal of the condensate labels from $  O(\bn_0)$ effectively imposes Dirichlet boundary conditions at the `ends' of the severed chain, and will require the use of a sine Fourier transform instead of the standard Fourier transform. In what follows, we quickly show how to diagonalise $\hat{H}_2^\mathcal{M}$ mainly emphasising the minor difference with the methods employed above for the rest of the modes. 

Due to the special structure of $\mathcal{M}$ (see figure \ref{fig:condensatechain}), we introduce the following parametrisation:
\begin{equation}
    \bn = \varepsilon (1+ 2 m) \bn_0 \,, \qquad m \in \mathbb{Z}_{>0}\,, \qquad \varepsilon\in \{-1,+1\}\,.
\end{equation}
Every mode $\bn \in \mathcal{M}$ is then uniquely identified with a pair $(m,\varepsilon)$, where the \textit{positive} integer $m\in \{1,2,3,\dots\}$ counts the number of steps in the chain and the binary parameter $\varepsilon$ specifies whether we are in the left or right branch (i.e., the overall sign). In these labels, we have $[\hat{\varphi}_{m,\varepsilon}, \hat{\pi}_{m',\varepsilon'}] = {\rm i} \delta_{m,m'}\delta_{\varepsilon,\varepsilon'}$ and $\hat{\varphi}_{ m,\varepsilon}^\dagger = \hat{\varphi}_{m,-\varepsilon}$ (note that conjugation simply relates the positive branch to the negative branch of the chain, cf.~figure \ref{fig:condensatechain}). Just like for the rest of the modes, one rewrites the corresponding Hamiltonian
\begin{equation}\label{eq:Mpluginhere}
    \hat{H}_2^\mathcal{M} = 	-\frac{1}{2}\sum_{m,\varepsilon} 
\left[ \left( \frac{|\hat{\pi}_{m,\varepsilon} |^2}{K_\mathcal{M}^{(2)}} +K_\mathcal{M}^{(0)} |\hat{\varphi}_{m,\varepsilon}|^2\right)+ \frac{\lambda}{2} \Phi^2  \big(2 |\hat{\varphi}_{m, \varepsilon}|^2 + \hat{\varphi}_{m,\varepsilon} \hat{\varphi}^\dagger_{m+\varepsilon,\varepsilon}+ \hat{\varphi}_{m,\varepsilon} \hat{\varphi}^\dagger_{m- \varepsilon,\varepsilon}\big)\right]   \,,
\end{equation}
showing explicitly that the last two terms encode the off-diagonal features in the $m$ index (but notice the structure is different than in \eqref{eq:pluginhere}). We now perform a Fourier {sine} transform for the field operator
\begin{equation}\label{eq:sineFT}
    \hat{\varphi}_\varepsilon(\theta)=\sqrt{\frac{2}{\pi}} \sum_{m=1}^\infty \sin(m \theta) \,\hat{\varphi}_{m,\varepsilon} \,, \qquad \hat{\varphi}_{m,\varepsilon} =\sqrt{\frac{2}{\pi}} \int_0^\pi \dd \theta\,\sin(m\theta)\, \hat{\varphi}_\varepsilon(\theta)\,,
\end{equation}
and similarly for the momentum operator, where $\theta \in (0,\pi)$ to reflect the vanishing of the basis functions at the boundaries. The transform preserves the canonical commutation relations $[\hat{\varphi}_\varepsilon(\theta) , \hat{\pi}_{\varepsilon'}(\theta')]= {\rm i} \delta_{\varepsilon,\varepsilon'}\delta(\theta-\theta')$, and lets us write the reality condition as $\hat{\varphi}^\dagger_\varepsilon(\theta) = \hat{\varphi}_{-\varepsilon}(\theta)$. By means of the inverse transforms we calculate\footnote{Using the completeness relation $\sum_{m=1}^\infty \sin(m\theta)\sin(m\theta') = \frac{\pi}{2} \delta(\theta-\theta')$, which holds as a distribution for $\theta, \theta' \in (0,\pi)$, and the identity $\sin \left[(m+\varepsilon)\theta\right]+ \sin \left[(m-\varepsilon) \theta\right] = 2 \sin (m \theta) \cos(\varepsilon\theta) = 2 \sin (m \theta) \cos(\theta)$, with $\varepsilon=\pm 1$.}
\begin{equation}
    \sum_{m=1}^\infty \hat{\varphi}_{m,\varepsilon}  \left(2\hat{\varphi}_{m,-\varepsilon}+  \hat{\varphi}_{m+\varepsilon,-\varepsilon} + \hat{\varphi}_{m-\varepsilon,-\varepsilon} \right)=2 \int_0^\pi \dd \theta \,(1 +  \cos \theta) \, \hat{\varphi}_\varepsilon(\theta)\hat{\varphi}_{-\varepsilon}(\theta)\,,
\end{equation}
so that \eqref{eq:Mpluginhere} can be rewritten as
\begin{equation}\label{eq:diagM}
    \hat{H}_2^\mathcal{M} =  -\frac{1}{2} \sum_{ \varepsilon}\int_0^{\pi} \dd \theta \, \left[
\frac{|\hat{\pi}_{\varepsilon}(\theta) |^2 }{K_\mathcal{M}^{(2)}} +\left( K_\mathcal{M}^{(0)}  
 +{\lambda}\Phi^2 f(\theta) \right) |\hat{\varphi}_{\varepsilon}(\theta) |^2\right]\,.
\end{equation}

Finally, we associate a coset label ${\mathbf{s}_0}$ to the contribution \eqref{eq:diagM}; this is done by renaming $K_\mathcal{M} = K_{\mathbf{s}_0}$ and by denoting the field and momentum operators for modes in $\mathcal{M}$ as $\hat{\varphi}_{{\mathbf{s}_0},\varepsilon}(\theta)$ and $\hat{\pi}_{{\mathbf{s}_0},\varepsilon}(\theta)$. Then, we can include $\hat{H}_2^\mathcal{M}= \hat{H}_2^{\mathbf{s}_0}$ together with the rest of the cosets contributions \eqref{eq:diagthetanotepsilon} in the total quadratic Hamiltonian \eqref{eq:H2Appendix}, obtaining
\begin{equation}\label{eq:finalHamprenu}
 \hat{H}_2=   -\frac{1}{2} \sum_{\mathbf{s}, \varepsilon}\int_{\mathcal{B}_{\mathbf{s}}}\dd \theta \, \left[
\frac{|\hat{\pi}_{\mathbf{s},\varepsilon}(\theta) |^2 }{K_\mathbf{s}^{(2)}} +\left( K_\mathbf{s}^{(0)}  
 +{\lambda}\Phi^2 f(\theta) \right) |\hat{\varphi}_{\mathbf{s},\varepsilon}(\theta) |^2\right]\,,
\end{equation}
where $\mathcal{B}_\mathbf{s} = [0,2\pi)$ for all generic bi-chains $\mathbf{s}\neq \mathbf{s}_0$, and $\mathcal{B}_{\mathbf{s}_0} = (0,\pi)$.

 In order to lighten the notation, we now define the multi-index $\pm{\boldsymbol{\nu}} :=\{\mathbf{s}\,,\theta\,, \pm\varepsilon\}$ which conveniently describes out-of-condensate modes in a way that is reminiscent of the old notation for the modes $\mathbf{n\neq \pm n_0}$. With this notation, the canonical structure and reality condition read $	[\hat{\varphi}_{\boldsymbol{\nu}},\hat{\pi}_{\boldsymbol{\nu'}}]
	={\rm i}\,\delta_{\boldsymbol{\nu},\boldsymbol{\nu'}}$ and $\hat{\varphi}_{\boldsymbol{\nu}}^\dagger=\hat{\varphi}_{-\boldsymbol{\nu}}$, respectively. Also, \eqref{eq:finalHamprenu} becomes
	\begin{equation}
			\hat{H}_2 =  \intsum\,
			\hat{H}_{\boldsymbol{\nu}} \,, \qquad \hat{H}_{\boldsymbol{\nu}}  = -\frac{1}{2}  \left(
		\frac{|\hat{\pi}_{\boldsymbol{\nu}} |^2}{K_{\boldsymbol{\nu}} ^{(2)}}  + \left( K_{\boldsymbol{\nu}} ^{(0)}
		+{\lambda}\Phi^2  f_{\boldsymbol{\nu}}  \right) |\hat{\varphi}_{\boldsymbol{\nu}} |^2\right)\,,
	\end{equation}
	where we introduced the ``sum-integral'' notation $\intsumTEXT_{\boldsymbol{\nu}} := \sum_{\mathbf{s},\varepsilon} \int_{\mathcal{B}_\mathbf{s}} \dd \theta $, and we denoted $f_{\boldsymbol{\nu}} = f(\theta)$ and $K_{\boldsymbol{\nu}} = K_\mathbf{s}$. This is nothing but \eqref{eq:HamiltonianNU} in the main text.
    
    This Hamiltonian is structurally identical to the free-theory scenario \eqref{eq:HfreeAppendix}, with the difference that $K_{\mathbf{n}}^{(0)}$ is replaced by the quantity $ K_{\boldsymbol{\nu}}^{(0)} 
+{\lambda}\Phi^2  f_{\boldsymbol{\nu}}$. 
Of course, by setting $\lambda=0$ one recovers the Hamiltonian \eqref{eq:HfreeAppendix}, used in section \ref{sec:QGcosmo} to derive the GFT condensate cosmology results. In fact, the label $\bnu$ itself reduces exactly to $\bn$ when $\lambda=0$. As we discuss in the main text, this diagonalisation procedure corresponds to a nontrivial Bogolyubov transformation which mixes the atomic ladder operators to define collective excitations.

{As a final remark, we note that while we have worked throughout with infinite chains, which simplifies the diagonalisation, this can be seen as an idealisation. In practice, only finitely many representation labels are usually retained in explicit GFT computations, so that the
chains are of finite length. The procedure above is unaffected: the Fourier transforms are replaced by their discrete counterpart, with $\theta$ restricted to the values $\theta_k=k\pi/(L+1)$ (with $k=1,\dots,L$, for a chain of $L$ sites), while the factor $f(\theta)=1+\cos\theta$ is unchanged. The idealisation therefore affects the technical details of the diagonalisation, but not its outcome. The same applies to any other restriction of the set of modes, such as
discarding degenerate configurations with a vanishing representation label, which
likewise cuts the chains into shorter segments and is handled in the same way.}

\section{Validity of neglecting interactions at the background level within GFT condensate cosmology}
\label{app:ApproximationVaidity}

In this appendix we carefully assess what conditions have to be satisfied in order to neglect the linear term \eqref{eq:H1nondiag} in the Hamiltonian for the fluctuations, which arises because we are expanding around a free-theory mean field, instead of expanding around a mean field of the full theory. We will present the argument first for a generic scalar field theory, and then particularise the conditions for GFT.

Let us start with an action for a real scalar field $\hat S[\hat\varphi]=\hat S_{F}[\hat\varphi]+\lambda \hat S_{I}[\hat\varphi]$. Here $\hat S_{\rm F}$ is the free theory action, quadratic in the fields, and $\hat S_{ I}$ is a generic interaction term containing beyond quadratic terms in the field, and controlled by the coupling $\lambda$. The dynamics of the theory is found from $\delta\hat S/\delta\hat\varphi=0$ which, taking expectation values for a quantum state of our choice gives
\begin{equation}
    \label{eq:ExpValDynamics}
    \left\langle{\frac{\delta \hat S_F}{\delta\hat\varphi}}\right\rangle+\lambda \left\langle{\frac{\delta \hat S_I}{\delta\hat\varphi}}\right\rangle=0 \,.
\end{equation}
We now take a mean field approximation by replacing $\hat\varphi\to\phi_0\coloneqq\ev{\hat\varphi}$ and we replace expectation values of nonlinear functions of $\hat\varphi$ by the same functions of $\phi_0$, namely $\langle\hat{F}[\hat\varphi]\rangle\to F[\phi_0]$. This is a good approximation of \eqref{eq:ExpValDynamics} for quantum states where the fluctuations can be neglected over the mean (e.g., coherent states), and leads to the mean field equations
\begin{equation}
    \label{eq:AppMeanField}
   \left.\frac{\delta S_F}{\delta\varphi}+\lambda \frac{\delta  S_I}{\delta\varphi}\right|_{\varphi=\phi_0}=0 \,.
\end{equation}
We are interested in computing quantum fluctuations on top of an approximated mean field solution $\varphi_0$, which does not solve the above equation, but rather the mean field equations of the free theory, namely \eqref{eq:AppMeanField} with $\lambda\to0$. To do this consistently, we need to make sure that the corrections to the free mean field configuration due to the interaction term are parametrically smaller than the quantum fluctuations on top of the free mean field that we aim to compute. This will occur only in a region of parameter space. To characterise it, let us write the interacting mean field as $\phi_0=\varphi_0+\delta\varphi_\lambda$. Since $\delta\varphi_\lambda$ vanishes in the limit $\lambda\to0$, and we are interested in weakly coupled theories where $\lambda$ is small, we can plug this ansatz for $\phi_0$ in \eqref{eq:AppMeanField} and expand in powers of $\delta\varphi_\lambda$. Solving formally the resulting equation for $\delta\varphi_\lambda$ leads to
\begin{equation}
\label{eq:SourceContribution}
    \delta\varphi_\lambda=\lambda D_F^{-1} J_I+ O\lr{\delta\varphi_\lambda^2,\lambda^2} \,,
\end{equation}
where we keep explicit the $\lambda$ dependence of neglected terms since we also want to do an expansion in small $\lambda$, and we have defined the effective source $J_I$ and the differential operators $D_F$ as 
\begin{equation}
    J_I\coloneqq -\frac{\delta S_I}{\delta\varphi}\Bigg|_{\varphi=\varphi_0}\,,\qquad D_F\coloneqq\frac{\delta^2 S_F}{\delta\varphi^2}\Bigg|_{\varphi=\varphi_0}\,.
\end{equation}
Note that we need to choose initial conditions for $\delta\varphi_\lambda$ to unambiguously define $D_F^{-1}$.
In general, we can approximate the true interacting mean field $\phi_0$ by the free mean field $\varphi_0$ if $|\delta\varphi_\lambda/\varphi_0|\ll1$, namely
\begin{equation}
    \left|\frac{\lambda D_F^{-1}J_I}{\varphi_0}\right|\ll1 \,.
    \label{eq:ValidityNeglectIntBack}
\end{equation}
We now proceed to compute the leading order quantum corrections to $\varphi_0$ and characterise the regime where $\delta\varphi_\lambda$ provides subleading contributions. To that end, we write $\hat\varphi=\varphi_0\hat{\mathbb{I}}+\delta\hat{\varphi}$. Expanding for small fluctuations we find 
\begin{equation}
    \hat S[\hat\varphi]=S[\varphi_0]\hat{\mathbb{I}}+\left.\frac{\delta S}{\delta\varphi}\right|_{\varphi=\varphi_0}\delta\hat\varphi+\frac{1}{2}\left.\frac{\delta^2S}{\delta\varphi^2}\right|_{\varphi=\varphi_0}\delta\hat\varphi^2+  O(\delta\hat\varphi^3) \,,
\end{equation}
and the linear dynamics for small quantum fluctuations around $\varphi_0$ is then given by
\begin{equation}
   (D_F+\lambda D_I)\delta\hat\varphi=\lambda J_I\hat{\mathbb{I}}+  O(\delta\hat\varphi^2)\,,
\end{equation}
where we have defined the operator
\begin{equation}
    D_I\coloneqq\frac{\delta^2 S_I}{\delta\varphi^2}\Bigg|_{\varphi=\varphi_0}\,.
\end{equation}
If $\varphi_0$ was an exact solution of the interacting mean field equations, the source term on the right hand side would vanish exactly. However, since it is only a solution of the free theory, we have ${\delta S}/{\delta\varphi}|_{\varphi=\varphi_0}=-\lambda J_I$ and the source term remains. We can now split the quantum fluctuations into a homogeneous piece and a sourced piece $\delta\hat\varphi=\delta\hat\varphi_h+\delta\hat\varphi_\lambda$, such that 
\begin{align}
   &\lr{D_F+\lambda D_I}\delta\hat{\varphi}_h=  O(\delta\hat\varphi_h^2,\delta\hat\varphi_h\delta\hat\varphi_\lambda)\,,
   \label{eq:PertFieldHomog}
   \\
   &D_F\delta\hat\varphi_\lambda=\lambda J_I\hat{\mathbb{I}}+ O\lr{\delta\hat\varphi_\lambda^2,\lambda^2}\,.
    \label{eq:PertFieldSource}
\end{align}
Note that $\delta\hat\varphi_\lambda$ is exactly the contribution to the quantum corrections that we want to neglect, which comes from the fact that we are expanding around an approximated mean field (the free mean field $\varphi_0$) instead of the true mean field $\phi_0$. Furthermore, we see that to leading order $\delta\hat\varphi_\lambda=\delta\varphi_\lambda\hat{\mathbb{I}}$, where $\delta\varphi_\lambda$ is the difference between the free mean field and the true interacting mean field. A sufficient condition for the existence of a regime in which $\delta\hat\varphi_\lambda$ can be neglected while meaningfully accounting for  $\delta\hat\varphi_h$ is given by $\sqrt{\ev{\delta\hat\varphi_\lambda^2}}\ll\sqrt{\ev{\delta\hat\varphi_h^2}}$, which we can write as
\begin{equation}
     \left|\frac{{\lambda}D_F^{-1} J_I}{\sqrt{\ev{\delta\hat\varphi_h^2}}}\right|\ll1\,.
     \label{eq:ValidityFreeMFPert}
\end{equation}
If the mean field approximation holds (i.e., $|\sqrt{\ev{\delta\hat\varphi^2}}/\varphi_0|\ll 1$), the above condition is more restrictive than the one required for the interacting mean field to be well approximated by the free mean field \eqref{eq:ValidityNeglectIntBack}. 

We now want to particularise these results to  our GFT with a quartic interaction potential. We start with the GFT action expanded in modes \eqref{eq:GFTFreeActionModesApp} and the interaction potential \eqref{eq:Vstill4}. Adapting the above formalism to the case of the real-valued GFT field, with complex modes satisfying $\varphi_{-\mathbf{n}} = \varphi_\mathbf{n}^*$, we write the free mean field as $\varphi_{0}=\varphi_{\pm\mathbf{n}_0}\to\Phi$ and
\begin{equation}
\begin{aligned}
    D_F&\to D_F^{\mathbf{nm}}(\chi,\chi')=\delta_{\mathbf{nm}}\lr{K^{(0)}_{\mathbf{n}}+K^{(2)}_{\mathbf{n}}\partial_\chi^2}\delta(\chi-\chi')\,,
    \\
    J_I&\to J_I^{\mathbf{n}}(\chi)=-\frac{\Phi^3(\chi)}{2}\lrsq{\delta_{\mathbf{n},{\bn}_0}+\delta_{\mathbf{n},-\mathbf{n}_0}+\frac{1}{3}\delta_{\mathbf{n},3\mathbf{n}_0}+\frac{1}{3}\delta_{\mathbf{n},-3\mathbf{n}_0}}\,.
\end{aligned}
\label{eq:ModeSource}
\end{equation}
In our case $J_I^{\mathbf{n}}$ vanishes for $\mathbf{n}\notin\{\pm \bn_0,\pm3\bn_0\}$, and it depends on relational time only through the GFT mean field $\Phi(\chi)$ of \eqref{eq:PhiFree}. 

We now need to find the unique solution to the corresponding differential equation,
\begin{equation}
\lr{K^{(0)}_{\mathbf{n}}+K^{(2)}_{\mathbf{n}}\partial_\chi^2} \delta\varphi^{\bn}_\lambda(\chi) = \lambda J_I^{\bn}(\chi)\,,
\end{equation}
with initial conditions dictated by \eqref{eq:state} (this fixes the ambiguity in the definition of $D_F^{-1}$ mentioned above). To that end, we apply Duhamel's principle to find the unique solutions for each mode (cf.~\eqref{eq:setsfreetheory}):
\begin{equation}
    \delta\varphi^{\bn}_\lambda=\frac{\lambda}{K_{\bn}^{(2)}\omega_{\bn}}\int_{0}^\chi S_{\bn}(\chi-t) J_I^{\bn}(t) \dd t\,,
    \qquad
    S_{\bn}(\chi-t)=
    \begin{cases}
        \sin(\omega_{\bn}(\chi-t))\,, &\text{for $\mathbf{n}\in   \mathfrak{N}^{\text{HO}}$}\,,
        \\
        \sinh(\omega_{\bn}(\chi-t))\,, &\text{for $\mathbf{n}\in   \mathfrak{N}^{\text{SQ}}$}\,,
    \end{cases}
\end{equation}
where $\omega_{\bn}$ is defined in \eqref{eq:omegafree}. Then, according to \eqref{eq:ModeSource}, since the only modes that are sourced are $\pm{\bn}_0$ and $\pm3{\bn}_0$, the only relevant upper bounds are given by
\begin{equation}
    |\delta\varphi_\lambda^{\pm\bn_0}|\leq|\lambda|\frac{\Phi^3(\chi)}{2}d_F^{\bn_0}(\chi) \,,
     \qquad
      |\delta\varphi_\lambda^{\pm3\bn_0}|\leq|\lambda|\frac{\Phi^3(\chi)}{6}d_F^{ 3\bn_0}(\chi)\,,
\end{equation}
where we have used that $\Phi(\chi)$ is positive and monotonically growing for $\chi>0$, and we have defined
\begin{equation}
\label{eq:DF}
    d^{\bn}_F(\chi)=\frac{1}{|K^{(2)}_{\bn}|\omega_{\bn}}\int_0^\chi|S_{\bn}(\chi-t)|\dd t
    = \frac{1}{|K_{\bn}^{(0)}|}\times
    \begin{cases}
        {2T^{\bn}_2(\chi)+1-\cos \big({\omega_{\bn}\chi-\pi T^{\bn}_2(\chi)}\big)} \,,&\text{for $\mathbf{n}\in   \mathfrak{N}^{\text{HO}}$}\,,
        \\
        {\cosh(\omega_{\bn}\chi)-1}\,, & \text{for $\bn\in\mathfrak{N}^\text{SQ}$}\,,
    \end{cases}
\end{equation}
with $T^{\bn}_2(\chi)=\lfloor\omega_{\bn}\chi/\pi\rfloor$ the floor (or integer part) of $\omega_{\bn}\chi/\pi$ (which basically counts the number of half-oscillation periods of the sine function over the integration interval $[0,\chi]$). Note that the function $d_F^{\bn}$ essentially grows linearly with $\chi$ for HO modes, whereas it grows exponentially in $\chi$ for SQ modes. 

 We can finally write sufficient conditions under which using the free mean field is a valid approximation in terms of the GFT parameters and the number of condensed quanta. By requiring that $|\delta\varphi_\lambda^{\pm\bn_0}| \ll \Phi(\chi)$ and $|\delta\varphi_\lambda^{\pm 3\bn_0}| \ll \Phi(\chi)$, we conclude that the {free} mean field is a good approximation to the interacting mean field at relational time $\chi$ provided the following conditions hold:
    \begin{equation}
    \label{eq:ValidityFreeMF}
        \left|\frac{d_F^{ \bn_0}(\chi)}{2}\lambda \Phi^2(\chi)\right|\ll1\,, \qquad \left|\frac{d_F^{ 3\bn_0}(\chi)}{6}\lambda\Phi^2(\chi)\right|\ll1 \,. 
    \end{equation}
Moreover, from $|\delta\varphi_\lambda^{\pm 3\bn_0}| \ll \sqrt{\ev{\hat\varphi_{\pm 3\bn_0}^2(\chi)}}$ we find that quantum fluctuations on top of the mean field can be computed while neglecting $\delta\hat\varphi_\lambda^{\bn}$ if
    \begin{equation}
    \label{eq:ValidityNeglectSource}
\left|\frac{1}{6}\frac{d_F^{3\bn_0}(\chi)}{\sqrt{\ev{\hat\varphi_{\pm3\bn_0}^2(\chi)}}}\lambda  \Phi^{3}(\chi)\right|\ll1 \,.
    \end{equation}
Using the exact bound \eqref{eq:PhiNcRatio}, these can also be turned into conditions involving the number of quanta $N_\text{c}(\chi)$. Notice how the conditions \eqref{eq:ValidityFreeMF} modify the standard diluteness condition $|\lambda| \Phi^2 (\chi) \ll 1$ with an integral function $d_F(\chi)$, which essentially captures how much the free-theory trajectory deviates from the true interacting mean field as a function of time. It is interesting to also note that since the condensate mode $\bn_0$ is of SQ type (and is assumed to have the largest squeezing rate), $d_F^{\bn_0}(\chi)$ outpaces $d_F^{3\bn_0}(\chi)$ at late times, regardless of whether the mode $3\bn_0$ is of SQ or HO type. As a result, the first condition in \eqref{eq:ValidityFreeMF} is the more restrictive one asymptotically, while the second is subleading (and redundant at late times).
 
Condition \eqref{eq:ValidityNeglectSource}, on the other hand, imposes a different physical constraint. Since the interactions act as a source for the modes $\pm3\bn_0$, whose mean-field background vanishes, \eqref{eq:ValidityNeglectSource} is simply a requirement that the sourced contribution remains smaller than the intrinsic quantum fluctuations of these modes.

\end{document}